\documentclass[11pt,a4paper]{article}
\pdfoutput=1

\usepackage{jheppub}

\usepackage{amsmath}
\usepackage{bbm}
\usepackage{float}
\usepackage{graphicx}	
\usepackage{hyperref}
\usepackage{mathtools}
\usepackage{cancel}

\usepackage[normalem]{ulem}	

\newcommand{\blue}[1]{\color{blue} #1 \color{black}}
\newcommand{\green}[1]{\color{cyan} #1 \color{black}}

\usepackage{multirow}
\usepackage{rotating}
\usepackage{subcaption}

\def\lsim{\raise0.3ex\hbox{$\;<$\kern-0.75em\raise-1.1ex
\hbox{$\sim\;$}}}
\def\gsim{\raise0.3ex\hbox{$\;>$\kern-0.75em\raise-1.1ex
\hbox{$\sim\;$}}}

\def\thetitle{ 
Sterile-active resonance: A global qualitative picture  \\
 \vspace{- 6mm}
}
\title{\thetitle}
\hypersetup{pdftitle={\thetitle}}
\author{Mark Brettell$^{a}$}
\author{Ivan Martinez-Soler$^{a}$}
\author{Hisakazu Minakata$^{b}$}
\affiliation{
$^{a}$Institute for Particle Physics Phenomenology, Durham University, South Road DH1 3LE, Durham, U.K. \\ 
$^{b}$Center for Neutrino Physics, Department of Physics, Virginia Tech, Blacksburg, Virginia 24061, USA \\ }

\emailAdd{mark.brettell@cern.ch,~ivan.j.martinez-soler@durham.ac.uk,~\\ hisakazu.minakata@gmail.com}
\date{\today}

\abstract{

In the $\nu$SM extended by adding an eV-scale sterile state, the $(3+1)$ model, the sterile-active level crossing entails the MSW resonance, here referred as the sterile-active (SA) resonance. In this paper, we construct an effective theory of SA resonance which involves only the sterile-active mixing angles and $\Delta m^2_{41}$, thanks to the given environment of high matter potential which freezes the $\nu$SM oscillations. We give our first attempt at an analytic treatment of the effective theory to illuminate the global picture of the SA resonance at a glance.  
We formulate a perturbative framework in which the structure of ``texture zeros'' of the $S$ matrix in the flavor space and the suppression by the small parameters $\sin \theta_{j 4}$ ($j=1,2,3$) allows us to reveal the flavor$-$event-type hierarchy of the resonance-effect strength in the probabilities. We have shown that the cascade events dominantly comes from the three paths through $P(\nu_{e} \rightarrow \nu_{e})$, $P(\bar{\nu}_{e} \rightarrow \bar{\nu}_{e})$, and $P(\bar{\nu}_{\mu} \rightarrow \bar{\nu}_{\tau})$, and a three-component fit is suggested to disentangle the SA resonance generation mechanisms. 
}

\newcommand{\Dmsqren}{\Delta m^2_{ \text{ren} }}

\begin{document} 

\maketitle

\section{Introduction}
\label{sec:introduction} 

There had been considerable amount of efforts devoted to understand the physics of neutrino flavor transformation~\cite{Kajita:2016cak,McDonald:2016ixn} in the standard three-flavor neutrino system in the neutrino-mass-embedded Standard Model ($\nu$SM). Thanks to the numerous works done in this context, by now, we feel that we have achieved a reasonable success in understanding the phenomena caused by the three neutrino system in vacuum and in matter. See, for example, refs.~\cite{Gonzalez-Garcia:2007dlo,Blennow:2013rca,Maltoni:2015kca} and the references cited therein. 

Possible existence of the SM gauge singlet neutral lepton, the sterile neutrino, signals opening of the new regime. In particular its eV-scale mass version has a long history since the first experimental claim by the LSND collaboration as an interpretation of the $\bar{\nu}_{e}$ excess in their stopped pion source experiment~\cite{LSND:2001aii}. It was followed by the MiniBooNE experiment which uses the booster beam and observed an excess in the both $\nu_{e}$ and $\bar{\nu}_{e}$ appearance channels, which is amount to the total significance of 4.7$\sigma$ CL~\cite{MiniBooNE:2018esg}. 
This reference reports that the confidence level of the combined LSND and MiniBooNE excesses is as high as 6.0$\sigma$. 
The sterile neutrino interpretation of the excess, however, suffers from the tension between the mixing parameter regions preferred by the appearance and disappearance channels, see e.g., refs.~\cite{Dentler:2018sju,Diaz:2019fwt,Dasgupta:2021ies}, and in the mass region~\cite{Hardin:2022muu}. Moreover, there are tensions with cosmological observations, although in this case, new interactions within the sterile sector might alleviate these issues, see e.g., refs.~\cite{Archidiacono:2016kkh,Song:2018zyl}. 

Another types of the neutrino anomalies are also discussed. The authors of ref.~\cite{Mention:2011rk} raised the issue of possible discrepancy between the re-calculated reactor anti-neutrino $\bar{\nu}_{e}$ flux~\cite{Mueller:2011nm,Huber:2011wv} and the experimentally measured value, the ``reactor anti-neutrino anomaly''. It appears that the latter is smaller by $\sim$5\% level. This discrepancy may be relieved by adopting the new measurement of the beta decay spectrum reported in ref.~\cite{Kopeikin:2021ugh}. It might largely settle the issue, see e.g., ref.~\cite{Giunti:2021kab}. 
However, recently, an improved version of the Ga source experiment BEST~\cite{Barinov:2022wfh} reported $4\sigma$ level deficit of $\nu_{e}$ flux from the $^{51}$Cr source. 

The reader may have noticed that our description of sterile search as well as referencing on sterile neutrino physics is minimal, missing almost all the enormous number of references. For more extensive reviews to see the progress in the sterile neutrino analyses, and the description of relevant experiments which are mentioned  only briefly or totally omitted above, we quote refs.~\cite{Diaz:2019fwt,Dasgupta:2021ies,Dentler:2018sju} and the references cited therein. As a guide for reaching the further references beyond, we just cite ref.~\cite{Acero:2022wqg}. 

What is the charm of the sterile neutrino hypothesis? By being a SM gauge singlet fermion, if this is a physical reality, it is likely to be the first alien particle which comes from outside the $\nu$SM. If so it will bring us valuable informations on physics beyond the $\nu$SM. However, as it stands, interpretation of the various anomalies mentioned above by the sterile neutrino hypothesis is far from established. Then, what should we do? For the time being we should make every possible test of the sterile neutrino hypothesis by taking the various concrete model, for example, the three active plus one sterile neutrinos model, the $(3+1)$ model. In this context it is of great value that the two new experiments for sterile search are either underway or coming soon, JSNS2 (an improved LSND-like setting at JPARC)~\cite{JSNS2:2021hyk}, and the Short-Baseline Neutrino Program at Fermilab~\cite{Machado:2019oxb}. 

In this paper we discuss yet another way of searching for the sterile neutrinos of eV-scale masses by using the resonance enhancement in matter, aiming at contributing to improved understanding of the phenomena. Our knowledge of the neutrino flavor transformation developed in the $\nu$SM tells us~\cite{Yasuda:2000xs,Nunokawa:2003ep} that an eV-scale massive sterile neutrino and some of the active ones have the level crossing at 1-10 TeV energies in the earth, producing the MSW resonance~\cite{Wolfenstein:1977ue,Mikheyev:1985zog} or the matter-enhanced vacuum oscillations~\cite{Barger:1980tf}. Importantly, the resonance can be detectable in Neutrino Telescopes, IceCube~\cite{IceCube:2014gqr} and KM3Net~\cite{KM3Net:2016zxf}. In fact, the authors of ref.~\cite{Nunokawa:2003ep} were the first to propose search for the ``sterile-active resonance'' at a Neutrino Telescope. 
The experimental search for the resonance has been pioneered by the IceCube group~\cite{IceCube:2016rnb}, who progressed to the eight years data~\cite{IceCube:2020tka}, and very recently accumulated almost eleven years of data set~\cite{IceCube:2024kel,IceCube:2024uzv}. In parallel, many authors contributed to understanding physics of the resonance~\cite{Nunokawa:2003ep,Yasuda:2000xs,Choubey:2007ji,Razzaque:2011ab,Razzaque:2012tp,Barger:2011rc,Esmaili:2012nz,Esmaili:2013cja,Esmaili:2013vza,Esmaili:2013fva,Lindner:2015iaa,Smithers:2021orb}. 

It is useful to introduce some simple terminologies because of their frequent usage in the rest of this paper. We use the term ``SA resonance'' for the sterile-active resonance with eV-scale massive sterile neutrino. For the sterile-active resonance region, the kinematical region where the SA resonance resides, we simply use ``SA resonance region'', or ``SA region'' for short. The limiting procedure to the SA resonance region is denoted as the ``SA limit''. These concepts will be explained in more details in sections~\ref{sec:SA-resonance} and~\ref{sec:effective-theory}. 

In this paper, we make our first attempt at analytic treatment of the SA resonance to facilitate a ``global overview at a glance'' by taking the $(3+1)$ model as a prototype. Foreseeing the progress to come in another ten years, it naturally suggests us to start working toward the multi-flavor and multi-event-type analyses with much more precise measurement to be available. 
The key feature of our treatment is to take into account the fact that the SA resonance moves around under the environment that the huge matter potential overwhelms the active three neutrino transformations, letting the $\nu$SM oscillation to freeze. This feature allows us to construct an effective theory of SA resonance, which involves only the sterile-active mixing angles and $\Delta m^2_{41}$. 

We treat the neutrino and anti-neutrino channels in the same footing, and analyze all the flavor oscillation channels at once, the novel feature of the analytic framework for the SA resonance, which has never been tried to our knowledge. With simple calculations it allows us to extract some interesting characteristic features of the resonance including the channel-dependent varying strengths of the resonance effect. As it offers the global bird-eye view of the SA resonance complex, it would allow us to give suggestions for the future Neutrino Telescope search. 
We assume the readers' familiarity with the $\nu$SM neutrino oscillations at a reasonable level, but beyond that our description will be pedagogical and self-contained.

\section{The ($3+1$) model}
\label{sec:(3+1)-model} 

We start by defining the ($3+1$) model. Though well studied as the simplest extension of the $\nu$SM with a single sterile state added, this section is meant to fix our notations and explain the necessary change in the convention for the flavor mixing matrices in section~\ref{sec:ATM-convention}. 

The neutrino evolution in the system of three active and one sterile neutrinos in matter, the ($3+1$) model, can be described by the Schr\"odinger equation in the flavor basis 
\begin{eqnarray}
i \frac{d}{dx} \nu = 
\frac{1}{2E} 
\left\{  
U_{(3+1)} 
\left[
\begin{array}{cccc}
0 & 0 & 0 & 0 \\
0 & \Delta m^2_{21} & 0 & 0 \\
0 & 0 & \Delta m^2_{31} & 0 \\
0 & 0 & 0 & \Delta m^2_{41} \\
\end{array}
\right] 
U_{(3+1)} ^{\dagger} 
+ 
\left[
\begin{array}{cccc}
a (x) & 0 & 0 & 0 \\
0 & 0 & 0 & 0 \\
0 & 0 & 0 & 0 \\
0 & 0 & 0 & b (x) \\
\end{array}
\right] 
\right\} 
\nu 
\equiv 
\frac{1}{2E} H_{(3+1)} \nu 
\nonumber \\
\label{evolution-flavor-basis-4nu}
\end{eqnarray}
where $\Delta m^2_{ji} \equiv m^2_{j} - m^2_{i}$ with the Latin indices $i, j$ denote the mass squared differences between the $j$-th and $i$-th eigenstate of neutrinos ($i, j=1,2,3,4$). In this paper we assume that $m^2_{4} \gg m^2_{k}$ ($k=1,2,3$). 

In eq.~\eqref{evolution-flavor-basis-4nu}, $U_{(3+1)}$ denotes the $4 \times 4$ flavor mixing matrix which relates the flavor basis to the mass eigenstate basis as $\nu_{\text{flavor}} = U_{(3+1)} \nu_{\text{mass}}$ in vacuum. 
Following the standard notation, we define $U_{(3+1)}$ as a simple $4 \times 4$ extension of the $\nu$SM $3\times3$ flavor mixing matrix~\cite{Maki:1962mu} $U_{(3\times3)} \vert_{\text{\tiny PDG} }$ of the Particle Data Group (PDG)~\cite{ParticleDataGroup:2024cfk} convention: 
\begin{eqnarray} 
&& 
U_{(3+1)} \vert_{\text{\tiny PDG} } 
\equiv 
U_{34} (\theta_{34}, \phi_{34} )
U_{24} (\theta_{24}, \phi_{24} )
U_{14} (\theta_{14} ) 
U_{23} (\theta_{23} ) 
U_{13} (\theta_{13}, \delta ) 
U_{12} (\theta_{12} ) \vert_{\text{\tiny PDG} } 
\nonumber \\
&=&
\left[
\begin{array}{cccc}
1 & 0 & 0 & 0\\
0 & 1 & 0 & 0\\
0 & 0 & c_{34} & e^{ - i \phi_{34} } s_{34} \\
0 & 0 & - e^{ i \phi_{34} } s_{34} & c_{34}
\end{array}
\right] 
\left[
\begin{array}{cccc}
1 & 0 & 0 & 0 \\
0 & c_{24} & 0 & e^{ - i \phi_{24} } s_{24} \\
0 & 0 & 1 & 0 \\
0 & - e^{ i \phi_{24} } s_{24} & 0 & c_{24} \\
\end{array}
\right] 
\left[
\begin{array}{cccc}
c_{14} & 0 & 0 & s_{14} \\
0 & 1 & 0 & 0 \\
0 & 0 & 1 & 0 \\
- s_{14} & 0 & 0 & c_{14} \\
\end{array}
\right]
\left[
\begin{array}{cc}
U_{(3\times3)} \vert_{\text{\tiny PDG} } & 0 \\ 
0 & 1 \\
\end{array}
\right], 
\nonumber \\
\label{U-PDG}
\end{eqnarray}
where the usual abbreviated notations are used, $c_{ij} \equiv \cos \theta_{ij}$ and $s_{ij} \equiv \sin \theta_{ij}$ . Therefore, the sterile and active states mixing angles and the CP phases are defined by eq.~\eqref{U-PDG} throughout this paper. The Schr\"odinger equation eq.~\eqref{evolution-flavor-basis-4nu} with use of $U_{(3+1)}$ in eq.~\eqref{U-PDG} defines the ($3+1$) model. We are left with defining the matter potential to place the $(3+1)$ model in matter, which we do immediately below. 

The functions $a(x)$ and $b(x)$ in eq.~\eqref{evolution-flavor-basis-4nu} denote the Wolfenstein matter potentials~\cite{Wolfenstein:1977ue} due to charged current (CC) and neutral current (NC) reactions, respectively, 
\begin{eqnarray} 
a(x) &=&  
2 \sqrt{2} G_F N_e E \approx 1.52 \times 10^{-4} \left( \frac{Y_e \rho}{\rm g\,cm^{-3}} \right) \left( \frac{E}{\rm GeV} \right) {\rm eV}^2, 
\nonumber \\
b(x) &=& \sqrt{2} G_F N_n E = \frac{1}{2} \left( \frac{N_n}{N_e} \right) a, 
\label{matt-potential}
\end{eqnarray}
where $G_F$ is the Fermi constant. $N_e$ and $N_n$ are the electron and neutron number densities in matter. $\rho$ and $Y_e$ denote, respectively, the matter density and number of electrons per nucleon in matter. These quantities except for $G_F$ are, in principle, position dependent. But for simplicity and transparency of our discussion in this paper we use the uniform matter density approximation. 


\subsection{ATM convention of the flavor mixing matrix}
\label{sec:ATM-convention} 

For convenience in using the existing formulas in the later part, starting in section~\ref{sec:effective-theory}, we switch to the so called ATM convention~\cite{Martinez-Soler:2018lcy} of the flavor mixing matrix. 
For this purpose we make the phase transformations of the states and the Hamiltonian in eq.~\eqref{evolution-flavor-basis-4nu}, 
\begin{eqnarray} 
&& 
\nu \rightarrow \text{diag} [ 1, 1, e^{ - i \delta}, 1 ] \nu 
\equiv \nu \vert_{\text{\tiny ATM} } 
\nonumber \\
&& 
H_{(3+1)} \vert_{\text{\tiny PDG} } 
\rightarrow \text{diag} [ 1, 1, e^{ - i \delta}, 1 ] H_{(3+1)} \vert_{\text{\tiny PDG} } ~\text{diag} [ 1, 1, e^{ i \delta}, 1 ] 
\equiv 
H_{(3+1)} \vert_{\text{\tiny ATM} } 
\label{PDG-ATM}
\end{eqnarray}
by which the Schr\"odinger equation remains the same form as in eq.~\eqref{evolution-flavor-basis-4nu}, but now $U_{(3+1)} \vert_{\text{\tiny PDG} }$ is replaced by $U_{(3+1)} \vert_{\text{\tiny ATM} } = \text{diag} [ 1, 1, e^{ - i \delta}, 1 ]~U_{(3+1)} \vert_{\text{\tiny PDG} }~\text{diag} [ 1, 1, e^{ i \delta}, 1 ]$. 
Of course, by being the rephasing of the wave function, there is no change in the observables. For simplicity of the notations we consistently abbreviate $\vert_{\text{\tiny ATM} }$ symbols hereafter because we exclusively use the ATM convention $U$ matrices in the rest of this paper. To appeal the feature that $H_{(3+1)} \vert_{\text{\tiny ATM} }$ is the flavor-basis Hamiltonian we simply denote it as $H_{\text{flavor}}$. 

If we write 
\begin{eqnarray} 
&&
U_{(3+1)} \vert_{\text{\tiny ATM} }
= 
U_{\text{sterile}} U_{\text{active}},
\nonumber \\
&&
U_{\text{sterile}} 
\equiv 
U_{34} (\theta_{34}, \delta_{34} )
U_{24} (\theta_{24}, \delta_{24} )
U_{14} (\theta_{14} ),  
\nonumber \\
&& 
U_{\text{active}} 
\equiv 
U_{23} (\theta_{23}, \delta ) U_{13} (\theta_{13} ) U_{12} (\theta_{12} ), 
\label{Usterile-active-ATM}
\end{eqnarray}
the explicit form of the $U$ matrices in eq.~\eqref{Usterile-active-ATM}, which are all of the ATM convention, are given by 
\begin{eqnarray} 
&& 
U_{34} (\theta_{34}, \phi_{34} ) 
= 
\left[
\begin{array}{cccc}
1 & 0 & 0 & 0\\
0 & 1 & 0 & 0\\
0 & 0 & c_{34} & e^{ - i ( \delta + \phi_{34} ) } s_{34} \\
0 & 0 & - e^{ i ( \delta + \phi_{34} ) } s_{34} & c_{34}
\end{array}
\right], 
\nonumber \\
&& 
U_{24} (\theta_{24}, \phi_{24} ) 
= 
\left[
\begin{array}{cccc}
1 & 0 & 0 & 0 \\
0 & c_{24} & 0 & s_{24} e^{ - i \phi_{24} } \\
0 & 0 & 1 & 0 \\
0 & - s_{24} e^{ i \phi_{24} } & 0 & c_{24} \\
\end{array}
\right], 
\hspace{6mm} 
U_{14} (\theta_{14} ) 
= 
\left[
\begin{array}{cccc}
c_{14} & 0 & 0 & s_{14} \\
0 & 1 & 0 & 0 \\
0 & 0 & 1 & 0 \\
- s_{14} & 0 & 0 & c_{14} \\
\end{array}
\right]. ~~~~~
\label{sterile-U-ATM}
\end{eqnarray}
\begin{eqnarray} 
&& 
\hspace{-8mm}
U_{23} (\theta_{23}, \delta )
= 
\left[
\begin{array}{cccc}
1 & 0 & 0 & 0\\
0 & c_{23} & e^{ i \delta} s_{23} & 0\\
0 & - e^{ - i \delta} s_{23} & c_{23} & 0\\
0 & 0 & 0 & 1 \\
\end{array}
\right], 
\hspace{5mm}
U_{13} (\theta_{13} ) 
= 
\left[
\begin{array}{cccc}
c_{13} & 0 & s_{13} & 0 \\
0 & 1 & 0 & 0 \\
- s_{13} & 0 & c_{13} & 0 \\
0 & 0 & 0 & 1 \\
\end{array}
\right], 
\hspace{5mm}
U_{12} (\theta_{12} ) 
= 
\left[
\begin{array}{cccc}
c_{12} & s_{12} & 0 & 0 \\
- s_{12} & c_{12} & 0 & 0 \\
0 & 0 & 1 & 0 \\
0 & 0 & 0 & 1 \\
\end{array}
\right]. 
\nonumber \\
\label{active-U-ATM}
\end{eqnarray}
The ATM convention flavor mixing matrix has been utilized e.g., in refs.~\cite{Minakata:2015gra,Denton:2016wmg,Martinez-Soler:2018lcy}. 

\section{Sterile-active (SA) resonance}
\label{sec:SA-resonance}  

It has been known since the early days~\cite{Yasuda:2000xs,Nunokawa:2003ep} that in the ($3+1$) model the sterile state which feels the matter potential with about a half strength of the $\nu_{e}$'s (in an appropriate phase convention, see eq.~\eqref{evolution-flavor-basis-4nu}) can produce new resonances at $\sim10$~TeV for the sterile neutrino mass of eV scale. Each one of the resonances is dubbed as the sterile-acvtive (SA) resonance in this paper. 
Given the energy scale of $\sim10$~TeV,\footnote{
At $E = 10$~TeV, $a / \Delta m^2_{41} \simeq 1.27$ for $\Delta m^2_{41} = 3~$eV$^2$, $\rho = 5$~g/cm$^3$. See eq.~\eqref{a/Dm2}. }
the most sensitive apparatus which can serve for its detection would be Neutrino Telescopes, IceCube~\cite{IceCube:2014gqr} and KM3Net~\cite{KM3Net:2016zxf}. Recently, the IceCube group reported the data from almost eleven years of search for the SA resonance~\cite{IceCube:2024kel,IceCube:2024uzv}. Interestingly they see a closed contour at 95\% CL in $\sin^2 2\theta_{24} - \Delta m^2_{41}$ plane, 
centered at $\sin^2 2\theta_{24} = 0.16$ and $\Delta m^2_{41} = 3.5$ eV$^2$, which may drop hints about possible structure. This feature certainly motivated us to pursue this investigation, although it is in tension with the results from the other experiments, such as MINOS+~\cite{MINOS:2017cae}. 
In this paper we aim at providing a hopefully comprehensive, bird-eye view of such resonance in the ($3+1$) model. A best hope would be that our study is able to give useful suggestions to formulate the experimental search strategy for the SA resonance in the coming precision era of the measurement. 

\begin{figure}[h!]
\begin{center}
\vspace{-22mm}
\includegraphics[width=1.1\textwidth]{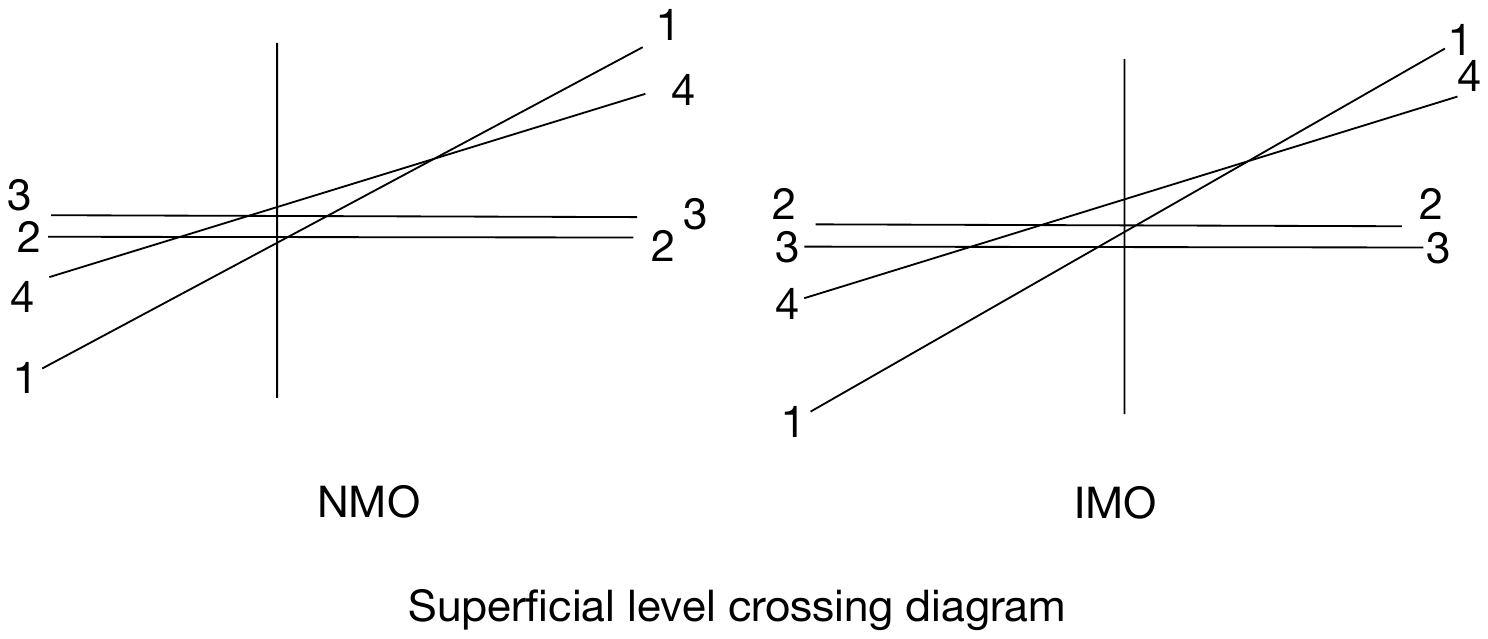}
\end{center}
\vspace{-16mm}
\caption{The superficial level crossing diagram without flavor mixing in the $(3+1)$ model. See the text for more information. The distance between the lines and the relative size between $\Delta m^2_{41}$ and $\Delta m^2_{31}$, for example, are very exaggerated. 
} 
\vspace{-2mm}
\label{fig:superficial-level-cross}
\end{figure}

The structure of the SA resonance in the ($3+1$) model can be revealed by drawing the {\it superficial level crossing diagram}, see Fig.~\ref{fig:superficial-level-cross}. In this figure, the flavor state mass-squared ($2E \times$ eigenvalues of $H_{(3+1)}$ in eq.~\eqref{evolution-flavor-basis-4nu}) are plotted as a function of the matter potential by ignoring the flavor mixing, $U_{(3+1)} = 1$. The left and right panels of Fig.~\ref{fig:superficial-level-cross} corresponds to the case of normal mass ordering (NMO) and inverted mass ordering (IMO), respectively. In the superficial level crossing diagram, therefore, the lines actually cross with each other, the feature lost when the flavor mixing is tuned on. While the term ``superficial'' is attached to indicate unrealistic feature of the diagram, we will learn shortly about its utility. 

In Fig.~\ref{fig:superficial-level-cross} the vertical line shows the vacuum, point of the zero matter potential, and therefore the right (left) side of vertical line is the neutrino (anti-neutrino) channel. In the case of NMO (left panel), in the neutrino channel, in addition to the 1-2 and 1-3 level crossings which correspond to the solar and atmospheric resonances in the $\nu$SM, there exists the unique SA resonance, associated with the 1-4 level crossing. In the anti-neutrino channel, there are two sterile-active resonances which correspond to the 2-4 and 3-4 level crossings. 
In the case of IMO (right panel), the feature of the SA resonance is essentially the same as the NMO case, apart from exchanging the role of 2-4 and 3-4 level crossings, as can be seen in the right panel of Fig.~\ref{fig:superficial-level-cross}. Therefore, in this paper we confine ourselves into the case of NMO, apart from brief remarks of possible interests in the case of IMO at the end of section~\ref{sec:conclusion}. A mild preference of the NMO over the IMO is reported in the atmospheric neutrino observation~\cite{Super-Kamiokande:2017yvm,IceCube:2019dyb} and the accelerator long-baseline experiments~\cite{T2K:2024wfn,NOvA:2021nfi,NOvA:2024-milano}. A tension between T2K and NOvA on the favorable region on CP also affects to what extent the NMO is preferred over IMO, see e.g., refs.~\cite{Capozzi:2018ubv,deSalas:2020pgw,Kelly:2020fkv,Esteban:2020cvm} in addition to the above. 

Here is the utility of the superficial level crossing diagram. In the ``real level crossing diagram''\footnote{
This is the one usually called the level crossing diagram, which however might be better denoted as the ``level nearing diagram'' as the levels never cross to each other. }
with flavor mixing turned on, the atmospheric-scale resonance correspond to the 2-3 level crossing in the NMO, but to the 1-3 level crossing in the IMO, see e.g. Fig.~1 in ref.~\cite{Denton:2016wmg}. 
But, the mixing angle which describes the resonance, after elevated to the matter-affected version, is $\theta_{13}$ in the both NMO and IMO. This fact can be understood by the superficial level crossing diagram which says that the atmospheric-scale crossing is always between the level 1 and 3, and hence the diagonalization must be performed by $U_{13} (\theta_{13})$, as done e.g., in refs.~\cite{Minakata:2015gra,Denton:2016wmg}.\footnote{
In the treatment in refs.~\cite{Minakata:2015gra,Denton:2016wmg}, $U_{23} (\theta_{23}, \delta)$ can be rotated away from the Hamiltonian and therefore cannot govern the dynamical properties such as the MSW resonances. This is consistent with our argument that generically the matter-affected atmospheric resonance angle should be $\theta_{13}$ irrespective of the mass orderings. 
On the other hand, it should be emphasized that the difference in the participating eigenvalues into the atmospheric-scale resonance between the NMO and IMO matters because the eigenvalue difference, $\lambda_{3} - \lambda_{2}$ or $\lambda_{3} - \lambda_{1}$, affects the discussion of adiabaticity. } 
Or, in other word, the superficial level crossing diagram dictates how the Hamiltonian is to be diagonalized.

\section{Effective theory of sterile-active (SA) resonance} 
\label{sec:effective-theory} 

\subsection{Freezing the $\nu$SM oscillations in the sterile-active resonance region} 
\label{sec:freezing-nuSM} 

Now we focus our discussion to the kinematical region where we could observe the effect of the sterile-active resonance, the ``SA resonance'', with eV-scale massive sterile neutrino. Again for simplicity, we denote the kinematical region where the SA resonance resides, $a / \Delta m^2_{41} \sim b / \Delta m^2_{41} \sim 1$, as the ``SA resonance region'', hereafter. 

That is, the matter effect is comparable to the effect of vacuum sterile neutrino oscillation, so that the SA resonance enhancement can take place. For an eV-scale sterile neutrino mass and the relevant matter density $\rho = 5$~g/cm$^3$ in the Earth, we obtain the reference neutrino energy of $\sim10$ TeV: 
\begin{eqnarray} 
\frac{ a }{ \Delta m^2_{41} } 
&=& 
1.27 
\left(\frac{ \Delta m^2_{41} }{ 3~\mbox{eV}^2}\right)^{-1} \left(\frac{\rho}{5.0 \,\text{g/cm}^3}\right) \left(\frac{E}{10~\mbox{TeV}}\right). 
\label{a/Dm2}
\end{eqnarray}
In eq.~\eqref{a/Dm2} we have used the reference values of $\Delta m^2_{41}$ consistent with the 95\% CL closed contour reported in refs.~\cite{IceCube:2024uzv,IceCube:2024kel}. Notice that $b / a = \frac{1}{2}$ in a charge-neutral medium, so that $b / \Delta m^2_{41}$ is also of order unity.

Then, by restricting our interest into the SA resonance region, abstracting out a much simpler theory from the ($3+1$) model is naturally suggested. As $\Delta m^2_{31} / a \simeq 6.3 \times 10^{-4} \ll 1$, 
the $\nu$SM flavor transformation is heavily suppressed. The freezing the $\nu$SM oscillation effect greatly simplifies the theory. 

To implement the strong matter effect background in the $(3+1)$ model in an analytic fashion, we call for the Denton {\it et al.} (DMP) perturbation theory~\cite{Denton:2016wmg} to embed it into the active $3 \times 3$ subspace of the $(3+1)$ model. However, we will argue later (see section~\ref{sec:some-remarks}) that taking this particular model is not essential. 

\subsection{Embedding DMP into the active neutrino part of the ($3+1$) model} 
\label{sec:embedding-DMP}

We define $H_{\text{active}}$, the active basis Hamiltonian, as 
$H_{\text{active}} = U_{\text{sterile}} ^{\dagger} H_{\text{flavor}} U_{\text{sterile}}$: 
\begin{eqnarray} 
H_{\text{active}} 
&=& 
\frac{1}{2E}
\left\{
U_{\text{active}} 
\left[
\begin{array}{cccc}
0 & 0 & 0 & 0 \\
0 & \Delta m^2_{21} & 0 & 0 \\
0 & 0 & \Delta m^2_{31} & 0 \\
0 & 0 & 0 & \Delta m^2_{41} \\
\end{array}
\right] 
U_{\text{active}} ^{\dagger} 
+ 
\left[
\begin{array}{cccc}
a & 0 & 0 & 0 \\
0 & 0 & 0 & 0 \\
0 & 0 & 0 & 0 \\
0 & 0 & 0 & 0 \\
\end{array}
\right] 
\right\} 
\nonumber \\
&+& 
\frac{1}{2E} 
\left\{ 
U_{\text{sterile}} ^{\dagger} 
\left[
\begin{array}{cccc}
a & 0 & 0 & 0 \\
0 & 0 & 0 & 0 \\
0 & 0 & 0 & 0 \\
0 & 0 & 0 & b \\
\end{array}
\right] 
U_{\text{sterile}} 
- 
\left[
\begin{array}{cccc}
a & 0 & 0 & 0 \\
0 & 0 & 0 & 0 \\
0 & 0 & 0 & 0 \\
0 & 0 & 0 & 0 \\
\end{array}
\right] 
\right\}. 
\label{Hactive-ATM} 
\end{eqnarray}
In the first line in eq.~\eqref{Hactive-ATM}, the active $3 \times 3$ part is nothing but the flavor basis Hamiltonian in the $\nu$SM. In DMP~\cite{Denton:2016wmg}, therefore, $H_{\text{active}}$ can be written by using the diagonalized bar basis Hamiltonian $\bar{H} (3 \times 3)$ in the $\nu$SM as 
\begin{eqnarray} 
H_{\text{active}}
&=& 
U_{23} (\theta_{23}, \delta ) 
U_{13} ( \widetilde{\theta}_{13} ) 
U_{12} ( \widetilde{\theta}_{12} ) 
\begin{bmatrix}
    \bar{H} (3 \times 3) & 0 \\
    0 & \frac{ \Delta m^2_{41} }{2E} \\
    \end{bmatrix}
U_{12} ( \widetilde{\theta}_{12} ) ^{\dagger} 
U_{13} ( \widetilde{\theta}_{13} ) ^{\dagger} 
U_{23} (\theta_{23}, \delta ) ^{\dagger} 
\nonumber \\
&+& 
\frac{1}{2E}
U_{\text{sterile}} ^{\dagger} 
\text{diag} \left( a, 0, 0, b \right) 
U_{\text{sterile}} 
- \frac{1}{2E} \text{diag} \left( a, 0, 0, 0 \right),
\label{Hactive-DMP} 
\end{eqnarray}
with $\bar{H} (3 \times 3)$ given by\footnote{
If not familiar to the form in eq.~\eqref{Hactive-DMP} with eq.~\eqref{barH-3x3} in DMP, see e.g., ref.~\cite{Minakata:2021dqh}. }
\begin{eqnarray}
&&
\bar{H} (3 \times 3) 
= 
\frac{1}{2E} 
\left[
\begin{array}{ccc}
\lambda_1 & 0 & 0 \\
0 & \lambda_2 & 0 \\
0 & 0 & \lambda_3 
\end{array}
\right] 
+ 
\epsilon c_{12} s_{12} s_{( \widetilde{\theta}_{13} - \theta_{13} )} \frac{ \Delta m^2_{ \text{ren} } }{2E}
\left[
\begin{array}{ccc}
0 & 0 & - \sin \widetilde{\theta}_{12} \\
0 & 0 & \cos \widetilde{\theta}_{12} \\
- \sin \widetilde{\theta}_{12} & \cos \widetilde{\theta}_{12} & 0 
\end{array}
\right]. ~~~~~~~
\label{barH-3x3} 
\end{eqnarray}
In this paper we introduce a unified nomenclature in which the matter-affected mixing angles are denoted as $\widetilde{\theta}_{ij}$, corresponding to the vacuum angle $\theta_{ij}$. Therefore, in eqs.~\eqref{Hactive-DMP} and \eqref{barH-3x3}, $\widetilde{\theta}_{13}$ and $\widetilde{\theta}_{12}$ ($\phi$ and $\psi$ in the original reference~\cite{Denton:2016wmg}) denote the matter-affected $\theta_{13}$ and $\theta_{12}$, respectively. 
$\lambda_{i}$ ($i=1,2,3$) are the eigenvalues of the unperturbed part of the Hamiltonian, and $s_{( \widetilde{\theta}_{13} - \theta_{13} )} \equiv \sin ( \widetilde{\theta}_{13} - \theta_{13} )$. $\epsilon$ is the unique expansion parameter in the DMP perturbation theory 
\begin{eqnarray} 
&&
\epsilon \equiv \frac{ \Delta m^2_{21} }{ \Delta m^2_{ \text{ren} } }, 
\hspace{10mm}
\Delta m^2_{ \text{ren} } \equiv \Delta m^2_{31} - s^2_{12} \Delta m^2_{21},
\label{epsilon-Dm2-ren-def}
\end{eqnarray}
defined by using $\Delta m^2_{ \text{ren} }$, the ``renormalized'' atmospheric $\Delta m^2$, as used in ref.~\cite{Minakata:2015gra}. 

Then, the flavor basis Hamiltonian of the $(3+1)$ model can be obtained from $H_{\text{active}}$ in \eqref{Hactive-DMP} as $H_{\text{flavor}} 
= U_{\text{sterile}} H_{\text{active}} U_{\text{sterile}} ^{\dagger}$, 
\begin{eqnarray} 
&&
\hspace{-4mm} 
H_{\text{flavor}} 
=  
U_{\text{sterile}} 
\biggl\{ 
U_{23} (\theta_{23}, \delta ) 
U_{13} ( \widetilde{\theta}_{13} ) 
U_{12} ( \widetilde{\theta}_{12} ) 
\begin{bmatrix}
    \bar{H} (3 \times 3) & 0 \\
    0 & \frac{ \Delta m^2_{41} }{2E} \\
    \end{bmatrix}
U_{12} ( \widetilde{\theta}_{12} ) ^{\dagger} 
U_{13} ( \widetilde{\theta}_{13} ) ^{\dagger} 
U_{23} (\theta_{23}, \delta ) ^{\dagger} 
\nonumber \\
&& 
\hspace{12mm} 
- 
\frac{1}{2E} 
\text{diag} \left( a, 0, 0, 0 \right) 
\biggr\} 
U_{\text{sterile}} ^{\dagger}
+ \frac{1}{2E}
\text{diag} \left( a, 0, 0, b \right), 
\label{Hflavor-DMP-embedded} 
\end{eqnarray}
with $\bar{H} (3 \times 3)$ given in eq.~\eqref{barH-3x3} and $U_{\text{sterile}}$  defined in eq.~\eqref{Usterile-active-ATM}. 

\subsection{Limiting procedure to the SA resonance region: An effective theory} 
\label{sec:IC-limit} 

We implement explicitly freezing of the $\nu$SM oscillation effect by the huge matter potential $\Delta m^2_{31} / a \sim 10^{-3} \ll 1$. The behavior of $\widetilde{\theta}_{13}$ and $\widetilde{\theta}_{12}$ in the SA resonance region can be illuminated by the asymptotic behavior of the matter-affected mixing angles at $a, b \rightarrow \pm \infty$. It is known in DMP that the both $\widetilde{\theta}_{13}$ and $\widetilde{\theta}_{12}$ approach quickly to the asymptotic values, $\widetilde{\theta}_{13} = \widetilde{\theta}_{12} = \frac{\pi}{2}$ in the neutrino channel ($a \rightarrow + \infty$), and $\widetilde{\theta}_{13} = \widetilde{\theta}_{12} = 0$ in the anti-neutrino channel ($a \rightarrow - \infty$). The asymptotic value is reached at around $Y_e \rho E \sim 100$ (g/cm$^3$)GeV for $\widetilde{\theta}_{13}$, and even faster for $\widetilde{\theta}_{12}$. See Fig.~1 in ref.~\cite{Denton:2016wmg}. 

The asymptotic behavior of $\widetilde{\theta}_{13}$ and $\widetilde{\theta}_{12}$ implies that the 1-3 and 1-2 rotations in the Hamiltonian in eq.~\eqref{Hflavor-DMP-embedded} become just the discrete rotations in the neutrino channel, 
\begin{eqnarray} 
&& 
\hspace{-4mm} 
U_{13} ( \widetilde{\theta}_{13} ) 
\rightarrow 
\left[
\begin{array}{cccc}
0 & 0 & 1 & 0 \\
0 & 1 & 0 & 0 \\
- 1 & 0 & 0 & 0 \\
0 & 0 & 0 & 1 \\
\end{array}
\right], 
\hspace{4mm} 
U_{12} ( \widetilde{\theta}_{12} ) 
\rightarrow 
\left[
\begin{array}{cccc}
0 & 1 & 0 & 0 \\
- 1 & 0 & 0 & 0 \\
0 & 0 & 1 & 0 \\
0 & 0 & 0 & 1 \\
\end{array}
\right], 
\hspace{4mm} 
U_{13} ( \widetilde{\theta}_{13} ) U_{12} ( \widetilde{\theta}_{12} ) 
\rightarrow  
\left[
\begin{array}{cccc}
0 & 0 & 1 & 0 \\
- 1 & 0 & 0 & 0 \\
0 & - 1 & 0 & 0 \\
0 & 0 & 0 & 1 \\
\end{array}
\right]. 
\nonumber \\
\label{13-12-freezed}
\end{eqnarray}
and 
$U_{13} ( \widetilde{\theta}_{13} )^* U_{12} ( \widetilde{\theta}_{12} )^* 
\rightarrow \text{diag} [1,1,1,1] $ in the anti-neutrino channel. By inserting eq.~\eqref{13-12-freezed} into eq.~\eqref{Hflavor-DMP-embedded}, we have reached the effective Hamiltonian for describing the SA resonance in the neutrino channel, 
\begin{eqnarray} 
H_{\text{flavor}} 
&=&
\frac{1}{2E} 
U_{\text{sterile}} 
\left[
\begin{array}{cccc}
( \lambda_3 - a ) & 0 & 0 & 0\\
0 & c^2_{23} \lambda_1 + s^2_{23} \lambda_2 & 
e^{ i \delta} c_{23} s_{23} ( \lambda_2 - \lambda_1 ) & 0\\
0 & e^{ - i \delta} c_{23} s_{23} ( \lambda_2 - \lambda_1 ) & 
s^2_{23} \lambda_1 + c^2_{23} \lambda_2 & 0\\
0 & 0 & 0 & \Delta m^2_{41} \\
\end{array}
\right] 
U_{\text{sterile}} ^{\dagger}
\nonumber \\
&+& 
\frac{1}{2E}
\text{diag} \left( a, 0, 0, b \right), 
\label{Hflavor-effective-IC-nu} 
\end{eqnarray}
where $U_{\text{sterile}} = U_{34} (\theta_{34}, \delta_{34} ) U_{24} (\theta_{24}, \delta_{24} ) U_{14} (\theta_{14} )$ as defined in eq.~\eqref{Usterile-active-ATM}. 
Similarly, in the anti-neutrino channel we obtain 
\begin{eqnarray} 
H_{\text{flavor}} 
&=&
\frac{1}{2E} 
U_{\text{sterile}} ^* 
\left[
\begin{array}{cccc}
\lambda_1 + |a| & 0 & 0 & 0 \\
0 & c^2_{23} \lambda_2 + s^2_{23} \lambda_3 & 
e^{ - i \delta} c_{23} s_{23} ( \lambda_3 - \lambda_2 ) & 0 \\
0 & e^{ i \delta} c_{23} s_{23} ( \lambda_3 - \lambda_2 ) & 
s^2_{23} \lambda_2 + c^2_{23} \lambda_3 & 0 \\
0 & 0 & 0 & \Delta m^2_{41} \\
\end{array}
\right] 
[ U_{\text{sterile}} ^* ] ^{\dagger} 
\nonumber \\
&+& 
\frac{1}{2E}
\text{diag} \left( - |a|, 0, 0, - |b| \right). 
\label{Hflavor-effective-IC-nubar} 
\end{eqnarray}
In reaching eqs.~\eqref{Hflavor-effective-IC-nu} and \eqref{Hflavor-effective-IC-nubar} we have ignored the terms of order $\epsilon \Dmsqren = \Delta m^2_{21}$. Notice that our notation is such that $a$ and $b$ in eq.~\eqref{matt-potential} are positive definite for the earth matter. However, to prevent any misunderstanding, we denote them as $- |a|$ and $- |b|$ in discussions of the anti-neutrino channel. 
 
It is interesting to note that the Hamiltonian in 
eqs.~\eqref{Hflavor-effective-IC-nu} and \eqref{Hflavor-effective-IC-nubar} 
have the structure akin to the flavor basis Hamiltonian in the $\nu$SM, apart from its $4 \times 4$ extension and inclusion of the NC matter potential. 
If we plug in the Hamiltonian~\eqref{Hflavor-effective-IC-nu}, or \eqref{Hflavor-effective-IC-nubar}, into the Schr\"odinger equation in eq.~\eqref{evolution-flavor-basis-4nu}, we call this theory as the {\it effective theory of SA resonance}. 

The DMP eigenvalues that appear in eqs.~\eqref{Hflavor-effective-IC-nu}, or \eqref{Hflavor-effective-IC-nubar}, simplify in the SA limit as 
\begin{eqnarray} 
&&
\lambda_{1} = \epsilon c^2_{12} \Dmsqren 
\nonumber \\
&& 
\lambda_{2} = ( c^2_{13}+ \epsilon s^2_{12}) \Dmsqren 
\nonumber \\
&& 
\lambda_{3} = a + (s^2_{13} + \epsilon s^2_{12}) \Dmsqren 
\label{DMP-eigenvalues-nu}
\end{eqnarray} 
in the neutrino channel, and 
\begin{eqnarray} 
&&
\lambda_{1} = 
( s^2_{13} + \epsilon s^2_{12} ) \Dmsqren - |a| 
\nonumber \\
&&
\lambda_{2} = 
\epsilon c^2_{12} \Dmsqren 
\nonumber \\
&&
\lambda_{3} = \lambda_{+} 
= 
( c^2_{13} + \epsilon s^2_{12}) \Dmsqren  
\label{DMP-eigenvalues-nubar}
\end{eqnarray} 
in the anti-neutrino channel. 

In the neutrino channel~\eqref{DMP-eigenvalues-nu}, $\lambda_1$, $\lambda_2$ and $( \lambda_3 - a )$ are all order $\sim \Dmsqren$ (or $\epsilon \Dmsqren$) and they can be neglected compared to $\Delta m^2_{41}$ in the same approximation as we made in taking the SA limit. The same statement holds for the anti-neutrino channel as $\lambda_1 + |a| \sim \Dmsqren$, see eq.~\eqref{DMP-eigenvalues-nubar}. But, up to now we have kept them, considering the possibility that such small quantity could be needed to resolve a subtle point. 

\subsection{Some remarks on validity of the effective theory of SA resonance etc.} 
\label{sec:some-remarks}

Now, we point out that our derivation of the effective theory of SA resonance is valid without relying on the particular way presented above, using DMP as the embedded $\nu$SM description. The derivation works as far as the validity of the SA limit is ensured. Namely, it goes through for much wider classes of the $\nu$SM description as far as the approximation $\widetilde{\theta}_{13} = \widetilde{\theta}_{12} = \frac{\pi}{2} (0)$ in the neutrino (anti-neutrino) channel holds. In fact we suspect that the SA limit can be justified in the treatment of the three-flavor system~\cite{Zaglauer:1988gz}, even though the algebra to show its validity might be slightly complicated. 

We note, however, that what is good for our derivation with DMP is that we can estimate quite easily how large the sub-asymptotic corrections are, by using the analytic expressions given in ref.~\cite{Denton:2016wmg}. In the neutrino channel the deviation from $\widetilde{c}_{13} = \widetilde{c}_{12} = 0$ occurs, in the leading order, 
\begin{eqnarray} 
&& 
\widetilde{c}_{13} = c_{13} s_{13} \frac{\Dmsqren}{a}, 
\hspace{10mm}
\widetilde{c}_{12} = \epsilon c_{12} s_{12} \frac{ s_{13} }{ c^2_{13} }, 
\label{sub-asymptotic}
\end{eqnarray} 
where we have ignored the quantities of order $\mathcal{O} [( \Dmsqren / a)^2 ]$ and $\mathcal{O} [ \epsilon^2 ]$ for $\widetilde{c}_{13}$ and $\widetilde{c}_{12}$, respectively. For the anti-neutrino channel the above sub-asymptotic correction is for $\widetilde{s}_{13}$. As indicated in eq.~\eqref{sub-asymptotic}, the way of approaching to the asymptotic is quite different between $\widetilde{c}_{13}$ and $\widetilde{c}_{12}$. For $\widetilde{c}_{13}$ smallness of $(\Dmsqren / a)$ matters, and for $\widetilde{c}_{12}$ smallness of $\epsilon$. Numerically $\widetilde{c}_{12} = \epsilon c_{12} s_{12} ( s_{13} / c^2_{13} ) = 2.07 \times 10^{-3}$, and in the SA resonance region $( \Dmsqren / a) \sim 10^{-3}$. Therefore, if the future measurement in the SA resonance region reaches the accuracy of $\sim 10^{-3}$ level, one has to keep such corrections in the matter-affected mixing angles in taking the SA limit. 

Without seeing any subtlety, in our subsequent treatment of the SA resonance effective theory, we propose to ignore these order $\sim \Dmsqren$ terms in the matrix part in eqs.~\eqref{Hflavor-effective-IC-nu} and \eqref{Hflavor-effective-IC-nubar} to solve the system, which is to be done in sections~\ref{sec:IC-theory-nu-channel} and~\ref{sec:IC-theory-nubar-channel}. Then, the Hamiltonian becomes $\text{diag} [0, 0, 0, \Delta m^2_{41}]$ in its matrix structure. One might wonder whether it could be derived by taking the naive limit $\Delta m^2_{41} \gg \Delta m^2_{k1} \rightarrow 0$ ($k=2,3$) from the beginning. If so the lengthy discussion of the SA limit turned out to be empty. However, one should notice that this naive limit effectively implies that all the $\nu$SM (or the active neutrino space) mixing angles vanish. On the other hand, in our SA limit 
$\widetilde{\theta}_{13} = \widetilde{\theta}_{12} = \frac{\pi}{2}$ in the neutrino channel. As we will see in the next section~\ref{sec:state-space}, the naive limit misses the correct physical state structure in the neutrino channel. 

A short comment is ready for the related work of the DMP-type treatment of the ($3+1$) model~\cite{Parke:2019jyu}. The authors discuss the effect of eV-scale sterile neutrino in the three active neutrino oscillation regime, $E \lsim$a few tens of GeV in ref.~\cite{Parke:2019jyu}. Whereas in this paper we address physics of the SA resonance at around $E \sim 10$ TeV. Each one of these two works covers the  different kinematical regions which are complementary to each other. 

\subsection{How accurate is the effective theory of SA resonance?} 
\label{sec:accuracy} 

\begin{figure}[h!]
\vspace{-4mm}
\begin{center}
\includegraphics[width=1.1\textwidth]{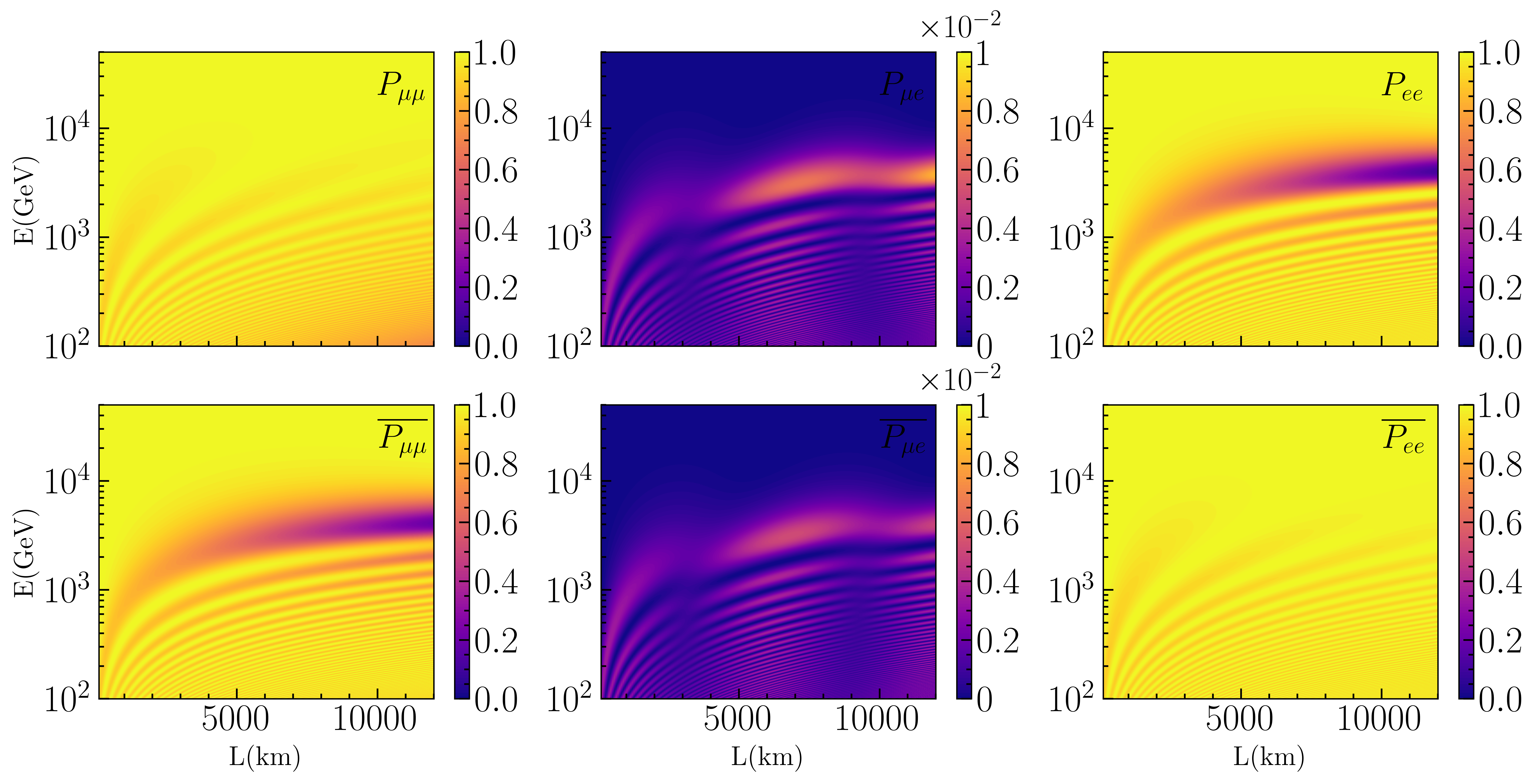} 
\includegraphics[width=1.1\textwidth]{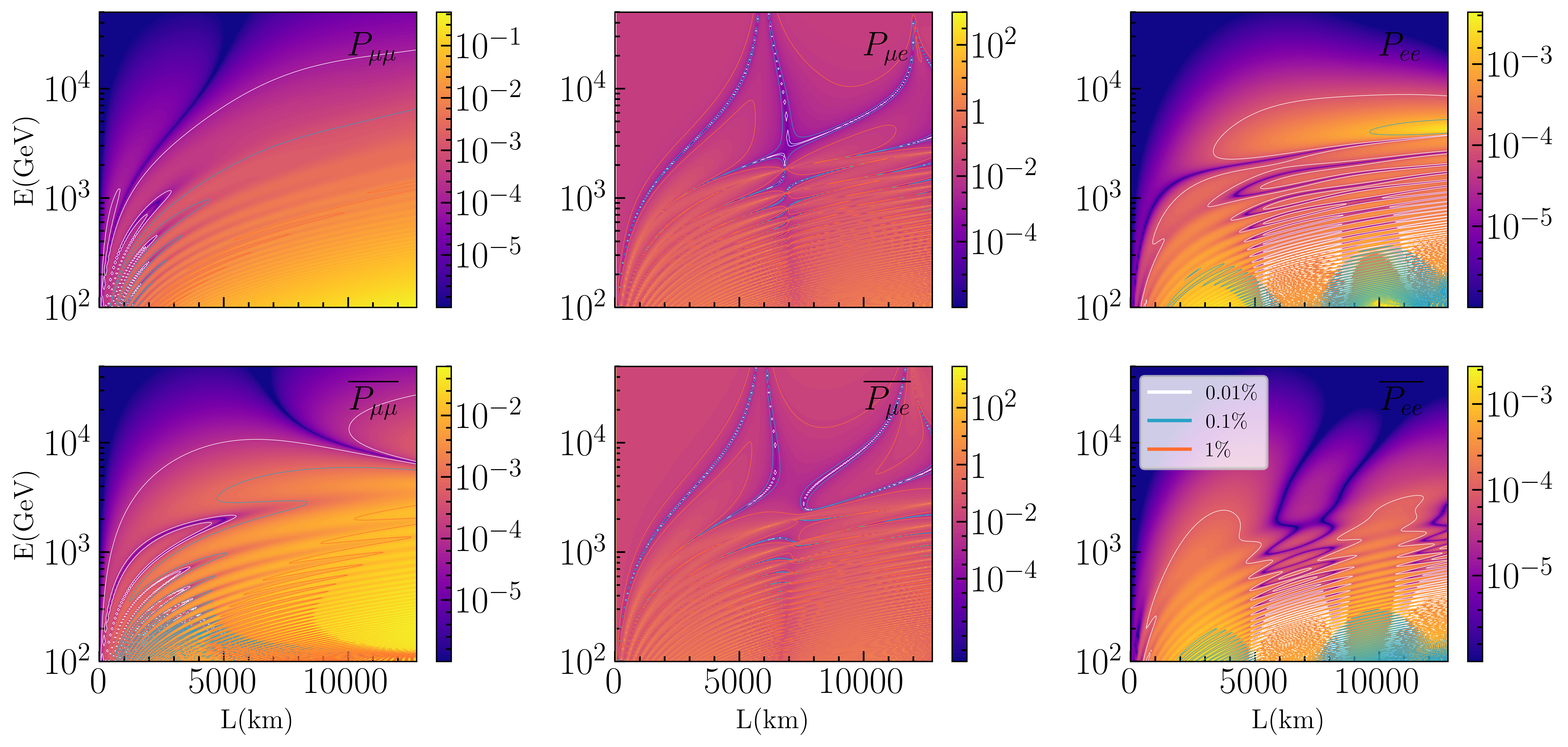}
\end{center}
\vspace{-4mm}
\caption{Presented in the upper six panels are the equi-contours of the probabilities in the $L-E$ space, computed using the $(3+1)$ model numerical code. From left to right $P(\nu_{\mu} \rightarrow \nu_{\mu})$, $P(\nu_{\mu} \rightarrow \nu_{e})$, and $P(\nu_{e} \rightarrow \nu_{e})$, and its anti-neutrino counterparts are drawn. 
Notice that the scale of the color variation is much different in $P(\nu_{\mu} \rightarrow \nu_{e})$ in the middle panel. 
In the lower six panels the fractional difference $F$ defined in eq.~\eqref{fractional-diff}, an indicator for the difference between the numerical code and the SA resonance effective theory results, are shown. See the text for more details. To guide the eye we have inserted 0.01\%, 0.1\%, and 1\% contours. We have used $\Delta m^2_{41} = 1~\text{eV}^2$, the matter density $\rho = 5.5 \,\text{g/cm}^3$, and $\theta_{14} = \theta_{24} = \theta_{34} = 10$ degree. 
} 
\vspace{-2mm}
\label{fig:exact-EFIC-comparison}
\end{figure}

It may be worthwhile to do a numerical test to know how accurate is the SA resonance effective theory after taking the SA limit and $\text{diag} [0, 0, 0, \Delta m^2_{41}]$ structure in the Hamiltonian. In Fig.~\ref{fig:exact-EFIC-comparison}, in the upper block with six panels, the equi-contours of the probability $P(\nu_{\beta} \rightarrow \nu_{\alpha}) \vert_{ \text{num-code} }$ computed using the $(3+1)$ model numerical code are presented in the $L-E$ (baseline - energy) space. This code includes the active neutrino mixing effect as well. 
In the lower six-panel block, to show the accuracy of the probability $P(\nu_{\beta} \rightarrow \nu_{\alpha}) \vert_{ \text{eff-th} }$ computed by the SA resonance effective theory, the fractional difference $F$, 
\begin{eqnarray} 
&& 
F \equiv \biggl |
\frac{ P(\nu_{\beta} \rightarrow \nu_{\alpha}) \vert_{ \text{num-code} } - P(\nu_{\beta} \rightarrow \nu_{\alpha}) \vert_{ \text{eff-th} } }{ P(\nu_{\beta} \rightarrow \nu_{\alpha}) \vert_{ \text{num-code} } }
\biggr |, 
\label{fractional-diff} 
\end{eqnarray}
is presented. In each block the upper three panels are for the neutrino channel, and the lower ones for the anti-neutrino channel. In each three panels, from left to right, $P(\nu_{\mu} \rightarrow \nu_{\mu})$, $P(\nu_{\mu} \rightarrow \nu_{e})$, and $P(\nu_{e} \rightarrow \nu_{e})$, and its anti-neutrino counterparts are displayed in focusing around the SA resonance region. 

In the disappearance channels, $P(\nu_{\mu} \rightarrow \nu_{\mu})$ and $P(\nu_{e} \rightarrow \nu_{e})$, the deviation from unity of the probability can be large, $\sim \mathcal{O} (1)$, and the agreement is generally good. For $P(\nu_{e} \rightarrow \nu_{e})$ (and its anti-neutrino counterpart) the agreement is $\sim 10^{-3}$ or better in most of the entire region of the $L-E$ space in Fig.~\ref{fig:exact-EFIC-comparison}. For $P(\nu_{\mu} \rightarrow \nu_{\mu})$ the agreement is better than $\sim 10^{-2}$ in the SA resonance region $E \gsim 1$ TeV assuming $1 \mbox{eV}^2 \lsim \Delta m^2_{41} \lsim 10 \mbox{eV}^2$. For $P(\nu_{\mu} \rightarrow \nu_{e})$ the situation is different. This probability is small, $\sim 10^{-2}$, due to suppression by a factor $s^2_{14} s^2_{24}$ ($s^2_{14}$) at off peak in the neutrino (anti-neutrino) channel, see Tables~\ref{tab:suppression-nu} and~\ref{tab:suppression-nubar} in section~\ref{sec:suppress-probability}. In most of the SA region the agreement is better than $\sim 10^{-2}$, but we observe a layer structure in part of which the agreement is worse.\footnote{
It is well known that if the probability itself is tiny the fractional difference can become large, even larger than unity. It happens, typically, when a small-amplitude variations of the two probabilities are off phase to each other. Dividing by a small number, the values of the fractional difference could be over-amplified, giving a misleading impression, which should not be trusted at a face value as an indicator of relative accuracy.  }

The effective theory works with the accuracy at a level of 1\%, or better, globally in the region of our interest. Then, it would be worthwhile to think about implementing the SA resonance effective theory into a numerical code, as it can accommodate variation of the earth matter density. This is an alternative way to utilize the effective theory without recourse to the further approximations to be discussed after section~\ref{sec:IC-theory-nu-channel}. 

In this paper we work under the uniform density approximation. Therefore, we cannot treat e.g., the shell structure of the earth, such as the crust, mantle, and core. However, we can introduce adiabatic density variation into our effective theory and then we could discuss the aspect of parametric enhancement of the SA resonance in our framework, as initiated in ref.~\cite{Esmaili:2013vza}. For introductory discussions of the parametric enhancement we quote ref.~\cite{Blennow:2013rca}, and for the extensive references including the originals refs.~\cite{Akhmedov:2006hb,Blennow:2013rca}. 

\section{State space of the sterile-active (SA) resonance effective theory} 
\label{sec:state-space}

In the $(3+1)$ model, the relation between the flavor and mass eigenstates are given in vacuum by 
\begin{eqnarray} 
&& 
\nu_{\text{flavor}} = 
U_{\text{sterile}} U_{\text{active}} 
\nu_{ \text{mass} }
\nonumber \\
&=& 
U_{34} (\theta_{34}, \phi_{34} ) U_{24} (\theta_{24}, \phi_{24} ) U_{14} (\theta_{14} ) 
U_{23} (\theta_{23}, \delta ) U_{13} (\theta_{13} ) U_{12} (\theta_{12} ) 
\nu_{ \text{mass} }.
\label{flavor-mass-vac} 
\end{eqnarray}
In matter, with the approximation of the leading-order DMP for the active sector, the flavor state - matter eigenstate relation is given by (denoting the latter as the ``bar'' state) 
\begin{eqnarray} 
&& 
\nu_{\text{flavor}} 
= 
U_{34} (\widetilde{\theta}_{34}, \widetilde{\phi}_{34} ) U_{24} (\widetilde{\theta}_{24}, \widetilde{\phi}_{24} ) U_{14} (\widetilde{\theta}_{14} ) 
U_{23} (\theta_{23}, \delta ) U_{13 } ( \widetilde{\theta}_{13} ) U_{12} ( \widetilde{\theta}_{12} ) 
\bar{\nu}_{\text{mass}} 
\label{flavor-mass-matt} 
\end{eqnarray}
where $\widetilde{\theta}_{ij}$ denotes the matter-affected mixing angles. Note that we have used the ATM convention for the $\nu$SM part of the flavor mixing matrix in the both eqs.~\eqref{flavor-mass-vac} and~\eqref{flavor-mass-matt}. 

In the above discussion we have taken the neutrino channel, but the similar relation can be written down for the flavor and matter eigenstates in the anti-neutrino system. It just requires the replacement $U_{ij} (\theta_{ij}) \rightarrow [ U_{ij} (\theta_{ij}) ]^*$ in vacuum, and in matter $U_{ij} (\widetilde{\theta}_{ij}) \rightarrow [ U_{ij} (\widetilde{\theta}_{ij}) ]^*$ and the sign changes in the matter potentials $a$ and $b$. 

To obtain the effective theory for the SA resonance from the $(3+1)$ model, we take the SA limit. In the neutrino system 
\begin{eqnarray} 
&& 
\left[
\begin{array}{c}
\nu_{e} \\
\nu_{\mu} \\
\nu_{\tau} \\
\nu_{S} \\
\end{array}
\right] 
= 
U_{34} (\widetilde{\theta}_{34}, \widetilde{\phi}_{34} ) U_{24} (\widetilde{\theta}_{24}, \widetilde{\phi}_{24} ) U_{14} (\widetilde{\theta}_{14} ) 
U_{23} (\theta_{23}, \delta ) 
\left[
\begin{array}{cccc}
0 & 0 & 1 & 0 \\
- 1 & 0 & 0 & 0 \\
0 & - 1 & 0 & 0 \\
0 & 0 & 0 & 1 \\
\end{array}
\right] 
\left[
\begin{array}{c}
\bar{\nu}_{1} \\
\bar{\nu}_{2} \\
\bar{\nu}_{3} \\
\bar{\nu}_{4} \\
\end{array}
\right] 
\nonumber \\
&=&
U_{34} (\widetilde{\theta}_{34}, \widetilde{\phi}_{34} ) 
U_{24} (\widetilde{\theta}_{24}, \widetilde{\phi}_{24} ) 
U_{14} (\widetilde{\theta}_{14} ) 
\left[
\begin{array}{c}
\bar{\nu}_{3} \\
- \left( c_{23} \bar{\nu}_{1} + e^{ i \delta} s_{23} \bar{\nu}_{2} \right) \\
e^{ - i \delta} s_{23} \bar{\nu}_{1} - c_{23} \bar{\nu}_{2} \\
\bar{\nu}_{4} \\
\end{array}
\right] 
\label{flavor-mass-IC-nu} 
\end{eqnarray}
where we have used that in the SA limit eq.~\eqref{13-12-freezed}. 

In section~\ref{sec:diagonalization-1-4} we will perform the 1-4 rotation to diagonalize the right-most 1-4 level crossing in the superficial level crossing diagram in Fig.~\ref{fig:superficial-level-cross}. From the physical state structure illuminated in eq.~\eqref{flavor-mass-IC-nu}, the $\widetilde{\theta}_{14}$ rotation acts onto the physical $\bar{\nu}_{3}$ and $\bar{\nu}_{4}$ states. 
If we turn the superficial diagram to the ``real level crossing diagram'' in which all the level crossings are diagonalized (an example in the $\nu$SM is displayed in Fig.~1 in ref.~\cite{Denton:2016wmg}), we are solving the physical 3-4 level crossing. 

In the anti-neutrino system, as the matter-affected $\widetilde{\theta}_{13}$ and $\widetilde{\theta}_{12}$ effectively vanish, so that $U_{13} ( \widetilde{\theta}_{13} ) ^* U_{12} ( \widetilde{\theta}_{12} )^* \rightarrow 1 \equiv \text{diag} [1,1,1,1]$. Therefore, in the anti-neutrino system no reshuffling of the matter-eigenstates (as in eq.~\eqref{flavor-mass-IC-nu}) occur. 

\section{Analysis of SA resonance effective theory: Neutrino channel} 
\label{sec:IC-theory-nu-channel} 

We have confirmed in section~\ref{sec:accuracy} that the effective theory of SA resonance can describe the resonance to a 1\% level accuracy. But, our goal in this paper is one step further, developing an analytic framework to handle the effective theory which allows us to obtain the global bird-eye view of the all-channel aspects of the SA resonance. We will carry out this task in sections~\ref{sec:IC-theory-nu-channel}-\ref{sec:probability-nu} and~\ref{sec:IC-theory-nubar-channel}-\ref{sec:probability-nubar} for the neutrino and anti-neutrino channels, respectively. 

We diagonalize the dominant part of the Hamiltonian by rotating it with the resonance angles. Here, we mean by the ``resonance angle'' the matter-effect-modified version of one of the mixing angles originally defined in vacuum. Such matter-affected mixing angle first appeared in the treatment of the MSW resonance enhancement of solar neutrino transformation~\cite{Wolfenstein:1977ue,Mikheyev:1985zog}. Which vacuum angle is to be elevated to the resonance angle is dictated by the superficial level crossing diagram, see Fig.~\ref{fig:superficial-level-cross}, as we have discussed in section~\ref{sec:SA-resonance} . 
To our understanding, this is nothing but the principle employed in the $\nu$SM to construct the atmospheric~\cite{Arafune:1996bt,Cervera:2000kp,Freund:2001pn,Akhmedov:2004ny,Minakata:2015gra} and the solar-resonance~\cite{Martinez-Soler:2019nhb} perturbation theories. It has been inherited to the DMP and AKT theories~\cite{Denton:2016wmg,Agarwalla:2013tza} which possess validities for the both solar and atmospheric resonances. 

\subsection{Effective Hamiltonian and introducing the tilde basis} 
\label{sec:tilde-basis-nu}

In the neutrino channel, following the discussion in section~\ref{sec:effective-theory}, we take the Hamiltonian 
\begin{eqnarray} 
\hspace{-3mm} 
2E H_{\text{flavor}} 
&=&
U_{34} (\theta_{34}, \delta_{34} ) 
U_{24} (\theta_{24}, \delta_{24} ) U_{14} (\theta_{14} ) 
\left[
\begin{array}{cccc}
0 & 0 & 0 & 0\\
0 & 0 & 0 & 0\\
0 & 0 & 0 & 0\\
0 & 0 & 0 & \Delta m^2_{41} \\
\end{array}
\right] 
U_{14} (\theta_{14} ) ^{\dagger} 
U_{24} (\theta_{24} \delta_{24} ) ^{\dagger} 
U_{34} (\theta_{34} \delta_{34} ) ^{\dagger} 
\nonumber \\
&+&
\left[
\begin{array}{cccc}
a & 0 & 0 & 0\\
0 & 0 & 0 & 0\\
0 & 0 & 0 & 0\\
0 & 0 & 0 & b \\
\end{array}
\right] 
\label{H-Flavor-effective-nu}
\end{eqnarray}
to diagonalize. To avoid frequent appearance of the cumbersome factor $(1/2E)$, we often write down the expression of $2E$ times Hamiltonian here and in what follows in this paper. 

The state space of the SA resonance effective theory is non-trivial after the SA limit is taken, through which the 1-3 and 1-2 level crossings both ``crash''. Therefore, we use the generic artificial state labels $i=1, 2, 3, 4$ with the eigenvalues $\bar{\lambda}_{i}$, and pay attention to how they are related to the energy eigenstates of the $(3+1)$ model. In this context, the important key feature is mentioned in section~\ref{sec:state-space}: We will perform the 1-4 state rotation in the artificial basis but it is the 3-4 rotation in the physical space, as illuminated in eq.~\eqref{flavor-mass-IC-nu}. 

Even after an enormous simplification of treating the matter-affected $\nu$SM part, a still remaining complexity of $H_{\text{flavor}}$ suggests introduction of the ``tilde basis'', \\ 
$\widetilde{H} = U_{34} (\theta_{34}, \phi_{34} ) ^{\dagger} H_{\text{flavor}} U_{34} (\theta_{34}, \phi_{34} )$, to diagonalize the dominant part of the Hamiltonian. We note that the way we define the tilde basis may not be unique. However, this is the natural choice if we remember our experiences in refs.~\cite{Minakata:2015gra,Denton:2016wmg}, in which we have rotated away the outmost $U_{23} (\theta_{23}, \delta)$ rotation matrix from the Hamiltonian to define the tilde basis. In our present case, however, the $U_{34} (\theta_{34}, \phi_{34} )$ does rotate the matter potential part of the Hamiltonian, yet still keeping it in a reasonably simple form.\footnote{
We note that in DMP with non-unitarity incorporated~\cite{Minakata:2021nii}, $\widetilde{H}$ includes the terms of $U_{23}$ rotated matter potential. } 

Our tilde-basis Hamiltonian then reads 
\begin{eqnarray} 
&&
2E \widetilde{H} 
\equiv  
U_{34} (\theta_{34}, \phi_{34} ) ^{\dagger} 
( 2E H_{\text{flavor}} ) 
U_{34} (\theta_{34}, \phi_{34} ) 
\nonumber \\
&=& 
\left[
\begin{array}{cccc}
\left( s^2_{14} \Delta m^2_{41} + a \right) & 0 & 0 & 
c_{24} c_{14} s_{14} \Delta m^2_{41} \\
0 & s^2_{24} c^2_{14} \Delta m^2_{41} & 0 & 0 \\
0 & 0 & s^2_{34} b & 0 \\
c_{24} c_{14} s_{14} \Delta m^2_{41} & 0 & 0 & 
\left( c^2_{24} c^2_{14} \Delta m^2_{41} + c^2_{34} b \right) \\
\end{array}
\right] 
\nonumber \\
&+& 
\left[
\begin{array}{cccc}
0 & 
e^{ i \phi_{24} } s_{24} c_{14} s_{14} \Delta m^2_{41} & 0 & 0 \\
e^{ - i \phi_{24} } s_{24} c_{14} s_{14} \Delta m^2_{41} & 
0 & 0 & 
e^{ - i \phi_{24} } c_{24} s_{24} c^2_{14} \Delta m^2_{41} \\
0 & 0 & 0 & - e^{ - i ( \delta + \phi_{34} ) } c_{34} s_{34} b \\
0 & e^{ i \phi_{24} } c_{24} s_{24} c^2_{14} \Delta m^2_{41} & 
- e^{ i ( \delta + \phi_{34} ) } c_{34} s_{34} b & 
0 \\
\end{array}
\right], 
\nonumber \\
&\equiv& 
2E \widetilde{H} _{0} + 2E \widetilde{H} _{1}, 
\label{tilde-H-nu}
\end{eqnarray}
where we have denoted $2E \widetilde{H}$ in a decomposed form into 
$2E \widetilde{H} _{0}$ and $2E \widetilde{H} _{1}$, the second and third lines in eq.~\eqref{tilde-H-nu}, respectively, expecting diagonalization of $2E \widetilde{H} _{0}$ by the 1-4 rotation. 

\subsection{Lack of the universal expansion parameter}
\label{sec:no-univ-param}

Before proceeding to the diagonalization of $\widetilde{H} _{0}$, we remark on some special feature of the perturbative framework, lack of the universal expansion parameter. This remark applies also to the anti-neutrino channel. In many of the perturbative frameworks to treat the $\nu$SM including DMP~\cite{Denton:2016wmg}, there exists a universal expansion parameter $\Delta m^2_{21} / \Delta m^2_{31} \simeq 0.03$ by which smallness of the correction terms is ensured, irrespective of the values of the mixing angles. In such theories one can address convergence of the perturbative series, see e.g., ref.~\cite{Minakata:2023ice}. 
Whereas in our framework, if we do not assume smallness of the sterile mixing angles, every element in $2E \widetilde{H}$ is of the same order of magnitude. It means that $\widetilde{H} _{0}$ and $\widetilde{H} _{1}$ have similar sizes to each other, so that, in general, perturbative smallness of the correction induced by $\widetilde{H} _{1}$ is not guaranteed.

Despite an apparent drawback of our analytic framework we will proceed to formulate the perturbation theory. This is a theoretically sound framework in the sense that unitarity is obeyed in order by order in perturbation theory. 
Then, the obvious question would be: Is such a framework useful? In the light of our objective of this study, and appealing the smallness of the sterile mixing angles, $\theta_{j 4}$ ($j =1-3$), we answer the question in the positive. In fact, the rather severe constraints on them are derived by the various phenomenological analyses~\cite{Dentler:2018sju,MINOS:2017cae,T2K:2019efw,NOvA:2017geg,IceCube:2024uzv}. 
In our qualitative discussions later, we often refer a rough guideline for the suppression factor we tentatively choose, for brevity, $s_{j 4} \sim 0.1$ for all $j$.\footnote{
A brief comment is ready on our rough guideline $s_{j 4} \sim 0.1$. Of course, one can improve the discussion to implement the bounds on sterile mixing angles obtained by various analyses assuming the $(3+1)$ model. However, at present, the preferred regions for the sterile neutrino interpretation by the various measurement did not appear to converge. The bound is sometimes derived under the treatment of ``one angle at one time'' approach. Therefore, we prefer to stay on qualitative remarks based on the above rough guideline, as it is in harmony with the level of qualitative validity of our analysis framework. }
We will present an organized discussion in section~\ref{sec:suppress-probability} of in what sense our framework is useful, and to what extent it can illuminate the global and qualitative features of the SA resonance. 

\subsection{Diagonalization of $\widetilde{H} _{0}$ by the 1-4 rotation: From the tilde to bar bases} 
\label{sec:diagonalization-1-4}

We have introduced the unified notation of the matter-affected mixing angles as $\widetilde{\theta}_{ij}$ corresponding to the vacuum angle $\theta_{ij}$. Therefore, the rotation matrices with the matter-affected mixing angles and phases have the same forms as in vacuum, eq.~\eqref{sterile-U-ATM}, but ($\theta_{j 4}$, $\phi_{j 4}$) replaced by ($\widetilde{\theta}_{j 4}$, $\widetilde{\phi}_{j 4}$) where $j=1,2,3$: That is, $U_{14} ( \widetilde{\theta}_{14} )$, 
$U_{24} (\widetilde{\theta}_{24}, \widetilde{\phi}_{24} )$, and 
$U_{34} ( \widetilde{\theta}_{34},  \widetilde{\phi}_{34} )$, where 
$\widetilde{c}_{14} \equiv \cos \widetilde{\theta}_{14}$ and 
$\widetilde{s}_{14} \equiv \sin \widetilde{\theta}_{14}$, etc. 
Notice that $\delta$ stays as it is, because it entered into our formulas due to change into the ATM convention of the $U$ matrix.  

Then, it is easy to diagonalize $2E \widetilde{H} _{0}$ to reach to the diagonalized basis which we call the ``bar'' basis: 
\begin{eqnarray} 
&& 
2E \bar{H} _{0}
= 
U_{14} ( \widetilde{\theta}_{14} ) ^{\dagger}
2E \widetilde{H} _{0}
U_{14} ( \widetilde{\theta}_{14} ) 
= 
\left[
\begin{array}{cccc}
\bar{\lambda}_{1} & 0 & 0 & 0 \\
0 & \bar{\lambda}_{2} & 0 & 0 \\
0 & 0 & \bar{\lambda}_{3} & 0 \\
0 & 0 & 0 & \bar{\lambda}_{4} \\
\end{array}
\right], 
\label{bar-H0-nu}
\end{eqnarray}
where the eigenvalues are given by 
\begin{eqnarray} 
&& 
\hspace{-10mm}
\bar{\lambda}_{1} 
= 
\frac{1}{2} 
\biggl\{
( s^2_{14} + c^2_{24} c^2_{14} ) \Delta m^2_{41} + ( a + c^2_{34} b ) 
- \sqrt{ \left[ ( c^2_{24} c^2_{14} - s^2_{14} ) \Delta m^2_{41} - ( a - c^2_{34} b ) \right] ^2 + \left( c_{24} \sin 2\theta_{14} \Delta m^2_{41} \right)^2 } 
\biggr\}, 
\nonumber \\
&&
\hspace{-10mm}
\bar{\lambda}_{2} 
= 
s^2_{24} c^2_{14} \Delta m^2_{41}, 
\nonumber \\
&&
\hspace{-10mm}
\bar{\lambda}_{3} 
= s^2_{34} b, 
\nonumber \\
&&
\hspace{-10mm}
\bar{\lambda}_{4} 
= 
\frac{1}{2} \biggl\{ 
( s^2_{14} + c^2_{24} c^2_{14} ) \Delta m^2_{41} + ( a + c^2_{34} b ) 
+ 
\sqrt{ \left[ ( c^2_{24} c^2_{14} - s^2_{14} ) \Delta m^2_{41} - ( a - c^2_{34} b ) \right] ^2 + \left( c_{24} \sin 2\theta_{14} \Delta m^2_{41} \right)^2 } 
\biggr\}. 
\nonumber \\
\label{eigenvalues-nu}
\end{eqnarray}
Note that $\bar{\lambda}_{4} > \bar{\lambda}_{3}$.  
The diagonalization condition leads to 
\begin{eqnarray} 
&& 
\cos 2\widetilde{\theta}_{14} 
= 
\frac{ ( c^2_{24} c^2_{14} - s^2_{14} ) \Delta m^2_{41} - ( a - c^2_{34} b ) }
{ \sqrt{ \left[ ( c^2_{24} c^2_{14} - s^2_{14} ) \Delta m^2_{41} - ( a - c^2_{34} b ) \right] ^2 + \left( c_{24} \sin 2\theta_{14} \Delta m^2_{41} \right)^2 } }, 
\nonumber \\
&& 
\sin 2\widetilde{\theta}_{14} 
= 
\frac{ c_{24} \sin 2\theta_{14} \Delta m^2_{41} }
{ \sqrt{ \left[ ( c^2_{24} c^2_{14} - s^2_{14} ) \Delta m^2_{41} - ( a - c^2_{34} b ) \right] ^2 + \left( c_{24} \sin 2\theta_{14} \Delta m^2_{41} \right)^2 } }. 
\label{matter-theta14}
\end{eqnarray}
The behavior of the matter-affected $\widetilde{\theta}_{14}$ implies a resonant enhancement at around the resonance point $( c^2_{24} c^2_{14} - s^2_{14} ) \Delta m^2_{41} = ( a - c^2_{34} b )$.

By the same 1-4 rotation we obtain $2E \bar{H} _{1}$ as 
\begin{eqnarray} 
&&
2E \bar{H} _{1} 
= 
U_{14} ( \widetilde{\theta}_{14} ) ^{\dagger}
2E \widetilde{H} _{1}
U_{14} ( \widetilde{\theta}_{14} ) 
= 
\left[
\begin{array}{cccc}
0 & A_{12} & A_{13} & 0 \\
A_{21} & 0 & 0 & A_{24} \\
A_{31} & 0 & 0 & A_{34} \\
0 & A_{42} & A_{43} & 0 \\
\end{array}
\right], 
\label{bar-H1-nu}
\end{eqnarray}
where the non-vanishing $A_{ij}$ elements are given by 
\begin{eqnarray} 
&&
A_{12}  
= 
e^{ i \phi_{24} } s_{24} \left( \widetilde{c}_{14} c_{14} s_{14} 
- \widetilde{s}_{14} c_{24} c^2_{14} \right) \Delta m^2_{41} 
= 
A_{21}^*, 
\nonumber \\
&& 
A_{13} 
= 
e^{ i ( \delta + \phi_{34} ) } \widetilde{s}_{14} c_{34} s_{34} b 
= 
A_{31}^*, 
\nonumber \\
&& 
A_{24} 
= 
e^{ - i \phi_{24} } s_{24} \left( \widetilde{s}_{14} c_{14} s_{14} 
+ \widetilde{c}_{14} c_{24} c^2_{14} \right) \Delta m^2_{41} 
= 
A_{42}^*, 
\nonumber \\
&& 
A_{34} 
= 
- e^{ - i ( \delta + \phi_{34} ) } \widetilde{c}_{14} c_{34} s_{34} b 
= 
A_{43}^*, 
\label{Aij-summary-nu}
\end{eqnarray}
in which hermiticity $A_{ij} = A_{ji}^*$ is used to simplify the presentation. 

\subsection{Computation of the flavor basis $S$ matrix}
\label{sec:flavor-basisS}

For calculating the $S$ matrix elements we utilize the well-known perturbative formulation, which we will sketch below. With $\bar{H} _{0}$ and $\bar{H} _{1}$ at hand in eqs.~\eqref{bar-H0-nu} and~\eqref{bar-H1-nu} it is straightforward to compute the bar basis $S$ matrix in the neutrino channel. The similar computation can be done in the anti-neutrino channel with $\bar{H} _{0}$ and $\bar{H} _{1}$ given in eqs.~\eqref{bar-H0-nubar} and \eqref{bar-H1-nubar}. 

\subsubsection{Calculation of the bar-basis $\bar{S}$ matrix to first order} 
\label{sec:calculation-barS}

To calculate $\bar{S}$ matrix we define $\Omega(x)$ as
\begin{eqnarray} 
\Omega(x) = e^{i \bar{H}_{0} x} \bar{S} (x).
\label{def-omega}
\end{eqnarray}
Using $i \frac{d}{dx} \bar{S} = \bar{H} (x) \bar{S}$, $\Omega(x)$ obeys the evolution equation
\begin{eqnarray} 
i \frac{d}{dx} \Omega(x) = H_{1} \Omega(x) 
\label{omega-evolution}
\end{eqnarray}
where
\begin{eqnarray} 
H_{1} \equiv e^{i \bar{H}_{0} x} \bar{H}^{(1)} e^{-i \bar{H}_{0} x}.
\label{def-H1}
\end{eqnarray}
Then, $\Omega(x)$ can be computed perturbatively as
\begin{eqnarray} 
\Omega(x) &=& 1 + 
(-i) \int^{x}_{0} dx' H_{1} (x') + 
(-i)^2 \int^{x}_{0} dx' H_{1} (x') \int^{x'}_{0} dx'' H_{1} (x'') 
+ \cdot \cdot \cdot,
\label{Omega-expansion}
\end{eqnarray}
and the $\bar{S}$ matrix is given by
\begin{eqnarray} 
\bar{S} (x) =  
e^{-i \bar{H}_{0} x} \Omega(x). 
\label{bar-Smatrix}
\end{eqnarray}
For simplicity of the expressions we use the notation ($j=1,2,3,4$) 
\begin{eqnarray} 
h_{j} = \frac{ \bar{\lambda}_{j} }{2E}, 
\label{hj-def}
\end{eqnarray}
which leads to 
\begin{eqnarray} 
&&
e^{ \pm i \bar{H}_{0} x } 
=
\left[
\begin{array}{cccc}
e^{ \pm i h_{1} x } & 0 & 0 & 0 \\
0 & e^{ \pm i h_{2} x } & 0 & 0 \\
0 & 0 & e^{ \pm i h_{3} x } & 0 \\
0 & 0 & 0 & e^{ \pm i h_{4} x } \\
\end{array}
\right]. 
\label{barS-0th}
\end{eqnarray}
As $\Omega=1$ in the leading order, the minus sign choice in eq.~\eqref{barS-0th} is nothing but the zeroth-order bar-basis $S$ matrix $\bar{S} (x) ^{(0)}$. 

Then, $H_{1} $ can be calculated as 
\begin{eqnarray} 
&& 
H_{1} \equiv e^{i \bar{H}_{0} x} \bar{H}^{(1)} e^{-i \bar{H}_{0} x} 
\nonumber \\
&=&
\frac{1}{2E}
\left[
\begin{array}{cccc}
0 & A_{12} e^{ - i ( h_{2} - h_{1} ) x } & A_{13} e^{ - i ( h_{3} - h_{1} ) x } & 0 \\
A_{21} e^{ i ( h_{2} - h_{1} ) x } & 0 & 0 & A_{24} e^{ - i ( h_{4} - h_{2} ) x } \\
A_{31} e^{ i ( h_{3} - h_{1} ) x } & 0 & 0 & A_{34} e^{ - i ( h_{4} - h_{3} ) x } \\
0 & A_{42} e^{ i ( h_{4} - h_{2} ) x } & A_{43} e^{ i ( h_{4} - h_{3} ) x } & 0 \\
\end{array}
\right]. 
\label{H1-result-nu}
\end{eqnarray} 
Using eq.~\eqref{Omega-expansion} one can easily calculate $\Omega (x)$ to first order, and then the first-order $\bar{S}$ matrix as  
\begin{eqnarray} 
&&
\bar{S} (x) ^{(1)} 
= e^{-i \bar{H}_{0} x} \Omega (x)^{(1)} 
\nonumber \\
&&
\hspace{-10mm}
= 
\frac{1}{2E} 
\left[
\begin{array}{cccc}
0 & A_{12} \frac{ e^{ - i h_{2} x } - e^{ - i h_{1} x } }{ ( h_{2} - h_{1} ) } & 
A_{13} \frac{ e^{ - i h_{3} x } - e^{ - i h_{1} x } }{ ( h_{3} - h_{1} ) } & 0 \\
A_{21} \frac{ e^{ - i h_{2} x } - e^{ - i h_{1} x } }{ ( h_{2} - h_{1} ) } & 0 & 0 & 
A_{24} \frac{ e^{ - i h_{4} x } - e^{ - i h_{2} x }  }{ ( h_{4} - h_{2} ) } \\
A_{31} \frac{ e^{ - i h_{3} x } - e^{ - i h_{1} x } }{ ( h_{3} - h_{1} ) } & 0 & 0 & 
A_{34} \frac{ e^{ - i h_{4} x } - e^{ - i h_{3} x } }{ ( h_{4} - h_{3} ) } \\
0 & A_{42} \frac{ e^{ - i h_{4} x } - e^{ - i h_{2} x } }{ ( h_{4} - h_{2} ) } & 
A_{43} \frac{ e^{ - i h_{4} x } - e^{ - i h_{3} x } }{ ( h_{4} - h_{3} ) } & 0 \\
\end{array}
\right] 
\equiv 
\left[
\begin{array}{cccc}
0 & \bar{S}_{12} ^{(1)} & \bar{S}_{13} ^{(1)} & 0 \\
\bar{S}_{21} ^{(1)} & 0 & 0 & \bar{S}_{24} ^{(1)} \\
\bar{S}_{31} ^{(1)} & 0 & 0 & \bar{S}_{34} ^{(1)} \\
0 & \bar{S}_{42} ^{(1)} & \bar{S}_{43} ^{(1)} & 0 \\
\end{array}
\right]. 
\nonumber \\
\label{barS-1st-nu}
\end{eqnarray}
Notice that $\bar{S} (x) ^{(1)}$ matrix elements satisfy the relation 
$\bar{S}_{ij} ^{(1)} = \left[ \bar{S}_{ji} ^{(1)} \right]^{T*}$, where the $T*$ operation stands for the transformation only on $A_{ij} \rightarrow A_{ji}$ (hermitian conjugation, see eq.~\eqref{Aij-summary-nu}), with the kinematical part $( e^{ - i h_{j} x } - e^{ - i h_{i} x } ) / ( h_{j} - h_{i} )$ untouched. It implies that the generalized $T$ invariance, or $T$-violation only by the complex number, see e.g., ref.~\cite{Fong:2017gke}.

\subsubsection{From the bar basis to the flavor basis: Neutrino channel} 
\label{sec:bar-to-flavorS}

The bar-basis Hamiltonian is related to the tilde-basis one as $\bar{H} = U_{14} ( \widetilde{\theta}_{14} )^{\dagger} \widetilde{H} U_{14} ( \widetilde{\theta}_{14} )$, and the tilde basis to the flavor basis as 
$\widetilde{H} = U_{34} (\theta_{34}, \phi_{34} )^{\dagger} H_{\text{flavor}} U_{34} (\theta_{34}, \phi_{34} )$. Then, the bar-basis to flavor-basis relation is given by 
\begin{eqnarray} 
&&
\bar{H} 
= 
U_{14} ( \widetilde{\theta}_{14} )^{\dagger}
U_{34} (\theta_{34}, \phi_{34} )^{\dagger} 
H_{\text{flavor}} 
U_{34} (\theta_{34}, \phi_{34} ) 
U_{14} ( \widetilde{\theta}_{14} ), 
\nonumber \\
&&
H_{\text{flavor}} 
= 
U_{34} (\theta_{34}, \phi_{34} ) U_{14} ( \widetilde{\theta}_{14} ) 
\bar{H} 
U_{14} ( \widetilde{\theta}_{14} )^{\dagger} U_{34} (\theta_{34}, \phi_{34} )^{\dagger}.  
\label{flavor-bar-H}
\end{eqnarray}
The same relation holds for the $S$ matrix: 
\begin{eqnarray} 
&&
S_{\text{flavor}} 
= 
U_{34} (\theta_{34}, \phi_{34} ) 
U_{14} ( \widetilde{\theta}_{14} ) 
\bar{S} 
U_{14} ( \widetilde{\theta}_{14} )^{\dagger} 
U_{34} (\theta_{34}, \phi_{34} )^{\dagger}. 
\label{flavor-bar-S}
\end{eqnarray}

Given the zeroth and the first-order bar basis $\bar{S}$ matrices (the minus sign choice in eq.~\eqref{barS-0th}, and in eq.~\eqref{barS-1st-nu}), it is now straightforward to compute the flavor basis $S$ matrix by performing the two rotations that appear in eq.~\eqref{flavor-bar-S}. 
It is customary to use the Greek characters for the subscript of the flavor-basis $S$ matrix elements, which we follow for the active neutrino species. For the sterile state we use the capital-$S$ for simplicity.  
To second order, the flavor-basis $S$ matrix has the following structure: 
\begin{eqnarray} 
&&
\hspace{-8mm}
S = 
\left[
\begin{array}{cccc}
S_{ee}  ^{(0)} & 0 & S_{e \tau} ^{(0)} & S_{e S} ^{(0)} \\
0 & S_{\mu \mu} ^{(0)} & 0 & 0\\
S_{\tau e} ^{(0)} & 0 & S_{\tau \tau} ^{(0)} & S_{\tau S} ^{(0)} \\
S_{S e} ^{(0)} & 0 & S_{S \tau} ^{(0)} & S_{S S} ^{(0)} \\
\end{array}
\right] 
+ 
\left[
\begin{array}{cccc}
0 & S_{e \mu} ^{(1)} & S_{e \tau} ^{(1)} & S_{e S} ^{(1)} \\
S_{\mu e} ^{(1)} & 0 & S_{\mu \tau} ^{(1)} & S_{\mu S} ^{(1)} \\
S_{\tau e} ^{(1)} & S_{\tau \mu} ^{(1)} & S_{\tau \tau} ^{(1)} & S_{\tau S} ^{(1)} \\
S_{S e} ^{(1)} & S_{S \mu} ^{(1)} & S_{S \tau} ^{(1)} & S_{S S} ^{(1)} \\
\end{array}
\right] 
+ 
\left[
\begin{array}{cccc}
S_{ee}  ^{(2)} & 0 & S_{e \tau} ^{(2)} & S_{e S} ^{(2)} \\
0 & S_{\mu \mu} ^{(2)} & S_{\mu \tau} ^{(2)} & S_{\mu S} ^{(2)} \\
S_{\tau e} ^{(2)} & S_{\tau \mu} ^{(2)} & S_{\tau \tau} ^{(2)} & S_{\tau S} ^{(2)} \\
S_{S e} ^{(2)} & S_{S \mu} ^{(2)} & S_{S \tau} ^{(2)} & S_{S S} ^{(2)} \\
\end{array}
\right].
\label{flavorS-nu-0th-2nd}
\end{eqnarray} 

To know the qualitative features of neutrino flavor transformation we need, in most of the channels, the $S$ elements up to the first order. For convenience for the readers who want to follow the calculation we collect the expressions of the flavor basis $S$ matrix elements up to the first order in Appendix~\ref{sec:flavor-S-nu}. But, as $S_{\mu \mu} ^{(1)} = S_{ee} ^{(1)} = 0$, the second order elements are required when we discuss the $\nu_{\mu} \rightarrow \nu_{\mu}$ and $\nu_{e} \rightarrow \nu_{e}$ channels. Taking the $\nu_{\mu} \rightarrow \nu_{\mu}$ channel as an example, a sketchy description of how to compute the second-order flavor basis $S$ matrix and the expression of $S_{\mu \mu}^{(2)}$ are given in Appendix~\ref{sec:S-second-order}. 
One may still ask why zeros only in $S_{\mu \mu} ^{(1)}$ and $S_{e e} ^{(1)}$. The answer is that the structure of diagonal zero in $\bar{S} (x) ^{(1)}$ matrix is preserved by the 1-4 and 3-4 rotations only in the first $2 \times 2$ sub-matrix. 

Now, we are ready to compute the oscillation probabilities in all the flavor oscillation channels. Before proceeding to this task, we make a general remark that all the probabilities computed respect $T$-invariance despite that the phases $\delta$ and $\phi_{34}$ show up in the ample place in the $S$ matrix elements, as can be seen in Appendix~\ref{sec:flavor-S-nu}. More comments on this $T$-invariance will be given in section~\ref{sec:T-invariance}. 

\section{Sterile-active resonance in the neutrino channel: Probabilities} 
\label{sec:probability-nu}

In this section we prefer to focus on the flavor channels which are most relevant for the observation of the SA resonance effects in Neutrino Telescopes. In energy region of $E = 1$-10 TeV, $\nu_{\mu}$ and $\bar{\nu}_{\mu}$ constitute the dominant component of the atmospheric neutrinos~\cite{Gaisser:2002jj}. Sub-leading $\nu_{e}$ and $\bar{\nu}_{e}$ component is estimated to be about 10\% or less~\cite{Honda:2004yz,Sponsler:2024iej}. The latter reference used the MCEq code~\cite{Fedynitch:2012fs} to solve for the neutrino fluxes analytically. For our qualitative discussions in this paper we assume the flavor ratio 
$( \nu_{e} + \bar{\nu}_{e} ) / ( \text{total neutrino flux} ) = 10$\% level (the rest is $\nu_{\mu} + \bar{\nu}_{\mu}$)  in the energy region $E \sim 1$ TeV, see Fig.~4 in ref.~\cite{Sponsler:2024iej}. 

In below we refer the terminology, ``track events'' and ``cascade events'', to classify the event topology. See ref.~\cite{Smithers:2021orb} for concise explanation of these experimental terminology, and for a recent comprehensive discussion of the cascade events. 

We note that it is unlikely that $\nu_{\mu}$ and $\bar{\nu}_{\mu}$ (or, $\nu_{e}$ and $\bar{\nu}_{e}$) components can be separated in an event-by-event basis in the IceCube experiment, at least for now. Therefore, a more elaborate discussion is needed to know how our separate neutrino anti-neutrino treatment could make a realistic sense. See section~\ref{sec:cascade-3-origins} and Tables~\ref{tab:suppression-nu} and~\ref{tab:suppression-nubar}.

\subsection{Track events and $P(\nu_{\mu} \rightarrow \nu_{\mu})$: Neutrino channel} 
\label{sec:track-nu}

Here, we first discuss the survival probability $P(\nu_{\mu} \rightarrow \nu_{\mu})$ to know how many $\nu_{\mu}$ survives in the earth matter, hoping that it may reveal the features useful to understand the characteristics of the experimental data. As discussed above $S_{\mu \mu} = S_{\mu \mu}^{(0)} + S_{\mu \mu}^{(2)}$, no first-order contribution. We notice that $S_{\mu \mu}^{(0)} = e^{ - i h_{2} x }$, and refer Appendix~\ref{sec:S-second-order} for $S_{\mu \mu}^{(2)}$. Then, we obtain the probability to second order 
\begin{eqnarray} 
&&
P(\nu_{\mu} \rightarrow \nu_{\mu}) 
= 
| S_{\mu \mu}^{(0)} |^2 + 2 \mbox{Re} \left[ \left\{ S_{\mu \mu}^{(0)} \right\}^* S_{\mu \mu}^{(2)} \right] 
\nonumber \\
&=& 
1 
- 4 s^2_{24} 
\left( \widetilde{c}_{14} c_{14} s_{14} - \widetilde{s}_{14} c_{24} c^2_{14} \right) ^2 
\left( \frac{ \Delta m^2_{41} }{2E} \right)^2 
\frac{ 1 }{ ( h_{2} - h_{1} )^2 } 
\sin^2 \frac{ ( h_{2} - h_{1} ) x }{2} 
\nonumber \\
&-&
4 s^2_{24} 
\left( \widetilde{s}_{14} c_{14} s_{14} + \widetilde{c}_{14} c_{24} c^2_{14} \right) ^2 
\left( \frac{ \Delta m^2_{41} }{2E} \right)^2 
\frac{ 1 }{ ( h_{4} - h_{2} )^2 } 
\sin^2 \frac{ ( h_{4} - h_{2} ) x }{2}.  
\label{Pmumu-2nd}
\end{eqnarray}
Notice that all the eigenvalues are of the similar order $\bar{\lambda}_{i} \sim \Delta m^2_{41}$, see eq.~\eqref{eigenvalues-nu}, as we are working in the SA resonance region, $\Delta m^2_{41} / 2E h_{i} = \Delta m^2_{41} / \bar{\lambda}_{i} \sim 1$. Then the kinematical factors in eq.~\eqref{Pmumu-2nd} are of order unity. 
As we stated in section~\ref{sec:tilde-basis-nu}, for the purpose of rough estimations, we will assume smallness of the sterile mixing angles, $s_{j 4} \sim 0.1$ for all $j=1,2,3$. Then, the second order correction terms in eq.~\eqref{Pmumu-2nd}
are suppressed by $s^2_{24} \simeq 10^{-2}$ under our assumption even at the resonance peak. 
Therefore our treatment predicts that, practically, $\nu_{\mu}$ travels in the Earth matter nearly unaffected. It suggests that the track events in Neutrino Telescopes in the neutrino channel may not be a good probe of the SA resonance effects. 

The absence of $S_{e \mu} ^{(0)}$, $S_{\tau \mu} ^{(0)}$ and $S_{S \mu} ^{(0)}$ in the zeroth-order $S$ matrix in eq.~\eqref{flavorS-nu-0th-2nd} implies that the other $\nu_{\mu}$-row probabilities, $P(\nu_{\mu} \rightarrow \nu_{X})$ ($X = e, \tau, S$), take the form of $| S_{X \mu}^{(1)} |^2$. It leads to the similar $s^2_{24}$ (or higher) suppressions. See Appendix~\ref{sec:probability-summary-nu} for the expressions of $P(\nu_{\mu} \rightarrow \nu_{e})$ and $P(\nu_{\mu} \rightarrow \nu_{\tau})$. 
In section~\ref{sec:suppress-probability} with Tables~\ref{tab:suppression-nu} and~\ref{tab:suppression-nubar}, we will give a more systematic discussion of the suppressions by smallness of the sterile mixing angles. 

\subsection{Cascade events via $P(\nu_{e} \rightarrow \nu_{e})$ and $P(\nu_{\mu} \rightarrow \nu_{X})$: Neutrino channel} 
\label{sec:cascade-nu}

Assuming the atmospheric neutrino flux of $\nu_{\mu}$ entering the earth, the cascade events may occur via $P(\nu_{\mu} \rightarrow \nu_{e})$ or $P(\nu_{\mu} \rightarrow \nu_{\tau})$. But, the estimation based on Appendix~\ref{sec:probability-summary-nu} tells us that $P(\nu_{\mu} \rightarrow \nu_{e})$ is suppressed by $s^2_{24}$, and $P(\nu_{\mu} \rightarrow \nu_{\tau})$ by $s^2_{24} s^2_{34}$ on resonance. See Table~\ref{tab:suppression-nu}. These $\nu_{\mu}$ initialed cascade events would therefore be suppressed to a 1\% level (at the probability) assuming $s_{24} \sim s_{34} \sim 0.1$. 

As mentioned above the atmospheric neutrino flux in the 1-10 TeV region includes 10\% level $\nu_{e}$. With this parent $\nu_{e}$ the cascade events can occur through the oscillations 
\begin{eqnarray} 
&& 
P(\nu_{e} \rightarrow \nu_{e}) 
= | S^{(0)}_{ee} |^2 
= 
1 - \sin^2 2\widetilde{\theta}_{14} \sin^2 \frac{ ( h_{4} - h_{1} ) x }{2}, 
\nonumber \\
&& 
P(\nu_{e} \rightarrow \nu_{\tau}) 
= | S^{(0)}_{\tau e} |^2 
= s^2_{34} 
\sin^2 2\widetilde{\theta}_{14} \sin^2 \frac{ ( h_{4} - h_{1} ) x }{2}, 
\label{Pee-etau}
\end{eqnarray}
where we have restricted to the leading order expressions.\footnote{
$P(\nu_{e} \rightarrow \nu_{S}) = c^2_{34} \sin^2 2\widetilde{\theta}_{14} \sin^2 \frac{ ( h_{4} - h_{1} ) x }{2}$. Given that $P(\nu_{e} \rightarrow \nu_{\mu})$ is of second order, it is easy to observe that $P(\nu_{e} \rightarrow \nu_{\tau}) + P(\nu_{e} \rightarrow \nu_{S})$ cancels the second term in $P(\nu_{e} \rightarrow \nu_{e})$ in eq.~\eqref{Pee-etau}, guaranteeing the zeroth-order unitarity. The $\nu_{e}$-row unitarity to the first order is shown to hold in Appendix~\ref{sec:nue-row}. }
For the expression of $P(\nu_{e} \rightarrow \nu_{e})$ with the second-order corrections, see eq.~\eqref{P-ee-nu} in Appendix~\ref{sec:nue-row}. 

It is obvious that the probabilities in all these three channels, including $P(\nu_{e} \rightarrow \nu_{S})$, enjoy the SA resonance enhancement described by the matter affected $\widetilde{\theta}_{14}$. Notice that $P(\nu_{e} \rightarrow \nu_{e})$ in eq.~\eqref{Pee-etau} has no suppression factor due to the sterile mixing angles on resonance, whereas $P(\nu_{e} \rightarrow \nu_{\tau})$ is suppressed by $s^2_{34}$. Therefore, in the neutrino channel, the dominant contribution to the cascade events comes from the atmospheric $\nu_{e}$ flux with resonance enhanced $P(\nu_{e} \rightarrow \nu_{e})$ in eq.~\eqref{Pee-etau}, which produce 10\% level contribution in flux times $P(\nu_{e} \rightarrow \nu_{e})$. 

\section{Analysis of SA resonance effective theory: Anti-neutrino channel} 
\label{sec:IC-theory-nubar-channel} 

In the anti-neutrino channel the Hamiltonian takes the form 
\begin{eqnarray} 
&& 
2E H_{\text{flavor}} 
\nonumber \\
&=& 
U_{34} (\theta_{34}, \delta_{34} ) ^* 
U_{24} (\theta_{24}, \delta_{24} ) ^* 
U_{14} (\theta_{14} ) ^* 
\left[
\begin{array}{cccc}
0 & 0 & 0 & 0\\
0 & 0 & 0 & 0\\
0 & 0 & 0 & 0\\
0 & 0 & 0 & \Delta m^2_{41} \\
\end{array}
\right] 
[ U_{14} (\theta_{14} )^*  ] ^{\dagger} 
[ U_{24} (\theta_{24} \delta_{24} )^* ] ^{\dagger} 
[ U_{34} (\theta_{34} \delta_{34} )^* ] ^{\dagger} 
\nonumber \\
&+& 
\left[
\begin{array}{cccc}
- |a| & 0 & 0 & 0\\
0 & 0 & 0 & 0\\
0 & 0 & 0 & 0\\
0 & 0 & 0 & - |b| \\
\end{array}
\right]. 
\label{H-Flavor-effective-nubar}
\end{eqnarray}
As in the neutrino case we move to the tilde basis 
$\widetilde{H} = [ U_{34} (\theta_{34}, \phi_{34} )^* ] ^{\dagger} H_{\text{flavor}}  U_{34} (\theta_{34}, \phi_{34} )^*$:  
\begin{eqnarray} 
2E \widetilde{H} 
&=&
\left[
\begin{array}{cccc}
\left( s^2_{14} \Delta m^2_{41} - |a|  \right) & 
e^{ - i \phi_{24} } s_{24} c_{14} s_{14} \Delta m^2_{41} & 0 & 
c_{24} c_{14} s_{14} \Delta m^2_{41} \\
e^{ i \phi_{24} } s_{24} c_{14} s_{14} \Delta m^2_{41} & 
s^2_{24} c^2_{14} \Delta m^2_{41} & 0 & 
e^{ i \phi_{24} } c_{24} s_{24} c^2_{14} \Delta m^2_{41} \\
0 & 0 & - s^2_{34} |b| & e^{ i ( \delta + \phi_{34} ) } c_{34} s_{34} |b| \\ 
c_{24} c_{14} s_{14} \Delta m^2_{41} & 
e^{ - i \phi_{24} } c_{24} s_{24} c^2_{14} \Delta m^2_{41} & e^{ - i ( \delta + \phi_{34} ) } c_{34} s_{34} |b| & 
\left( c^2_{24} c^2_{14} \Delta m^2_{41} - c^2_{34} |b| \right) \\
\end{array}
\right]. 
\nonumber \\
\label{tilde-H-nubar}
\end{eqnarray}

\subsection{Diagonalization of the Hamiltonian: Local resonance approximation}
\label{sec:LRA} 

An important characteristics of the anti-neutrino channels in our SA resonance effective theory, as compared to the neutrino channel, is that it has the two resonance enhancements at the 2-4 and 3-4 crossings, as indicated by the superficial level crossing diagram in Fig.~\ref{fig:superficial-level-cross}. The structure is akin to the one in the neutrino channel in the $\nu$SM (NMO). We thus follow DMP~\cite{Denton:2016wmg}, which does the atmospheric 1-3 rotation first and then the solar 1-2 rotation second. That is, diagonalize the high-energy crossing first and then low-energy one. Thus we do the 2-4 rotation first, and then the 3-4. 

It should be noticed that our procedure implicitly assumes that the diagonalization can be done one by one, the local resonance approximation. We warn that this approximation may not hold in a good accuracy, given that the two resonances can overlap in a good fraction of the parameter space so that they may interact with each other. But, this is the situation we never met in the $\nu$SM neutrino propagation, as the atmospheric 1-3 and the solar 1-2 resonances are far apart to each other. While finding a way to overcome the local resonance approximation is a burning request it eludes us at this moment. 
Then, one may ask: In what sense our analytic approach useful? We will answer this question in section~\ref{sec:characteristic-feature}.  

\subsection{Diagonalization of $\widetilde{H}$ by the 2-4 and then 3-4 rotations} 
\label{sec:diagonalization-24-34}

We proceed to perform first the 2-4 rotation. We decompose $\widetilde{H}$ into $\widetilde{H} _{0}$ and $\widetilde{H} _{1}$, 
\begin{eqnarray} 
2E \widetilde{H} 
&=& 
\left[
\begin{array}{cccc}
\left( s^2_{14} \Delta m^2_{41} - |a|  \right) & 0 & 0 & 0 \\
0 & s^2_{24} c^2_{14} \Delta m^2_{41} & 0 & 
e^{ i \phi_{24} } c_{24} s_{24} c^2_{14} \Delta m^2_{41} \\
0 & 0 & - s^2_{34} |b| & 0 \\ 
0 & e^{ - i \phi_{24} } c_{24} s_{24} c^2_{14} \Delta m^2_{41} & 0 & 
\left( c^2_{24} c^2_{14} \Delta m^2_{41} - c^2_{34} |b| \right) \\
\end{array}
\right] 
\nonumber \\
&+& 
\left[
\begin{array}{cccc}
0 & e^{ - i \phi_{24} } s_{24} c_{14} s_{14} \Delta m^2_{41} & 0 & 
c_{24} c_{14} s_{14} \Delta m^2_{41} \\
e^{ i \phi_{24} } s_{24} c_{14} s_{14} \Delta m^2_{41} & 0 & 0 & 0 \\
0 & 0 & 0 & e^{ i ( \delta + \phi_{34} ) } c_{34} s_{34} |b| \\ 
c_{24} c_{14} s_{14} \Delta m^2_{41} & 
0 & e^{ - i ( \delta + \phi_{34} ) } c_{34} s_{34} |b| & 0 \\
\end{array}
\right] 
\nonumber \\
&\equiv& 
2E \widetilde{H} _{0} + 2E \widetilde{H} _{1}, 
\label{tilde-H-nubar-decompose}
\end{eqnarray}
where the last line defines the two matrices in the first and second lines in eq.~\eqref{tilde-H-nubar-decompose}. 

We diagonalize the 2-4 subspace of $\widetilde{H} _{0}$ by the 2-4 rotation to the hat-basis $\hat{H} _{0}$ by using $U_{24} (\widetilde{\theta}_{24}, \widetilde{\phi}_{24} )$, the rotation matrix in matter, which is defined in eq.~\eqref{sterile-U-ATM} but the vacuum parameters ($\theta_{24}$, $\phi_{24}$) replaced by the matter-affected ones ($\widetilde{\theta}_{24}$, $\widetilde{\phi}_{24}$). 
\begin{eqnarray} 
&& 
2E \hat{H} _{0} 
\equiv 
[ U_{24} ( \widetilde{\theta}_{24},  \widetilde{\phi}_{24} ) ^* ] ^{\dagger}
2E \widetilde{H} _{0}
U_{24} ( \widetilde{\theta}_{24},  \widetilde{\phi}_{24} ) ^*
= 
\left[
\begin{array}{cccc}
\hat{\lambda}_{1} & 0 & 0 & 0 \\
0 & \hat{\lambda}_{2} & 0 & 0 \\
0 & 0 & \hat{\lambda}_{3} & 0 \\
0 & 0 & 0 & \hat{\lambda}_{4} \\
\end{array}
\right]. 
\label{hat-H0-nubar}
\end{eqnarray}
The eigenvalues can be obtained as in the usual eigenvalue problem of the 2-4 crossing as 
\begin{eqnarray} 
&& 
\hat{\lambda}_{1} 
= \left( s^2_{14} \Delta m^2_{41} - |a|  \right), 
\nonumber \\
&&
\hat{\lambda}_{2} 
= 
\frac{1}{2} 
\biggl\{
( c^2_{14} \Delta m^2_{41} - c^2_{34} |b| ) 
- \sqrt{ \left[ \cos 2\theta_{24} c^2_{14} \Delta m^2_{41} - c^2_{34} |b| 
\right]^2 + \left( \sin 2\theta_{24} c^2_{14} \Delta m^2_{41} \right)^2 } 
\biggr\}, 
\nonumber \\
&&
\hat{\lambda}_{3} 
= 
- s^2_{34} |b|, 
\nonumber \\
&&
\hat{\lambda}_{4} 
= 
\frac{1}{2} 
\biggl\{
( c^2_{14} \Delta m^2_{41} - c^2_{34} |b| ) 
+ \sqrt{ \left[ \cos 2\theta_{24} c^2_{14} \Delta m^2_{41} - c^2_{34} |b| 
\right]^2 + \left( \sin 2\theta_{24} c^2_{14} \Delta m^2_{41} \right)^2 } 
\biggr\}, 
\nonumber \\
\label{eigenvalues-hat-basis}
\end{eqnarray}
taking $\hat{\lambda}_{4} > \hat{\lambda}_{2}$. The diagonalization condition determines the cosine and sine of $2 \widetilde{\theta}_{24}$: 
\begin{eqnarray} 
&&
\cos 2 \widetilde{\theta}_{24} 
= 
\frac{ \left[ \cos 2\theta_{24} c^2_{14} \Delta m^2_{41} - c^2_{34} |b| 
\right] } 
{ \sqrt{ \left[ \cos 2\theta_{24} c^2_{14} \Delta m^2_{41} - c^2_{34} |b| 
\right]^2 + \left( \sin 2\theta_{24} c^2_{14} \Delta m^2_{41} \right)^2 } }, 
\nonumber \\
&&
\sin 2 \widetilde{\theta}_{24} 
= 
\frac{ \sin 2\theta_{24} c^2_{14} \Delta m^2_{41} } 
{ \sqrt{ \left[ \cos 2\theta_{24} c^2_{14} \Delta m^2_{41} - c^2_{34} |b| 
\right]^2 + \left( \sin 2\theta_{24} c^2_{14} \Delta m^2_{41} \right)^2 } }, 
\label{matter-theta24}
\end{eqnarray}
which describes the resonance at $\cos 2\theta_{24} c^2_{14} \Delta m^2_{41} = c^2_{34} |b|$. 

It may be better to mention that, as in the case of neutrino channel, there is a $\pm$ sign ambiguity in $\cos 2 \widetilde{\theta}_{24}$. The real and the imaginary parts of the diagonalization condition lead to 
\begin{eqnarray} 
&&
\tan 2 \widetilde{\theta}_{24} 
= 
\frac{ \sin 2\theta_{24} c^2_{14} \Delta m^2_{41} 
\cos ( \widetilde{\phi}_{24} - \phi_{24} ) }
{ \left[ \cos 2\theta_{24} c^2_{14} \Delta m^2_{41} - c^2_{34} |b| 
\right]  }, 
~~~~
\text{and}
~~~~
\sin ( \widetilde{\phi}_{24} - \phi_{24} ) = 0, 
\label{sign-ambiguity} 
\end{eqnarray}
which means $\cos ( \widetilde{\phi}_{24} - \phi_{24} ) = \pm 1$. In the above, in eq.~\eqref{matter-theta24}, we have taken the plus sign, which leads to $\hat{\lambda}_{4} > \hat{\lambda}_{2}$.  
If we took the minus sign the eigenvalue ordering reverses. With the plus (minus) solution $\hat{\lambda}_{2} \rightarrow - \infty$ ($\hat{\lambda}_{2} \rightarrow$ finite) as $|b| \rightarrow + \infty$. Therefore, we confirm that the plus sign solution is physical. 
We will return to the implication of the relationship between the matter-affected and the original phases in section~\ref{sec:T-invariance}. 

By the same 2-4 rotation $\widetilde{H}_{1}$ transforms to $\hat{H} _{1}$ as 
\begin{eqnarray} 
&&
2E \hat{H} _{1} 
\equiv 
[ U_{24} ( \widetilde{\theta}_{24},  \widetilde{\phi}_{24} ) ^* ] ^{\dagger}
2E \widetilde{H} _{1}
U_{24} ( \widetilde{\theta}_{24},  \widetilde{\phi}_{24} ) ^*
= 
\left[
\begin{array}{cccc}
0 & \hat{A}_{12} & 0 & \hat{A}_{14} \\
\hat{A}_{21} & 0 & \hat{A}_{23} & 0 \\
0 & \hat{A}_{32} & 0 & \hat{A}_{34} \\
\hat{A}_{41} & 0 & \hat{A}_{43} & 0 \\
\end{array}
\right], 
\label{hat-H1-nubar}
\end{eqnarray}
where the non-vanishing $\hat{A}_{ij}$ elements are given by 
\begin{eqnarray} 
&&
\hat{A}_{12} 
= 
- e^{ - i \phi_{24} } c_{14} s_{14} \sin ( \widetilde{\theta}_{24} - \theta_{24} ) 
\Delta m^2_{41} 
=
\hat{A}_{21} ^*, 
\nonumber \\
&&
\hat{A}_{14} 
= 
c_{14} s_{14} \cos ( \widetilde{\theta}_{24} - \theta_{24} ) \Delta m^2_{41} 
=
\hat{A}_{41} ^*, 
\nonumber \\
&&
\hat{A}_{23} 
= 
- \widetilde{s}_{24} e^{ i \widetilde{\phi}_{24} } e^{ - i ( \delta + \phi_{34} ) } c_{34} s_{34} |b| 
=
\hat{A}_{32} ^*,  
\nonumber \\
&&
\hat{A}_{34} 
= 
\widetilde{c}_{24} e^{ i ( \delta + \phi_{34} ) } c_{34} s_{34} |b| 
=
\hat{A}_{43} ^*,  
\label{hat-Aij-summary-nubar}
\end{eqnarray}
where we have used $\cos ( \widetilde{\phi}_{24} - \phi_{24} ) = 1$, or $\widetilde{\phi}_{24} - \phi_{24} = 0$. 

Now, we perform the final 3-4 rotation. We re-decompose $\hat{H}$ as 
\begin{eqnarray} 
&&
2E \hat{H} 
= 
\left[
\begin{array}{cccc}
\hat{\lambda}_{1} & 0 & 0 & 0 \\
0 & \hat{\lambda}_{2} & 0 & 0 \\
0 & 0 & \hat{\lambda}_{3} & \hat{A}_{34} \\
0 & 0 & \hat{A}_{43} & \hat{\lambda}_{4} \\
\end{array}
\right] 
+ 
\left[
\begin{array}{cccc}
0 & \hat{A}_{12} & 0 & \hat{A}_{14} \\
\hat{A}_{21} & 0 & \hat{A}_{23} & 0 \\
0 & \hat{A}_{32} & 0 & 0 \\
\hat{A}_{41} & 0 & 0 & 0 \\
\end{array}
\right] 
\equiv 
2E \hat{H} _{0} + 2E \hat{H} _{1}, 
\label{hat-H-decomposition}
\end{eqnarray}
This procedure is necessary to give the 3-4 level crossing an enhanced, usually ``resonant'', character. We make a 3-4 rotation with the matter-affected angle $\widetilde{\theta}_{34}$ to diagonalize the 3-4 subspace of $\hat{H} _{0}$: 
\begin{eqnarray} 
&& 
2E \bar{H} _{0} 
\equiv 
[ U_{34} ( \widetilde{\theta}_{34},  \widetilde{\phi}_{34} ) ^* ] ^{\dagger}
2E \hat{H} _{0}
U_{34} ( \widetilde{\theta}_{34},  \widetilde{\phi}_{34} ) ^*
= 
\left[
\begin{array}{cccc}
\bar{\lambda}_{1} & 0 & 0 & 0 \\
0 & \bar{\lambda}_{2} & 0 & 0 \\
0 & 0 & \bar{\lambda}_{3} & 0 \\
0 & 0 & 0 & \bar{\lambda}_{4} \\
\end{array}
\right], 
\label{bar-H0-nubar}
\end{eqnarray}
where the eigenvalues can be obtained from the 3-4 eigenvalue equation as 
\begin{eqnarray} 
\bar{\lambda}_{1} 
&=&
\hat{\lambda}_{1}, 
\nonumber \\
\bar{\lambda}_{2} 
&=&
\hat{\lambda}_{2}, 
\nonumber \\
\bar{\lambda}_{3} 
&=&
\frac{1}{2} 
\biggl\{ 
( \hat{\lambda}_{3} + \hat{\lambda}_{4} ) 
- \sqrt{ ( \hat{\lambda}_{4} - \hat{\lambda}_{3} )^2 + ( \widetilde{c}_{24} \sin 2\theta_{34} |b| )^2 } 
\biggr\}, 
\nonumber \\
\bar{\lambda}_{4} 
&=& 
\frac{1}{2} 
\biggl\{ 
( \hat{\lambda}_{3} + \hat{\lambda}_{4} ) 
+ \sqrt{ ( \hat{\lambda}_{4} - \hat{\lambda}_{3} )^2 + ( \widetilde{c}_{24} \sin 2\theta_{34} |b| )^2 } 
\biggr\}. 
\label{eigenvalues-bar-basis}
\end{eqnarray} 
The eigenvalues in the hat basis are given in eq.~\eqref{eigenvalues-hat-basis}. 
The diagonalization condition determines the matter-affected mixing angle 
$\widetilde{\theta}_{34}$ as 
\begin{eqnarray} 
&& 
\cos 2\widetilde{\theta}_{34} 
= 
\frac{ ( \hat{\lambda}_{4} - \hat{\lambda}_{3} ) } 
{ \sqrt{ ( \hat{\lambda}_{4} - \hat{\lambda}_{3} )^2 + ( \widetilde{c}_{24} \sin 2\theta_{34} |b| )^2 } }, 
\nonumber \\
&& 
\sin 2\widetilde{\theta}_{34} 
= 
\frac{ \widetilde{c}_{24} \sin 2\theta_{34} |b| } 
{ \sqrt{ ( \hat{\lambda}_{4} - \hat{\lambda}_{3} )^2 + ( \widetilde{c}_{24} \sin 2\theta_{34} |b| )^2 } }. 
\label{matter-theta34}
\end{eqnarray}

Similarly to the 2-4 diagonalization, there exists the $\pm$ sign problem on $\cos ( \widetilde{\phi}_{34} - \phi_{34} )$ attached on $\cos 2\widetilde{\theta}_{34}$. Here we solve it without assuming the asymptotic behavior of the eigenvalues. Through the diagonalization procedure the eigenvalues can be obtained as 
\begin{eqnarray} 
&&
\bar{\lambda}_{3} 
=
\widetilde{c}^2_{34} \hat{\lambda}_{3} + \widetilde{s}^2_{34} \hat{\lambda}_{4} 
- e^{ - i ( \delta + \widetilde{\phi}_{34} ) } \widetilde{c}_{34} \widetilde{s}_{34} \hat{A}_{34} 
- e^{ i ( \delta + \widetilde{\phi}_{34} ) } \widetilde{c}_{34} \widetilde{s}_{34} \hat{A}_{43} 
\nonumber \\
&&
\bar{\lambda}_{4} 
= 
\widetilde{s}^2_{34} \hat{\lambda}_{3} + \widetilde{c}^2_{34} \hat{\lambda}_{4} 
+ e^{ - i ( \delta + \widetilde{\phi}_{34} ) } \widetilde{c}_{34} \widetilde{s}_{34} \hat{A}_{34} 
+ e^{ i ( \delta + \widetilde{\phi}_{34} ) } \widetilde{c}_{34} \widetilde{s}_{34} \hat{A}_{43} 
\label{eigenvalues-computed}
\end{eqnarray}
If we use the plus sign for $\cos 2\widetilde{\theta}_{34}$, as taken in eq.~\eqref{matter-theta34}, we recover the eigenvalues eq.~\eqref{eigenvalues-bar-basis}. But if we place a minus sign on $\cos 2\widetilde{\theta}_{34}$, $\bar{\lambda}_{3}$ in eq.~\eqref{eigenvalues-computed} becomes $\bar{\lambda}_{4}$ in eq.~\eqref{eigenvalues-bar-basis}, and vice versa. But, physically it must be the case that $\bar{\lambda}_{4} > \bar{\lambda}_{3}$, as one can confirm by drawing the ``real level crossing diagram''. Thus, we must choose the plus sign on $\cos 2\widetilde{\theta}_{34}$, as done in eq.~\eqref{matter-theta34}. 

By the same 3-4 rotation $\hat{H}_{1}$ in eq.~\eqref{hat-H-decomposition} transforms to $\bar{H} _{1}$ as 
\begin{eqnarray} 
&&
2E \bar{H} _{1} 
\equiv 
[ U_{34} ( \widetilde{\theta}_{34},  \widetilde{\phi}_{34} ) ^* ] ^{\dagger}
2E \hat{H} _{1}
U_{34} ( \widetilde{\theta}_{34},  \widetilde{\phi}_{34} ) ^*
\nonumber \\
&=&
\left[
\begin{array}{cccc}
0 & \hat{A}_{12} & - e^{ - i ( \delta + \widetilde{\phi}_{34} ) } \widetilde{s}_{34} \hat{A}_{14} & \widetilde{c}_{34} \hat{A}_{14} \\
\hat{A}_{21} & 0 & \widetilde{c}_{34} \hat{A}_{23} & e^{ i ( \delta + \widetilde{\phi}_{34} ) } \widetilde{s}_{34} \hat{A}_{23} \\
- e^{ i ( \delta + \widetilde{\phi}_{34} ) } \widetilde{s}_{34} \hat{A}_{41} & 
\widetilde{c}_{34} \hat{A}_{32}  & 0 & 0 \\
\widetilde{c}_{34} \hat{A}_{41}  & e^{ - i ( \delta + \widetilde{\phi}_{34} ) } \widetilde{s}_{34} \hat{A}_{32} & 0 & 0 \\
\end{array}
\right], 
\label{bar-H1-nubar}
\end{eqnarray}
where the $\hat{A}_{ij}$ elements are given in eq.~\eqref{hat-Aij-summary-nubar}. 

The qualitative behaviors of $\widetilde{\theta}_{24}$ and $\widetilde{\theta}_{34}$ which can be read off from eqs.~\eqref{matter-theta24} and~\eqref{matter-theta34} are as follow: At the small-$|b|$ region ($|b| \ll \Delta m^2_{41}$) $\widetilde{\theta}_{24} \approx \theta_{24}$ and $\widetilde{\theta}_{34} \approx 0$. Then, moving toward positive asymptotically large $|b|$ ($|b| \gg \Delta m^2_{41}$), $\widetilde{\theta}_{24}$ behaves as $\theta_{24} \rightarrow \frac{\pi}{4} \rightarrow \frac{\pi}{2}$, and $\widetilde{\theta}_{34}$ behaves as $0 \rightarrow \widetilde{\theta}_{34} \vert_{\text{max}} \rightarrow 0$, where $\widetilde{\theta}_{34} \vert_{\text{max}} \approx 30$~degree. $\widetilde{\theta}_{34}$ does not reach to $\frac{\pi}{4}$ because $\sin 2 \widetilde{\theta}_{34}$ cannot vanish, see eq.~\eqref{matter-theta34}. 
Notice that the $|b| \rightarrow \infty$ behavior of $\widetilde{\theta}_{34}$ is different from the one of $\widetilde{\theta}_{24}$, as $\sin 2\widetilde{\theta}_{34} \approx 0$ due to $\widetilde{c}_{24} \approx 0$ there. $\widetilde{\theta}_{24}$ describe the usual MSW resonance and $\widetilde{\theta}_{34}$ shows the typical behavior of the enhanced ``vacuum oscillation''. 
Despite our picture of local resonance approximation, the ``one by one'' diagonalization procedure in fact keeps influence of the 2-4 diagonalization onto the 3-4. Thus, our treatment retain, at least partly, some ``interaction'' between the two overlapping resonances.

\subsection{Computation of the flavor basis $S$ matrix and the probability: \\ Anti-neutrino channel}
\label{sec:flavor-basisS}

With $\bar{H} _{0}$ and $\bar{H} _{1}$ at hand in eqs.~\eqref{bar-H0-nubar} and~\eqref{bar-H1-nubar} it is easy to repeat the computation of the bar basis $S$ matrix in the anti-neutrino channel. Calculation of the bar-basis $\bar{S}$ matrix elements can be done as described in section~\ref{sec:calculation-barS}. The zeroth order $\bar{S}$ matrix takes the same form as in eq.~\eqref{barS-0th} with the minus sign choice, but the eigenvalues $\bar{\lambda}_{j}$ replaced by the ones given in eq.~\eqref{eigenvalues-bar-basis}. 
The first order $\bar{S}$ matrix can be calculated by the same method that leads us to eq.~\eqref{barS-1st-nu} in the neutrino channel. In the anti-neutrino channel $\bar{S} (x) ^{(1)}$ has the following matrix structure 
\begin{eqnarray} 
&&
\bar{S} (x) ^{(1)} 
= e^{-i \bar{H}^{(0)} x} \Omega (x)^{(1)} 
= 
\left[
\begin{array}{cccc}
0 & \bar{S}_{12} & \bar{S}_{13} & \bar{S}_{14} \\
\bar{S}_{21} & 0 & \bar{S}_{23} & \bar{S}_{24} \\
\bar{S}_{31} & \bar{S}_{32} & 0 & 0 \\
\bar{S}_{41} & \bar{S}_{42} & 0 & 0 \\
\end{array}
\right]. 
\label{barS-1st-nubar}
\end{eqnarray} 
For convenience of the readers who want to verify the calculations, we present the explicit forms of the bar-basis (as well as the flavor-basis, see below) $S$ matrix elements in Appendix~\ref{sec:flavor-S-nubar}. 

Similarly to the treatment in section~\ref{sec:bar-to-flavorS} in the neutrino channel, the flavor-basis $S$ matrix can be computed using the relation between the bar- and flavor-bases in the anti-neutrino channel as 
\begin{eqnarray} 
&&
\hspace{-5mm}
S_{\text{flavor}} 
= 
U_{34} (\theta_{34}, \phi_{34} )^* 
U_{24} (\widetilde{\theta}_{24}, \widetilde{\phi}_{24} ) ^* 
U_{34} ( \widetilde{\theta}_{34},  \widetilde{\phi}_{34} ) ^* 
\bar{S} 
[ U_{34} ( \widetilde{\theta}_{34},  \widetilde{\phi}_{34} ) ^* ] ^{\dagger}
[ U_{24} (\widetilde{\theta}_{24}, \widetilde{\phi}_{24} ) ^* ] ^{\dagger} 
[ U_{34} (\theta_{34}, \phi_{34} )^* ]^{\dagger}. 
\nonumber \\
\label{flavor-bar-S-nubar}
\end{eqnarray}
It can be shown that the flavor-basis $S$ matrix in the anti-neutrino channel has the following structure, to first order, as 
\begin{eqnarray} 
&&
S = 
\left[
\begin{array}{cccc}
S_{ee}  ^{(0)} & 0 & 0 & 0 \\
0 & S_{\mu \mu} ^{(0)} & S_{\mu \tau} ^{(0)} & S_{\mu S} ^{(0)} \\
0 & S_{\tau \mu} ^{(0)} & S_{\tau \tau} ^{(0)} & S_{\tau S} ^{(0)} \\
0 & S_{S \mu} ^{(0)} & S_{S \tau} ^{(0)} & S_{S S} ^{(0)} \\
\end{array}
\right] 
+ 
\left[
\begin{array}{cccc}
0 & S_{e \mu} ^{(1)} & S_{e \tau} ^{(1)} & S_{e S} ^{(1)} \\
S_{\mu e} ^{(1)} & S_{\mu \mu} ^{(1)} & S_{\mu \tau} ^{(1)} & S_{\mu S} ^{(1)} \\
S_{\tau e} ^{(1)} & S_{\tau \mu} ^{(1)} & S_{\tau \tau} ^{(1)} & S_{\tau S} ^{(1)} \\
S_{S e} ^{(1)} & S_{S \mu} ^{(1)} & S_{S \tau} ^{(1)} & S_{S S} ^{(1)} \\
\end{array}
\right]. 
\label{flavorS-1st-nubar}
\end{eqnarray} 
The most notable feature of the $S$ matrix in eq.~\eqref{flavorS-1st-nubar} is that, in the leading order, $\bar{\nu}_{e}$ row (and column) decouples from the remaining part of the $S$ matrix. This is the feature which is pointed out and extensively used in the analyses done in refs.~\cite{Razzaque:2011ab,Razzaque:2012tp}.\footnote{
Of course, omission of the $\nu_{e}$ row should not be applied to the neutrino channel, as we lose the chance of illuminating the resonance-enhanced $P(\nu_{e} \rightarrow \nu_{e})$, see eq.~\eqref{Pee-etau}. }

Now, the computation of the flavor-basis $S$ matrix elements can be done straightforwardly. Then, one can calculate the probability as we have done in the neutrino channel. To make our presentation compact as much as possible, we summarize the bar-basis and the flavor basis $S$ matrices in the anti-neutrino channel in Appendix~\ref{sec:flavor-S-nubar}, and the results of the probabilities in several relevant channels in Appendix~\ref{sec:probability-summary-nubar}. 

We are ready to discuss the probability in the physically relevant channels, as was done in section~\ref{sec:probability-nu} for the neutrino channels. Most naturally we start from the $\bar{\nu}_{\mu}$ disappearance probability $P(\bar{\nu}_{\mu} \rightarrow \bar{\nu}_{\mu})$, which is studied most vigorously in the literature for its importance in the experimental analysis for the SA resonance. See~\cite{IceCube:2024kel,IceCube:2024uzv,Nunokawa:2003ep,Choubey:2007ji,Razzaque:2011ab,Razzaque:2012tp,Barger:2011rc,Esmaili:2012nz,Esmaili:2013cja,Esmaili:2013vza,Esmaili:2013fva}, and the references cited therein. 

\section{Sterile-active resonance in the anti-neutrino channel: Probabilities} 
\label{sec:probability-nubar}

\subsection{Track events and $P(\bar{\nu}_{\mu} \rightarrow \bar{\nu}_{\mu})$: Anti-neutrino channel} 
\label{sec:track-nubar}

Most probably the highest sensitivity channel in the neutrino observation in the ice or deep ocean would be the ``track'' event caused by muon, either created by $\bar{\nu}_{\mu}$ interaction in the detector, or produced in the nearby detector in the earth and penetrating the detector volume. As the atmospheric (anti-) neutrino flux in the TeV region is dominated by $\nu_{\mu}$ and $\bar{\nu}_{\mu}$, we focus on $P(\bar{\nu}_{\mu} \rightarrow \bar{\nu}_{\mu})$ first. The probability reads 
\begin{eqnarray} 
&&
P(\bar{\nu}_{\mu} \rightarrow \bar{\nu}_{\mu}) 
= 
\vert S_{\mu \mu}^{(0)} \vert^2 
+ 2 \mbox{Re} \left[ ( S_{\mu \mu}^{(0)} )^* S_{\mu \mu}^{(1)} \right] 
\nonumber \\
&=&
1 - \widetilde{s}^4_{24} \sin^2 2\widetilde{\theta}_{34} \sin^2 \frac{ ( h_{4} - h_{3} ) x }{2} 
- \sin^2 2\widetilde{\theta}_{24} 
\biggl\{ 
\widetilde{c}^2_{34} \sin^2 \frac{ ( h_{4} - h_{2} ) x }{2} 
+ \widetilde{s}^2_{34} \sin^2 \frac{ ( h_{3} - h_{2} ) x }{2} 
\biggr\} 
\nonumber \\
&-&
\sin 2\theta_{34} \widetilde{s}_{24} 
\sin 2\widetilde{\theta}_{24} \sin 2\widetilde{\theta}_{34} 
\frac{ |b| }{2E} 
\nonumber \\
&\times& 
\biggl\{ 
\frac{ 1 }{ ( h_{3} - h_{2} ) } 
\left[ 
( \widetilde{c}^2_{24} - \widetilde{s}^2_{24} \widetilde{s}^2_{34} ) 
\sin^2 \frac{ ( h_{3} - h_{2} ) x }{2} 
+ \widetilde{s}^2_{24} \widetilde{c}^2_{34} 
\biggl\{ \sin^2 \frac{ ( h_{4} - h_{3} ) x }{2} - \sin^2 \frac{ ( h_{4} - h_{2} ) x }{2} \biggr\} 
\right] 
\nonumber \\
&-&
\frac{ 1 }{ ( h_{4} - h_{2} ) } 
\left[ 
( \widetilde{c}^2_{24} - \widetilde{s}^2_{24} \widetilde{c}^2_{34} ) 
\sin^2 \frac{ ( h_{4} - h_{2} ) x }{2} 
+ \widetilde{s}^2_{24} \widetilde{s}^2_{34} 
\biggl\{ \sin^2 \frac{ ( h_{4} - h_{3} ) x }{2} - \sin^2 \frac{ ( h_{3} - h_{2} ) x }{2} \biggr\} 
\right] 
\biggr\}. 
\nonumber \\
\label{Pmumu-1st-nubar}
\end{eqnarray}
The first line in eq.~\eqref{Pmumu-1st-nubar} comes from the zeroth order $S_{\mu \mu}^{(0)}$ matrix elements. On the resonance peak, this leading contribution carry no suppression factor by the smallness of the sterile mixing angles, as the both $\theta_{24}$ and $\theta_{34}$ are elevated to the matter enhanced $\widetilde{\theta}_{24}$ and $\widetilde{\theta}_{34}$. This is the  ``bona fide'' IceCube resonance enhancement in the SA resonance region. On the contrary, at off-peak, the leading term in $P(\bar{\nu}_{\mu} \rightarrow \bar{\nu}_{\mu})$ has the suppression factor $s^2_{24}$, and the first-order terms $s^2_{24} s^2_{34}$. See Table~\ref{tab:suppression-nubar}. For the definitions of the terms on- and off-peak, though customary, see section~\ref{sec:suppress-probability} and Appendix~\ref{sec:understanding-tables}. 

Track events can also be generated by utilizing 10\% level atmospheric $\bar{\nu}_{e}$ flux and the conversion probability $P(\bar{\nu}_{e} \rightarrow \bar{\nu}_{\mu})$. Due to $T$-invariance (see section~\ref{sec:T-invariance}), the conversion probability is equal to 
$P(\bar{\nu}_{\mu} \rightarrow \bar{\nu}_{e})$, which is given in eq.~\eqref{P-mue-2nd-nubar}. As $S_{e \mu}^{(0)} = 0$, the lowest-order term in the probability is of second order, and is given by $| S_{e \mu}^{(1)} |^2$. Then, there is a characteristic suppression factor $s^2_{14}$, which makes this contribution smaller by $\sim10^{-3}$ (including 0.1 flux fraction) compared to the $\bar{\nu}_{\mu}$ flux originated contribution. 

Therefore, it appears that the dominant mechanism for producing the SA resonance effect is by utilizing 90\% level atmospheric $\bar{\nu}_{\mu}$ flux with the disappearance channel probability $P(\bar{\nu}_{\mu} \rightarrow \bar{\nu}_{\mu})$. This is the ``golden channel'' for the SA resonance, which has been analyzed by many authors as mentioned above. 

\subsection{Cascade events, $P(\bar{\nu}_{\mu} \rightarrow \bar{\nu}_{\tau})$ and $P(\bar{\nu}_{e} \rightarrow \bar{\nu}_{e})$: Anti-neutrino channel} 
\label{sec:cascade-nubar}

We first examine the scenario of utilizing the 90\% level dominant atmospheric $\bar{\nu}_{\mu}$ flux and $P(\bar{\nu}_{\mu} \rightarrow \bar{\nu}_{X})$ ($X= e, \tau$) which are appropriate for generating the cascade events. Due to the structure of the $S$ matrix in eq.~\eqref{flavorS-1st-nubar}, $P(\bar{\nu}_{\mu} \rightarrow \bar{\nu}_{e})$ starts at the second order, and is suppressed by $s^2_{14}$. Therefore, the contribution is small, $\sim 1$\% level. 
On the other hand, $P(\bar{\nu}_{\mu} \rightarrow \bar{\nu}_{\tau})$ has a promising feature. As shown in eq.~\eqref{Pmutau-1st-nubar}, see Appendix~\ref{sec:probability-summary-nubar}, its leading-order term include the one with no suppression factor (on-peak) as 
\begin{eqnarray} 
&&
P (\bar{\nu}_{\mu} \rightarrow \bar{\nu}_{\tau}) 
= 
c^2_{34} \widetilde{s}^2_{24} \sin^2 2\widetilde{\theta}_{34} 
\sin^2 \frac{ ( h_{4} - h_{3} ) x }{2}. 
\label{Pmutau-nubar}
\end{eqnarray}
Therefore, the atmospheric $\bar{\nu}_{\mu}$ flux with $P(\bar{\nu}_{\mu} \rightarrow \bar{\nu}_{\tau})$ would offer a sensitive probe for the SA resonance search with using the cascade events. 

Finally we examine the cascade events that caused by 
10\%-level atmospheric $\bar{\nu}_{e}$ flux and $P(\bar{\nu}_{e} \rightarrow \bar{\nu}_{X})$ ($X= e, \tau$). We note that $P(\bar{\nu}_{e} \rightarrow \bar{\nu}_{\tau})$ starts at the second order, and suppressed by $s^2_{14} \sim10^{-2}$. Therefore, the dominant contribution to cascade events is from the original $\bar{\nu}_{e}$ flux without modification (no SA resonance effect) of the oscillation, $P(\bar{\nu}_{e} \rightarrow \bar{\nu}_{e}) = | S_{e e}^{(0)} |^2 =1$, up to the second-order corrections which is computed as in eq.~\eqref{P-ee-2nd-nubar}. 

\subsection{Three origins of the cascade events and their possible separation}
\label{sec:cascade-3-origins}

Our discussions on the origin of the cascade events in sections~\ref{sec:cascade-nu} and~\ref{sec:cascade-nubar} reveal an interesting feature which may leads to deeper understanding of physics in the SA resonance region.

\begin{itemize}

\item 
In the neutrino channel the cascade events dominantly come from the 10\%-level atmospheric $\nu_{e}$ flux and the unsuppressed $P(\nu_{e} \rightarrow \nu_{e})$ in eq.~\eqref{Pee-etau}. The SA resonance effect causes depletion of the number of events in the resonance affected region. 

\item 
In the anti-neutrino channel the cascade events have the two origins, one from the dominant 90\%-level $\bar{\nu}_{\mu}$ flux which is converted to $\bar{\nu}_{\tau}$, see eq.~\eqref{Pmutau-nubar}. The SA resonance effect produces an enhancement of the event number distribution. 
The other way of yielding the cascade events is to utilize 10\% $\bar{\nu}_{e}$ flux which is practically unaffected by the SA resonance, as $P(\bar{\nu}_{e} \rightarrow \bar{\nu}_{e}) =1$ in a good approximation. 

\end{itemize}

Thus, it may be possible to do a three-component fit of the data of the cascade events which differ in the various factors: (1) number of events with the azimuthal angle dependence, and (2) enhancement, depletion, and/or having no effect in the event number distribution by the SA resonance effects. 
If feasible with the real data, the three-component cascade analysis could establish the nature of the enhancement as manifestation of the SA resonance. At the same time it offers another tool to diagnose atmospheric neutrinos in the multi-TeV energy regions, even though an event by event discrimination of the neutrino vs. anti-neutrino induced events would not be tenable.

\section{Interesting characteristic features of our analytic framework}
\label{sec:characteristic-feature} 

So far we have formulated the SA resonance effective theory, and given its first analytic treatment. 
Due to its novel approach, however, the framework shows the mixed features, intriguing, unexpected, and some not completely understood. It would be worthwhile to present clarifying discussions on the so far poorly understood aspects of the framework with its interesting new features, to facilitate the better analytic treatment to emerge. 

\subsection{Sterile mixing angle suppression of the probability}
\label{sec:suppress-probability} 

As we noted our framework lacks the universal expansion parameter such as $\epsilon \simeq \Delta m^2_{21} / \Delta m^2_{31}$. On the other hand, the suppression by the sterile mixing angles does occur in a channel dependent and/or (perturbative) order dependent manner. For phenomenology discussions, it is important to understand the rich varieties of the way how the $s_{j 4} \equiv \sin \theta_{j 4}$ ($j=1,2,3$) suppression takes place. In Tables~\ref{tab:suppression-nu} and~\ref{tab:suppression-nubar}, therefore, we give a compact summary of the $s_{j 4}$ suppression of the probability in the neutrino and anti-neutrino channels, respectively. 

\begin{table}[h!]
\vglue 0.2cm
\begin{center}
\caption{In some of the oscillation channels relevant for our discussions, the suppression factors on the probabilities due to the sterile mixing angles $s_{j 4} \equiv \sin \theta_{j 4}$ ($j=1,2,3$) are tabulated. The factors are given, at on-resonance peak and off-resonance peak, separately. Only the least severe suppression factor is quoted in each category. In the disappearance channels, we count how many powers of $s_{j 4}$ is placed on each portion of $1 - P(\nu_{\alpha} \rightarrow \nu_{\alpha})$. In the first column,  the oscillation channel is given with the equation number for the probability expression. This Table \ref{tab:suppression-nu} is for the neutrino channel. 
}
\label{tab:suppression-nu}
\vglue 0.2cm
\begin{tabular}{c|c|c|c}
\hline 
$\nu$ channel & 
On resonance peak & 
Off peak, leading (higher) & 
Next to leading order 
\\
\hline 
\hline 
$P(\nu_{\mu} \rightarrow \nu_{\mu})$~\eqref{Pmumu-2nd}& \green{$s^2_{24}$} & \green{$s^2_{24}$}  & 0th$\times$2nd 
\\
\hline 
$P(\nu_{\mu} \rightarrow \nu_{e})$~\eqref{P-mue-nu} 
& \green{$s^2_{24}$} & $s^2_{14} s^2_{24}$ & (1st order)$^2$ 
\\
\hline
$P(\nu_{\mu} \rightarrow \nu_{\tau})$~\eqref{P-mutau-nu} 
& $s^2_{24} s^2_{34}$ & $s^2_{24} s^2_{34}$ & (1st order)$^2$  
\\
\hline 
$P(\nu_{e} \rightarrow \nu_{e})$~\eqref{P-ee-nu}
& \blue{no suppression} & $s^2_{14} s^2_{24}$, $s^2_{14} s^2_{34}$, $s^4_{14}$ & 0th$\times$2nd 
\\
\hline 
$P(\nu_{e} \rightarrow \nu_{\tau})$~\eqref{P-etau-nu} & \green{$s^2_{34}$} & 
$s^2_{14} s^2_{34}$ & 0th$\times$1st
\\
\hline 
\end{tabular}
\end{center}
\vglue -0.2cm 
\end{table}

\begin{table}[h!]
\begin{center}
\caption{The same as in Table~\ref{tab:suppression-nu}, but for the anti-neutrino channels. }
\label{tab:suppression-nubar}
\vglue 0.2cm
\begin{tabular}{c|c|c|c}
\hline 
Anti-$\nu$ channel & 
On resonance peak & 
Off peak, leading (higher) & 
Next to leading order 
\\
\hline 
\hline 
$P (\bar{\nu}_{\mu} \rightarrow \bar{\nu}_{\mu})$~\eqref{Pmumu-1st-nubar} 
& \blue{no suppression} & \green{$s^2_{24}$} or \green{$s^2_{34}$} ($s^2_{24} s^2_{34}$) & ~0th$\times$1st
\\
\hline 
$P (\bar{\nu}_{\mu} \rightarrow \bar{\nu}_{\tau})$ \eqref{Pmutau-1st-nubar} 
& \blue{no suppression} & 
$s^2_{24} s^2_{34}$ & 0th$\times$1st
\\
\hline
$P (\bar{\nu}_{\mu} \rightarrow \bar{\nu}_{e})$ \eqref{P-mue-2nd-nubar} & \green{$s^2_{14}$} & \green{$s^2_{14}$} & (1st order)$^2$ 
\\
\hline 
$P (\bar{\nu}_{e} \rightarrow \bar{\nu}_{e})$ \eqref{P-ee-2nd-nubar} & \green{$s^2_{14}$} & \green{$s^2_{14}$} & ~0th$\times$2nd  
\\
\hline 
$P (\bar{\nu}_{e} \rightarrow \bar{\nu}_{\tau})$~\eqref{P-etau-2nd-nubar} & 
\green{$s^2_{14}$} & $s^2_{14} s^2_{34}$ & (1st order)$^2$
\\
\hline 
%
%
\end{tabular}
\end{center}
\vglue -0.4cm 
\end{table}

The various clarifying comments are needed to understand the contents of Tables~\ref{tab:suppression-nu} and~\ref{tab:suppression-nubar}. In the disappearance channels, we count how many powers of $s_{j 4}$ is placed on each term of $1 - P(\nu_{\alpha} \rightarrow \nu_{\alpha})$. The suppression factor is different depending upon we are ``on-resonance peak'' or ``off-resonance peak''. Simply all the relevant mixing angles becomes matter-affected and close to the maximal at on-resonance. The suppression factor at on-resonance peak is given in the second column of Tables~\ref{tab:suppression-nu} and~\ref{tab:suppression-nubar}. The nomenclatures on- and off-peak are customary (we believe), but otherwise it will be explained more explicitly in Appendix~\ref{sec:understanding-tables}. 
In the third column the suppression factor at off-peak is tabulated in the leading-order and the correction terms in parenthesis, in the sense of perturbation theory formulated in section~\ref{sec:flavor-basisS}. In many cases the suppression factors for the leading- and next to leading-order are identical, and the factor in parenthesis is abbreviated in such case.

The fourth column is a little involved. It is to describe the nature of the next to leading order contribution to the probability. Typically they are the interference terms between the zeroth and the first-order terms, denoted as ``0th$\times$1st''. When the first-order term is absent in the particular channel, it takes the form ``0th$\times$2nd''. Sometimes it occurs that the zeroth-order $S$ matrix element vanishes, see eqs.~\eqref{flavorS-nu-0th-2nd} and~\eqref{flavorS-1st-nubar}. In that case the leading order term is given by the first-order $S$ matrix element squared, denoted as ``(1st order)$^2$''. In this case the fourth column shows the nature of the leading-order term, and the suppression factor always exists even at on-resonance peak. 

To focus on the dominant structure of the SA resonance we have colored the unsuppressed (blue) and $s^2_{j 4}$-suppressed (cyan) to emphasize that they will be the dominant observables in the foreseeable feature, assuming $s_{j 4} \sim 0.1$. Obviously, the majority of the colored items are at on-resonance peak, as indicated in the second columns of Tables~\ref{tab:suppression-nu} and~\ref{tab:suppression-nubar}. These are nothing but the principal target of our phenomenology discussions that are borne out from our analytic treatment of the SA resonance effective theory. 

Here are some features that are worth to be highlighted (See Tables~\ref{tab:suppression-nu} and~\ref{tab:suppression-nubar}): 
\begin{itemize}

\item 
There exist only three channels with the unsuppressed SA resonance effect at on-peak, $P(\nu_{e} \rightarrow \nu_{e})$, $P (\bar{\nu}_{\mu} \rightarrow \bar{\nu}_{\mu})$, and $P (\bar{\nu}_{\mu} \rightarrow \bar{\nu}_{\tau})$, which include the unique neutrino channel $\nu_{e} \rightarrow \nu_{e}$ and the unique appearance one $\bar{\nu}_{\mu} \rightarrow \bar{\nu}_{\tau}$ in the anti-neutrino. 
\item 
There exist a sharp contrast between $P(\nu_{\mu} \rightarrow \nu_{\mu})$ and $P (\bar{\nu}_{\mu} \rightarrow \bar{\nu}_{\mu})$. In the former the resonance effect is nearly invisible, which can be understood by $s^2_{24}$ suppression at on- and off-peak, as it comes from the second-order corrections. 

\item 
Another interesting contrast is the both unsuppressed resonance effects in $P (\bar{\nu}_{\mu} \rightarrow \bar{\nu}_{\mu})$ and $P (\bar{\nu}_{\mu} \rightarrow \bar{\nu}_{\tau})$, and a big disparity between $P(\nu_{\mu} \rightarrow \nu_{\tau})$ and $P (\bar{\nu}_{\mu} \rightarrow \bar{\nu}_{\tau})$. Essentially no effect in $P(\nu_{\mu} \rightarrow \nu_{\tau})$ because of $s^2_{24} s^2_{34}$ suppression at on- and off-peak. 

\end{itemize}
We remark that the features emphasized above may have been closely related with the ones mentioned in the various analyses done previously. The authors of ref.~\cite{Esmaili:2013cja} observed that $P (\bar{\nu}_{\mu} \rightarrow \bar{\nu}_{\tau})$ can be enhanced if the both $\theta_{24}$ and $\theta_{34}$ are nonzero, and raised the possibility of cascade analysis which shed light to the $\theta_{34}$ measurement due to the spectral distortion. The authors of ref.~\cite{Smithers:2021orb} paid attention to the features described in the third item above, and described the strategy of joint analysis of the $\bar{\nu}_{\mu}$-induced track events and cascade due to the $\bar{\nu}_{\tau}$ appearance. 
Given that our framework offers a simple bird-eye view of the SA resonance enhancement with the varying strengths, we hope, it can contribute to further advance formulation of the strategy for exploration of the sterile sector. Our own proposal is already described in section~\ref{sec:cascade-3-origins}. 

\subsection{Mechanism of sterile mixing angle suppression} 
\label{sec:mechanism} 

As we observe, the pattern of $s_{j 4}$ suppression is not simple and the oscillation-channel dependent, as shown in Tables~\ref{tab:suppression-nu} and~\ref{tab:suppression-nubar}. It depends on the flavor structure of the $S$ matrix, such as distribution of zeros in the zeroth- and first-order $S$ matrices, see eqs.~\eqref{flavorS-nu-0th-2nd} and~\eqref{flavorS-1st-nubar}. It also comes from the elements of the Hamiltonian, and alternative choice of picking up $s_{34}$ (or not) from the final rotation to obtain the flavor basis $S$ matrix. 

Here, we discuss only one example to give a feeling on how many $s_{j 4}$ power suppression comes about at on-peak leading, off-peak leading, and off-peak higher order terms in $P(\bar{\nu}_{\mu} \rightarrow \bar{\nu}_{\mu})$. At on-peak the first line in eq.~\eqref{Pmumu-1st-nubar} is unsuppressed by the vacuum sterile mixing angles, as they are replaced by the matter affected angles, which are enhanced on peak. In the off-peak leading order term, $s^2_{34}$ or $s^2_{24}$ suppression appear, see second and third terms in the first line. In the off-peak first order terms, the second to the fifth lines in eq.~\eqref{Pmumu-1st-nubar}, are suppressed by the pre-factor 
\begin{eqnarray} 
&&
\sin 2\theta_{34} \widetilde{s}_{24}
\sin 2 \widetilde{\theta}_{24} \sin 2 \widetilde{\theta}_{34} 
= 
\sin^2 2\theta_{34} \sin^2 2\theta_{24} 
\nonumber \\
&\times&
\frac{1}{2} 
\frac{ \left( c^2_{14} \Delta m^2_{41} \right)^2 } 
{ \left[ \cos 2\theta_{24} c^2_{14} \Delta m^2_{41} - c^2_{34} |b| 
\right]^2 + \left( \sin 2\theta_{24} c^2_{14} \Delta m^2_{41} \right)^2 } 
\frac{ |b| } 
{ \sqrt{ ( \hat{\lambda}_{4} - \hat{\lambda}_{3} )^2 + ( \widetilde{c}_{24} \sin 2\theta_{34} |b| )^2 } } 
\nonumber \\
\label{prefactor-mumu}
\end{eqnarray}
in which we have used the expressions of $\sin 2 \widetilde{\theta}_{24}$ and $\sin 2 \widetilde{\theta}_{34}$ given in eqs.~\eqref{matter-theta24} and ~\eqref{matter-theta34}, respectively. We note that, the kinematical factors, the second line in eq.~\eqref{prefactor-mumu}, are of order unity in the SA resonance region. Therefore, the first order terms in $P(\bar{\nu}_{\mu} \rightarrow \bar{\nu}_{\mu})$ is suppressed by the factor $s^2_{24} s^2_{34} \simeq 10^{-4}$, assuming that $s_{24} = s_{34} \simeq 0.1$. 

We may make a comment on the leading-order term in $P(\nu_{e} \rightarrow \nu_{\tau})$ in eq.~\eqref{P-etau-nu}. In the both zeroth- and first-order terms in the probability, we have the identical suppression factor $s^2_{34}$ at on-resonance peak. It appears that this occurs by accident, as explained in Appendix~\ref{sec:strange-looking}. 
In general, the off-peak leading and higher-order terms have different $s_{j 4}$ power suppressions, but in many cases they have the identical-in-total $s_{j 4}$ suppression. 

\subsection{$T$-invariance of the probability} 
\label{sec:T-invariance}

At the end of section~\ref{sec:IC-theory-nu-channel} we have remarked that the probability in the neutrino channel respects $T$-invariance. Our computation of the anti-neutrino probability shows the same feature. For this to be true, it must overcome presence of the additional phases from the matter-affected $\widetilde{\theta}_{24}$ and $\widetilde{\theta}_{34}$ rotations which involve the new matter-affected phases $\widetilde{\phi}_{24}$ and $\widetilde{\phi}_{34}$. 
(Notice that there is no $\widetilde{\phi}_{14}$ in our convention.) However, we have learnt in section~\ref{sec:IC-theory-nubar-channel} that the diagonalization of the Hamiltonian by the 2-4 and 3-4 rotations requires $\widetilde{\phi}_{24} = \phi_{24}$ and $\widetilde{\phi}_{34} = \phi_{34}$. That is, the effects of the extra phases are systematically cancelled by their vacuum counterparts. 

Notice that the dynamical ($\widetilde{\phi}_{j 4}$) or the built-in in theory ($\phi_{j 4}$) phases appear independently to each other in the bar-basis and flavor-basis $S$ matrices, as one can easily confirm in Appendix~\ref{sec:flavor-S-nubar}. But, when one computes the oscillation probability, they disappear thanks to the relation $\widetilde{\phi}_{j 4} = \phi_{j 4}$, and all the probabilities becomes free from the phases. Thus, complete cancellation between the phases is highly nontrivial feature. $T$-invariance of the probability is partially displayed in Appendices~\ref{sec:probability-summary-nu} and~\ref{sec:probability-summary-nubar}.\footnote{
Without the phases the oscillation probability is invariant under $T$, the time reversal. Another characterization of $T$-invariance is that the probability does not involve odd power of $x$~\cite{Cabibbo:1977nk}. In a recent analysis this property is utilized to evaluate the sensitivity to $T$-violation in the next-generation long-baseline neutrino experiments~\cite{Schwetz:2021cuj}. The $T$-even property is evident in the probabilities given in Appendices~\ref{sec:probability-summary-nu} and~\ref{sec:probability-summary-nubar}. }

Nonetheless, a careful reader may ask whether this property is the artifact of our treatment. That is, the 2-4 and 3-4 level crossings are diagonalized one by one without treating them fully ``interacting'' even though they overlap. It is a tantalizing question to settle, whether $T$-invariance is the general feature of the effective theory of SA resonance, or it holds only because of our treatment which is based on the local resonance approximation. 

\section{Concluding remarks}
\label{sec:conclusion}

In this paper we have created the sterile-active (SA) resonance effective theory by taking into account the environments of high matter potential, $a, b \sim \Delta m^2_{41} \sim 10$ eV$^2$ $\gg \Delta m^2_{31}$, freezing the $\nu$SM oscillations in the relevant kinematical region. We believe that the effective theory itself retains all the important properties of the SA resonance that are built-in in the $(3+1)$ model. Then, we have given our first attempt at an analytic treatment of the theory to illuminate the global, qualitative features of the SA resonance. To our knowledge this is done for the first time in the manner that covers the both neutrino and anti-neutrino channels in the same footing, and illuminates the flavor- and event-type dependent hierarchy of the SA resonance in the probabilities. The variety of features of the $s_{j 4} \equiv \sin \theta_{j 4}$ suppression of the probabilities provides the core of such phenomenology discussions, revealing the flavor channel structure of the ``track'' and ``cascade'' events. 

We have described each steps of our analysis in a self-contained way, such that every interested reader can follow. 
As our physics focus is on the global and qualitative multi-flavor properties of the SA resonance, we carried out only a spot check of the numerical accuracy of our analytic formulas. Globally, the agreement is about $\sim$1-2 \% level in comparison to the numerical calculation using the effective theory Hamiltonian, which is sufficient for our use in this paper.  

\subsection{Brief summary of our phenomenological discussion} 
\label{sec:pheno-summary}

Foreseeing the progress in the KM3Net~\cite{KM3Net:2016zxf} construction and the further development of the ``IceCube-Gen2''~\cite{IceCube:2014gqr} project, with capabilities of detecting the SA resonance, it is now an urgent question to ask which set of the experimental observables could testify in a robust way that the SA resonance is discovered. In this paper we tried to give a partial answer to this question. The resonance will leave the trace in the various flavor channels, so that the multi-channel, multi-event type analysis will be crucial. We hope that the feature of the global and qualitative multi-flavor properties of the SA resonance uncovered in our framework can boost the move toward such advanced observations.

In this paper we have made a few observations by utilizing our framework: 
\begin{itemize}

\item 
In the neutrino channel the track events may not be a good probe of the SA resonance effects. Instead, the cascade events from the sub-dominant 10\% level atmospheric $\nu_{e}$ flux with the resonance enhanced, $s_{j 4}$ unsuppressed $P(\nu_{e} \rightarrow \nu_{e})$ in eq.~\eqref{Pee-etau} could give the clear observable signature of the SA resonance. 

\item 
In the anti-neutrino channel the track events with the dominant $\bar{\nu}_{\mu}$ atmospheric flux would provide the best opportunity with the SA resonance enhanced, unsuppressed $P(\bar{\nu}_{\mu} \rightarrow \bar{\nu}_{\mu})$, as analyzed by the many authors. 
At the same time, $\bar{\nu}_{\mu}$ atmospheric flux origin cascade events through $P(\bar{\nu}_{\mu} \rightarrow \bar{\nu}_{\tau})$ in eq.~\eqref{Pmutau-nubar} have the similar SA resonance-enhanced feature, which provide the promising chance for the future observation of the cascade events. 

\item 
As stated above, the cascade events have the three origins: They can be produced by the 90\%-level fraction atmospheric $\bar{\nu}_{\mu}$ flux with $P (\bar{\nu}_{\mu} \rightarrow \bar{\nu}_{\tau})$, in which the SA resonance effect enhances the number of events. They also come from the 10\%-level atmospheric $\nu_{e}$ (and $\bar{\nu}_{e}$) flux through $P(\nu_{e} \rightarrow \nu_{e})$ and 
$P(\bar{\nu}_{e} \rightarrow \bar{\nu}_{e})$, in the former (latter) the resonance effect deplete (does not affect) the number of events. A possibility of three component fit is suggested to distinguish between these three components. 

\end{itemize}

\subsection{Theoretical aspect of our framework: Status summary} 
\label{sec:theory-summary}

We have found that our analytic treatment of the effective theory of SA resonance enables us to draw a global and qualitative overview of the SA resonance in the $(3+1)$ model. Moreover, we have argued that its construction can be done even beyond the DMP-embedding approximation of the $\nu$SM part, as far as we stay in the region $\Delta m^2_{31} / a \sim 10^{-3} \ll 1$. 

However, we have encountered the various theoretical issues in developing our analytic treatment. They include: (1) In the both neutrino and anti-neutrino channels, decomposition of the Hamiltonian into the zeroth and first order terms is not done in reference to the universal expansion parameter, such as $\Delta m^2_{21} / \Delta m^2_{31}$, because it is not available in this theory. Then, there is no guarantee that the first-order correction is smaller than the zeroth order term. (2) In the anti-neutrino channel, there exist the two SA resonances, corresponding to the 2-4 and 3-4 level crossings, which overlap in a wide range in the parameter space. Then, the one by one diagonalization method which we have employed in our treatment may not be a good approximation. 

At this stage, we are not able to answer all these questions completely. We briefly state here what we did and what should be done. For the problem (1) we feel we have presented a reasonable solution: Using the structure of flavor space ``texture zeros'' of the $S$ matrix and the suppression by the small parameters $s_{j 4} \equiv \sin \theta_{j 4}$ ($j=1,2,3$), we can elucidate the hierarchy of the probabilities, in which oscillation channel the effect of the SA resonance is large or small. 
Look at Tables~\ref{tab:suppression-nu} and~\ref{tab:suppression-nubar}: The blue- and cyan-colored items are unsuppressed and least severely suppressed terms, respectively, where we tentatively set $s_{j 4} = 0.1$ for all $j=1$-3 for brevity. In the case that the suppression is not universal, our analytic formulas offer the estimate of such effect for the given $s_{j 4}$. Thus our analytic framework can identify  the ``large'' effects, offering a basis for phenomenology discussions. 

How to handle the dynamics of the overlapping two MSW resonances in an analytic fashion is a difficult problem to solve. It never occurs in the $\nu$SM neutrino propagation, with the atmospheric 1-3 and the solar 1-2 resonances being positioned far apart to each other. Clearly, we need a new, powerful diagonalization procedure for such system. Possible candidates would include the exact method~\cite{Zaglauer:1988gz,Kimura:2002wd} and the fast converging ``rotation'' method~\cite{Denton:2018fex}. When a good candidate theory is identified we can ask a further question of whether it embodies powerfulness and a reasonable simplicity in treating the effective theory of SA resonance. Our attitude in this paper is that before such a satisfactory solution is uncovered, the ``one by one'' procedure may be considered as a way of approximating the combined effects of the two overlapping resonances. 

In this paper we have confined ourselves into the case of NMO, the normal mass ordering. Which change do we expect if we discuss the IMO? As indicated in Fig.~\ref{fig:superficial-level-cross} the visible change in the level crossing diagram is an exchange between the second and third states. It means that there is no significant change between the IMO and NMO, except that the crossing involving the fourth state occurs (from left most part, $|a|, |b| \rightarrow \infty$) first in the 3-4 and then 2-4 levels in the IMO. Then, if necessary, one can repeat the anti-neutrino analysis by doing the 3-4 rotation first and then 2-4. 
Our attitude not to enter into this task is that the IMO discussion is crucial if we address the mass ordering determination by using the SA resonance. Since it requires elaborate analysis of the accurately measured data set, we can wait until this situation becomes realistic.

\appendix 

\section{Flavor basis $S$ matrix elements in the neutrino channels}
\label{sec:flavor-S-nu}

First we give a note for compact presentation of the $S$ matrix elements, in general. The computed results of the $S$ matrix elements in the $k$-th order satisfies $S_{\alpha \beta} ^{(k)} = \left[ S_{\beta \alpha} ^{(k)} \right]^{T*}$, where $T*$ operation is introduced in section~\ref{sec:calculation-barS}. It is taking hermitian conjugate\footnote{
Here is an example to prevent misunderstanding of the $T*$ operation: 
$S_{\tau \mu} ^{(1)} = \left[ S_{\mu \tau} ^{(1)} \right]^{T*}$ implies that 
$S_{\mu \tau} ^{(1)} = e^{ i ( \delta + \phi_{34} ) } s_{34} \left( - \widetilde{s}_{14} \bar{S}_{21} ^{(1)} + \widetilde{c}_{14} \bar{S}_{24} ^{(1)} \right)$, and  
$S_{\tau \mu} ^{(1)} = e^{ - i ( \delta + \phi_{34} ) } s_{34} 
\left( - \widetilde{s}_{14} \bar{S}_{12} ^{(1)} + \widetilde{c}_{14} \bar{S}_{42} ^{(1)} \right)$. }
leaving the kinematical factor $( e^{ - i h_{j} x } - e^{ - i h_{i} x } ) / ( h_{j} - h_{i} )$ untouched. This reflects the generalized $T$-invariance~\cite{Fong:2017gke}. 

The non-vanishing $S_{\alpha \beta}^{(0)}$ matrix elements in the zeroth-order in the neutrino channels are given by: 
\begin{eqnarray} 
&&
S_{ee}  ^{(0)}
= 
\left( \widetilde{s}^2_{14} e^{ - i h_{4} x } + \widetilde{c}^2_{14} e^{ - i h_{1} x } \right), 
\nonumber \\
&&
S_{\mu \mu} ^{(0)} 
= e^{ - i h_{2} x }, 
\nonumber \\
&&
S_{\tau \tau}^{(0)} 
= 
c^2_{34} e^{ - i h_{3} x }  
+ s^2_{34} \left( \widetilde{c}^2_{14} e^{ - i h_{4} x } + \widetilde{s}^2_{14} e^{ - i h_{1} x } \right), 
\nonumber \\
&&
S_{S S} ^{(0)} 
= 
s^2_{34} e^{ - i h_{3} x } 
+ c^2_{34} \left( \widetilde{c}^2_{14} e^{ - i h_{4} x }  + \widetilde{s}^2_{14} e^{ - i h_{1} x } \right). 
\label{S-elements-0th-nu-1}
\end{eqnarray} 
\begin{eqnarray} 
&&
S_{e \tau} ^{(0)} 
= 
e^{ i ( \delta + \phi_{34} ) } s_{34} \widetilde{c}_{14} \widetilde{s}_{14} \left( e^{ - i h_{4} x } - e^{ - i h_{1} x } \right) 
= 
\left[ S_{\tau e} ^{(0)} \right]^{T*}, 
\nonumber \\
&& 
S_{e S} ^{(0)} 
= 
c_{34} \widetilde{c}_{14} \widetilde{s}_{14} \left( e^{ - i h_{4} x } - e^{ - i h_{1} x } \right) 
= 
\left[ S_{S e} ^{(0)} \right]^{T*}, 
\nonumber \\
&&
S_{\tau S} ^{(0)} 
= 
e^{ - i ( \delta + \phi_{34} ) } c_{34} s_{34} 
\left[ - e^{ - i h_{3} x } 
+ \left( \widetilde{c}^2_{14} e^{ - i h_{4} x } + \widetilde{s}^2_{14} e^{ - i h_{1} x } \right) \right] 
= 
\left[ S_{S \tau} ^{(0)} \right]^{T*}. 
\label{S-elements-0th-nu-2}
\end{eqnarray}

The non-vanishing $S_{\alpha \beta}^{(1)}$ matrix elements are given, using the $\bar{S}_{ij}^{(1)}$ matrix elements given in eq.~\eqref{barS-1st-nu}, as 
\begin{eqnarray} 
&& 
S_{\tau \tau} ^{(1)} 
= 
e^{ i ( \delta + \phi_{34} ) } c_{34} s_{34} 
\left( - \widetilde{s}_{14} \bar{S}_{31} ^{(1)} + \widetilde{c}_{14} \bar{S}_{34} ^{(1)} \right) 
+ e^{ - i ( \delta + \phi_{34} ) } c_{34} s_{34} 
\left( - \widetilde{s}_{14} \bar{S}_{13} ^{(1)} + \widetilde{c}_{14} \bar{S}_{43} ^{(1)} \right), 
\nonumber \\
&&
S_{S S} ^{(1)} 
= 
- e^{ i ( \delta + \phi_{34} ) } c_{34} s_{34} 
\left( - \widetilde{s}_{14} \bar{S}_{31} ^{(1)} + \widetilde{c}_{14} \bar{S}_{34} ^{(1)} \right) 
- e^{ - i ( \delta + \phi_{34} ) } c_{34} s_{34} 
\left( - \widetilde{s}_{14} \bar{S}_{13} ^{(1)} + \widetilde{c}_{14} \bar{S}_{43} ^{(1)} \right).
\nonumber \\
\label{S-elements-1st-nu-1}
\end{eqnarray}
\begin{eqnarray} 
&& 
S_{e \mu} ^{(1)} 
= 
\left( \widetilde{c}_{14} \bar{S}_{12} ^{(1)} + \widetilde{s}_{14} \bar{S}_{42}^{(1)} \right) 
= 
\left[ S_{\mu e} ^{(0)} \right]^{T*}, 
\nonumber \\
&&
S_{e \tau} ^{(1)} 
= 
c_{34} \left( \widetilde{c}_{14} \bar{S}_{13} ^{(1)} + \widetilde{s}_{14} \bar{S}_{43} ^{(1)} \right) 
= 
\left[ S_{\tau e} ^{(0)} \right]^{T*}, 
\nonumber \\
&&
S_{e S} ^{(1)} 
= 
- e^{ - i ( \delta + \phi_{34} ) } s_{34} \left( \widetilde{c}_{14} \bar{S}_{13} ^{(1)} + \widetilde{s}_{14} \bar{S}_{43} ^{(1)} \right) 
= 
\left[ S_{S e} ^{(0)} \right]^{T*}, 
\nonumber \\
&&
S_{\mu \tau} ^{(1)} 
= 
e^{ i ( \delta + \phi_{34} ) } s_{34} \left( - \widetilde{s}_{14} \bar{S}_{21} ^{(1)} + \widetilde{c}_{14} \bar{S}_{24} ^{(1)} \right) 
= 
\left[ S_{\tau \mu} ^{(0)} \right]^{T*}, 
\nonumber \\
&&
S_{\mu S} ^{(1)} 
= 
c_{34} \left( - \widetilde{s}_{14} \bar{S}_{21} ^{(1)} + \widetilde{c}_{14} \bar{S}_{24} ^{(1)} \right) 
= 
\left[ S_{S \mu} ^{(0)} \right]^{T*}, 
\nonumber \\
&&
S_{\tau S} ^{(1)} 
= 
e^{ - i ( \delta + \phi_{34} ) } 
\left[ 
e^{ i ( \delta + \phi_{34} ) } c^2_{34} \left( - \widetilde{s}_{14} \bar{S}_{31} ^{(1)} + \widetilde{c}_{14} \bar{S}_{34} ^{(1)} \right) 
- e^{ - i ( \delta + \phi_{34} ) } s^2_{34} \left( - \widetilde{s}_{14} \bar{S}_{13} ^{(1)} + \widetilde{c}_{14} \bar{S}_{43} ^{(1)} \right) 
\right] 
= 
\left[ S_{S \tau} ^{(0)} \right]^{T*}. 
\nonumber \\
\label{S-elements-1st-nu-2}
\end{eqnarray}

\section{Second order computation of the $S$ matrix}
\label{sec:S-second-order}

To compute the $S$ matrix to second order, one can just follow the same method as the first order calculation of the flavor-basis $S$ matrix described in sections~\ref{sec:calculation-barS} and \ref{sec:bar-to-flavorS}. To obtain $\Omega^{(2)}$ we compute the third term in eq.~\eqref{Omega-expansion} using the expression of $H_{1}$ in eq.~\eqref{def-H1}, with paying attention to the matrix multiplication. The $\bar{S}$ matrix is then given by eq.~\eqref{bar-Smatrix}. By performing the two rotations as in \eqref{flavor-bar-S}, we reach the second order flavor-basis $S$ matrix. 

We just quote here the result which is necessary to compute the second order term of $P(\nu_{\mu} \rightarrow \nu_{\mu})$ as done in eq.~\eqref{Pmumu-2nd}: 
\begin{eqnarray} 
&&
\hspace{-8mm}
S_{\mu \mu}^{(2)} = 
\bar{S} ^{(2)} _{22} = 
\frac{ A_{21} A_{12} }{ (2E)^2 }
\left[ - \frac{ e^{ - i h_{2} x } - e^{ - i h_{1} x } }{ ( h_{2} - h_{1} )^2 } + \frac{ (-i x ) e^{ - i h_{2} x } }{ ( h_{2} - h_{1} ) } \right] 
+ \frac{ A_{24} A_{42} }{ (2E)^2 } 
\left[ \frac{ e^{ - i h_{4} x } - e^{ - i h_{2} x } }{ ( h_{4} - h_{2} )^2 }  - \frac{ (-i x ) e^{ - i h_{2} x } }{ ( h_{4} - h_{2} ) } \right], 
\nonumber \\
\label{S-mumu-2nd-nu}
\end{eqnarray}
where the $A_{ij}$ elements are given in eq.~\eqref{Aij-summary-nu}.

\section{The oscillation probability in the neutrino channel} 
\label{sec:probability-summary-nu}

In sections~\ref{sec:probability-summary-nu} (neutrino channel) and~\ref{sec:probability-summary-nubar} (anti-neutrino channel), the explicit expressions of the probabilities in some of the flavor oscillation channels are presented. When we mention about unitarity in the $\nu_{\alpha}$-row ($\bar{\nu}_{\alpha}$-row) we often implies about first- or second-order unitarity, as unitarity in the zeroth order has been proven in all the channels. 

\subsection{The oscillation probability in the $\nu_{\mu}$ row} 
\label{sec:numu-row}

$P(\nu_{\mu} \rightarrow \nu_{\mu})$ is given to second order in eq.~\eqref{Pmumu-2nd} in section~\ref{sec:track-nu}. The leading second-order probabilities in the remaining three channels are given below: 
\begin{eqnarray} 
&&
P(\nu_{\mu} \rightarrow \nu_{e}) 
= 
| S^{(1)}_{e \mu} |^2 
\nonumber \\
&=&
s^2_{24} 
\left( \frac{ \Delta m^2_{41} }{2E} \right)^2 
\biggl[ 
4 \widetilde{c}^2_{14} 
\left( \widetilde{c}_{14} c_{14} s_{14} - \widetilde{s}_{14} c_{24} c^2_{14} \right)^2 
\frac{1}{ ( h_{2} - h_{1} )^2 } \sin^2 \frac{ ( h_{2} - h_{1} ) x }{2} 
\nonumber \\
&+& 
4 \widetilde{s}^2_{14}  
\left( \widetilde{s}_{14} c_{14} s_{14} + \widetilde{c}_{14} c_{24} c^2_{14} \right) ^2 
\frac{1}{ ( h_{4} - h_{2} )^2 } \sin^2 \frac{ ( h_{4} - h_{2} ) x }{2} 
\nonumber \\
&-& 
2 \sin 2\widetilde{\theta}_{14} 
\left( \widetilde{c}_{14} c_{14} s_{14} - \widetilde{s}_{14} c_{24} c^2_{14} \right) 
\left( \widetilde{s}_{14} c_{14} s_{14} + \widetilde{c}_{14} c_{24} c^2_{14} \right) 
\nonumber \\
&\times& 
\frac{ 1 }{ ( h_{2} - h_{1} ) ( h_{4} - h_{2} ) } 
\biggl\{ \sin^2 \frac{ ( h_{4} - h_{2} ) x }{2}
- \sin^2 \frac{ ( h_{4} - h_{1} ) x }{2}
+ \sin^2 \frac{ ( h_{2} - h_{1} ) x }{2} 
\biggr\} 
\biggr]. 
\label{P-mue-nu}
\end{eqnarray}
\begin{eqnarray} 
&& 
P(\nu_{\mu} \rightarrow \nu_{\tau}) 
= 
| S^{(1)}_{\tau \mu} |^2 
\nonumber \\
&=&
s^2_{24} s^2_{34} 
\left( \frac{ \Delta m^2_{41} }{2E} \right)^2 
\biggl[ 
4 \widetilde{s}^2_{14} 
\left( \widetilde{c}_{14} c_{14} s_{14} - \widetilde{s}_{14} c_{24} c^2_{14} \right)^2 
\frac{1}{ ( h_{2} - h_{1} )^2 } 
\sin^2 \frac{ ( h_{2} - h_{1} ) x }{2} 
\nonumber \\
&+& 
4 \widetilde{c}^2_{14} 
\left( \widetilde{s}_{14} c_{14} s_{14} + \widetilde{c}_{14} c_{24} c^2_{14} \right)^2 
\frac{1}{ ( h_{4} - h_{2} )^2 } 
\sin^2 \frac{ ( h_{4} - h_{2} ) x }{2} 
\nonumber \\
&+& 
2 \sin 2\widetilde{\theta}_{14} 
\left( \widetilde{c}_{14} c_{14} s_{14} - \widetilde{s}_{14} c_{24} c^2_{14} \right) 
\left( \widetilde{s}_{14} c_{14} s_{14} + \widetilde{c}_{14} c_{24} c^2_{14} \right) 
\nonumber \\
&\times&
\frac{1}{ ( h_{2} - h_{1} ) ( h_{4} - h_{2} ) } 
\biggl\{ \sin^2 \frac{ ( h_{4} - h_{2} ) x }{2}
- \sin^2 \frac{ ( h_{4} - h_{1} ) x }{2}
+ \sin^2 \frac{ ( h_{2} - h_{1} ) x }{2} 
\biggr\} 
\biggr]. 
\label{P-mutau-nu}
\end{eqnarray} 
The expression of $P(\nu_{\mu} \rightarrow \nu_{S})$ is proportional to $P(\nu_{\mu} \rightarrow \nu_{\tau})$: \\
$P(\nu_{\mu} \rightarrow \nu_{S}) 
= ( c^2_{34} / s^2_{34} ) P(\nu_{\mu} \rightarrow \nu_{\tau})$. One can prove the $\nu_{\mu}$-row unitarity using these results. 

\subsection{The oscillation probability in the $\nu_{e}$ row} 
\label{sec:nue-row}

Among the probabilities in the $\nu_{e}$ row, only 
$P (\nu_{e} \rightarrow \nu_{\tau})$ and $P (\nu_{e} \rightarrow \nu_{S})$ possess the first-order corrections: 
\begin{eqnarray} 
&&
P (\nu_{e} \rightarrow \nu_{\tau}) 
= 
s^2_{34} 
\sin^2 2\widetilde{\theta}_{14} \sin^2 \frac{ ( h_{4} - h_{1} ) x }{2} 
\nonumber \\
&+& 
c^2_{34} s^2_{34} \sin^2 2\widetilde{\theta}_{14} 
\frac{ b }{2E} 
\frac{ 1 }{ ( h_{3} - h_{1} ) } 
\biggl\{ 
- \sin^2 \frac{ ( h_{4} - h_{3} ) x }{2} 
+ \sin^2 \frac{ ( h_{4} - h_{1} ) x }{2} 
+ \sin^2 \frac{ ( h_{3} - h_{1} ) x }{2} 
\biggr\} 
\nonumber \\
&-& 
c^2_{34} s^2_{34} \sin^2 2\widetilde{\theta}_{14} 
\frac{ b }{2E} 
\frac{ 1 }{ ( h_{4} - h_{3} ) } 
\biggl\{ 
\sin^2 \frac{ ( h_{4} - h_{3} ) x }{2} 
+ \sin^2 \frac{ ( h_{4} - h_{1} ) x }{2} 
- \sin^2 \frac{ ( h_{3} - h_{1} ) x }{2} 
\biggr\}. 
\nonumber \\
\label{P-etau-nu}
\end{eqnarray}
\begin{eqnarray} 
&&
P (\nu_{e} \rightarrow \nu_{S}) 
= 
c^2_{34} 
\sin^2 2\widetilde{\theta}_{14} \sin^2 \frac{ ( h_{4} - h_{1} ) x }{2} 
\nonumber \\
&-& 
c^2_{34} s^2_{34} \sin^2 2\widetilde{\theta}_{14} 
\frac{ b }{2E} 
\frac{ 1 }{ ( h_{3} - h_{1} ) } 
\biggl\{ 
- \sin^2 \frac{ ( h_{4} - h_{3} ) x }{2} 
+ \sin^2 \frac{ ( h_{4} - h_{1} ) x }{2} 
+ \sin^2 \frac{ ( h_{3} - h_{1} ) x }{2} 
\biggr\} 
\nonumber \\
&+& 
c^2_{34} s^2_{34} \sin^2 2\widetilde{\theta}_{14} 
\frac{ b }{2E} 
\frac{ 1 }{ ( h_{4} - h_{3} ) } 
\biggl\{ 
\sin^2 \frac{ ( h_{4} - h_{3} ) x }{2} 
+ \sin^2 \frac{ ( h_{4} - h_{1} ) x }{2} 
- \sin^2 \frac{ ( h_{3} - h_{1} ) x }{2} 
\biggr\}. 
\nonumber \\
\label{P-eS-nu}
\end{eqnarray}
The expression of $P(\nu_{e} \rightarrow \nu_{\tau})$ in eq.~\eqref{P-etau-nu} is, in fact, very peculiar in the sense that the zeroth-order and first-order terms have comparable sizes because they have the same suppression factor $s^2_{34}$. This feature occurs in an accidental way, as explained in Appendix~\ref{sec:strange-looking}.  

The corrections to $P (\nu_{e} \rightarrow \nu_{e})$ starts at the second order: 
\begin{eqnarray} 
&& 
P(\nu_{e} \rightarrow \nu_{e}) 
= 
| S^{(0)}_{ee} |^2 + 2 \mbox{Re} \left[ \left\{ S_{ee}^{(0)} \right\}^* S_{ee}^{(2)} \right] 
\nonumber \\
&=& 
1 - \sin^2 2\widetilde{\theta}_{14} \sin^2 \frac{ ( h_{4} - h_{1} ) x }{2} 
\nonumber \\
&-& 
4 \widetilde{c}^2_{14} 
s^2_{24} \left( \widetilde{c}_{14} c_{14} s_{14} - \widetilde{s}_{14} c_{24} c^2_{14} \right)^2  
\left( \frac{ \Delta m^2_{41} }{ 2E } \right)^2 
\nonumber \\
&\times&
\frac{ 1 }{ ( h_{2} - h_{1} )^2 } 
\biggl[ 
\widetilde{c}^2_{14} 
\sin^2 \frac{ ( h_{2} - h_{1} ) x }{2} 
+ \widetilde{s}^2_{14} 
\left\{ \sin^2 \frac{ ( h_{4} - h_{2} ) x }{2} - \sin^2 \frac{ ( h_{4} - h_{1} ) x }{2} \right\}  
\biggr] 
\nonumber \\
&-&
4 \widetilde{s}^2_{14} 
s^2_{24} \left( \widetilde{s}_{14} c_{14} s_{14} + \widetilde{c}_{14} c_{24} c^2_{14} \right)^2 
\left( \frac{ \Delta m^2_{41} }{ 2E } \right)^2 
\nonumber \\
&\times&
\frac{ 1 }{ ( h_{4} - h_{2} )^2 } 
\biggl[ 
\widetilde{s}^2_{14} 
\sin^2 \frac{ ( h_{4} - h_{2} ) x }{2} 
- \widetilde{c}^2_{14} 
\left\{ \sin^2 \frac{ ( h_{4} - h_{1} ) x }{2} - \sin^2 \frac{ ( h_{2} - h_{1} ) x }{2} \right\} 
\biggr] 
\nonumber \\
&-&
\sin^2 2 \widetilde{\theta}_{14} 
c^2_{34} s^2_{34} \left( \frac{b}{ 2E } \right)^2 
\frac{ 1 }{ ( h_{3} - h_{1} )^2 } 
\biggl[ 
\widetilde{c}^2_{14} \sin^2 \frac{ ( h_{3} - h_{1} ) x }{2} 
+ \widetilde{s}^2_{14} 
\left\{ \sin^2 \frac{ ( h_{4} - h_{3} ) x }{2} - \sin^2 \frac{ ( h_{4} - h_{1} ) x }{2} \right\} 
\biggr] 
\nonumber \\
&-& 
\sin^2 2 \widetilde{\theta}_{14} 
c^2_{34} s^2_{34} \left( \frac{b}{ 2E } \right)^2 
\frac{ 1 }{ ( h_{4} - h_{3} )^2 } 
\biggl[ 
\widetilde{s}^2_{14} 
\sin^2 \frac{ ( h_{4} - h_{3} ) x }{2} 
- \widetilde{c}^2_{14} 
\left\{ \sin^2 \frac{ ( h_{4} - h_{1} ) x }{2} - \sin^2 \frac{ ( h_{3} - h_{1} ) x }{2} \right\}  
\biggr] 
\nonumber \\
&-& 
2 \widetilde{c}^2_{14} \widetilde{s}^2_{14} 
\biggl\{ 
s^2_{24} \left( \widetilde{c}_{14} c_{14} s_{14} - \widetilde{s}_{14} c_{24} c^2_{14} \right)^2 
\left( \frac{ \Delta m^2_{41} }{ 2E } \right)^2 
\frac{ 1 }{ ( h_{2} - h_{1} ) } 
+ \widetilde{s}^2_{14} c^2_{34} s^2_{34} 
\left( \frac{ b }{ 2E } \right)^2 
\frac{ 1 }{ ( h_{3} - h_{1} ) } 
\nonumber \\
&+&
s^2_{24} \left( \widetilde{s}_{14} c_{14} s_{14} + \widetilde{c}_{14} c_{24} c^2_{14} \right)^2 
\left( \frac{ \Delta m^2_{41} }{ 2E } \right)^2 
\frac{ 1 }{ ( h_{4} - h_{2} ) } 
+ \widetilde{c}^2_{14} c^2_{34} s^2_{34} \left( \frac{ b }{ 2E } \right)^2 
\frac{ 1 }{ ( h_{4} - h_{3} ) } 
\biggr\} 
x \sin ( h_{4} - h_{1} ) x 
\nonumber \\
&+& 
2 \sin 2 \widetilde{\theta}_{14} s^2_{24} c^2_{14} 
\left\{ \left( s^2_{14}  - c^2_{24} c^2_{14} \right) \sin 2 \widetilde{\theta}_{14} + \sin 2\theta_{14} c_{24} \cos 2 \widetilde{\theta}_{14} \right\} 
\left( \frac{ \Delta m^2_{41} }{2E} \right)^2 
\nonumber \\
&\times&
\biggl[ 
\frac{ 1 }{ ( h_{4} - h_{2} ) ( h_{2} - h_{1} ) } 
\biggl\{ 
\widetilde{s}^2_{14} \sin^2 \frac{ ( h_{4} - h_{2} ) x }{2} 
+ \widetilde{c}^2_{14} \sin^2 \frac{ ( h_{2} - h_{1} ) x }{2} 
\biggr\} 
\nonumber \\
&& 
\hspace{8mm}
- \biggl\{ 
\frac{ \widetilde{s}^2_{14} }{ ( h_{4} - h_{1} ) ( h_{2} - h_{1} ) } 
+ \frac{ \widetilde{c}^2_{14} }{ ( h_{4} - h_{1} ) ( h_{4} - h_{2} ) } 
\biggr\} 
\sin^2 \frac{ ( h_{4} - h_{1} ) x }{2} 
\biggr] 
\nonumber \\
&-& 
2 \sin^2 2 \widetilde{\theta}_{14} c^2_{34} s^2_{34} 
\left( \frac{ b }{2E} \right)^2 
\biggl[ 
\frac{ 1 }{ ( h_{4} - h_{3} ) ( h_{3} - h_{1} ) } 
\biggl\{ 
\widetilde{s}^2_{14} \sin^2 \frac{ ( h_{4} - h_{3} ) x }{2} 
+ \widetilde{c}^2_{14} \sin^2 \frac{ ( h_{3} - h_{1} ) x }{2} 
\biggr\} 
\nonumber \\
&& 
\hspace{8mm}
- \biggl\{ 
\frac{ \widetilde{s}^2_{14} }{ ( h_{4} - h_{1} ) ( h_{3} - h_{1} ) } 
+ \frac{ \widetilde{c}^2_{14} }{ ( h_{4} - h_{1} ) ( h_{4} - h_{3} ) } 
\biggr\} 
\sin^2 \frac{ ( h_{4} - h_{1} ) x }{2} 
\biggr]. 
\label{P-ee-nu}
\end{eqnarray}

$P(\nu_{e} \rightarrow \nu_{\mu})$ starts from the second order, and by $T$-invariance, $P(\nu_{e} \rightarrow \nu_{\mu}) = P(\nu_{\mu} \rightarrow \nu_{e})$, whose latter is given in \eqref{P-mue-nu}. $P(\nu_{e} \rightarrow \nu_{e})$ has the  interesting resonance-enhanced zeroth-order term, see eq.~\eqref{Pee-etau}, but we have computed the second order terms to complete Table~\ref{tab:suppression-nu}. 
We have verified unitarity in the $\nu_{e}$-row in the zeroth order in section~\ref{sec:cascade-nu}. Since the first-order terms in $P(\nu_{e} \rightarrow \nu_{\tau})$ and $P(\nu_{e} \rightarrow \nu_{S})$ add up to zero, as can be seen in eqs.~\eqref{P-etau-nu} and~\eqref{P-eS-nu}, unitarity in the $\nu_{e}$-row holds to first order. The second order unitarity is left for the readers' exercise.

\section{Flavor basis $S$ matrix elements in the anti-neutrino channels}
\label{sec:flavor-S-nubar} 

\subsection{The bar-basis $S$ matrix elements}
\label{sec:bar-S-nubar}

The expressions of non-vanishing $\bar{S}$ matrix elements are given by 
\begin{eqnarray} 
&&
\bar{S}_{12} = 
- e^{ - i \phi_{24} } c_{14} s_{14} \sin ( \widetilde{\theta}_{24} - \theta_{24} ) 
\frac{ \Delta m^2_{41} }{2E}
\frac{ e^{ - i h_{2} x } - e^{ - i h_{1} x } }{ ( h_{2} - h_{1} ) } 
= 
\left[ S_{21} \right]^{T*}, 
\nonumber \\
&& 
\bar{S}_{13} 
= 
- e^{ - i ( \delta + \widetilde{\phi}_{34} ) } c_{14} s_{14} \widetilde{s}_{34} 
\cos ( \widetilde{\theta}_{24} - \theta_{24} ) \frac{ \Delta m^2_{41} }{2E} 
\frac{ e^{ - i h_{3} x } - e^{ - i h_{1} x }  }{ ( h_{3} - h_{1} ) } 
= 
\left[ S_{31} \right]^{T*}, 
\nonumber \\
&& 
\bar{S}_{14} 
= 
c_{14} s_{14} \widetilde{c}_{34} \cos ( \widetilde{\theta}_{24} - \theta_{24} ) 
\frac{ \Delta m^2_{41} }{2E} 
\frac{ e^{ - i h_{4} x } - e^{ - i h_{1} x } }{ ( h_{4} - h_{1} ) } 
= 
\left[ S_{41} \right]^{T*}, 
%
\nonumber \\
&& 
\bar{S}_{23} 
= 
- e^{ i \widetilde{\phi}_{24} } e^{ - i ( \delta + \phi_{34} ) } 
c_{34} s_{34} \widetilde{s}_{24} \widetilde{c}_{34} 
\frac{ |b| }{2E} 
\frac{ e^{ - i h_{3} x } - e^{ - i h_{2} x }  }{ ( h_{3} - h_{2} ) } 
= 
\left[ S_{32} \right]^{T*}, 
\nonumber \\
&& 
\bar{S}_{24} 
= 
- e^{ i ( \widetilde{\phi}_{34} - \phi_{34} ) } e^{ i \widetilde{\phi}_{24} } 
c_{34} s_{34} \widetilde{s}_{24} \widetilde{s}_{34} \frac{ |b| }{2E} 
\frac{ e^{ - i h_{4} x } - e^{ - i h_{2} x } }{ ( h_{4} - h_{2} ) } 
= 
\left[ S_{42} \right]^{T*}, 
%
%
%
%
%
\label{barS-elements-1st-nubar}
\end{eqnarray}
where the explicit expressions of $\hat{A}_{ij}$ elements are implemented. 

\subsection{Flavor basis $S$ matrix elements: Zeroth and first order} 
\label{sec:flavor-S-nubar-0th-1st} 

In the anti-neutrino channel, the notation such as $S_{\bar{\mu} \bar{\mu}}$ is necessary for the $S$ matrix element describing $\bar{\nu}_{\mu} \rightarrow \bar{\nu}_{\mu}$. However, for simplicity we just denote it as $S_{\mu \mu}$ with understanding that we discuss the anti-neutrino channels in this and the other particular sections that are devoted to the anti-neutrinos. 
The non-vanishing zeroth-order $S_{\alpha \beta}^{(0)}$ matrix elements are given below. In the presentation given below, for simplicity of the expressions, we implement $e^{ \pm i ( \widetilde{\phi}_{24} - \phi_{24} ) } = e^{ \pm i ( \widetilde{\phi}_{34} - \phi_{34} ) } = 1$. 

\begin{eqnarray} 
S_{ee}  ^{(0)}
&=&
e^{ - i h_{1} x },  
\nonumber \\
S_{\mu \mu} ^{(0)} 
&=&
\left[ \widetilde{c}^2_{24} e^{ - i h_{2} x } 
+ \widetilde{s}^2_{24} 
\left( \widetilde{c}^2_{34} e^{ - i h_{4} x } + \widetilde{s}^2_{34} e^{ - i h_{3} x } \right) \right], 
\nonumber \\
S_{\tau \tau} ^{(0)} 
&=& 
c^2_{34} 
\left( \widetilde{s}^2_{34} e^{ - i h_{4} x } + \widetilde{c}^2_{34} e^{ - i h_{3} x } \right)  
+ s^2_{34} 
\left[ \widetilde{s}^2_{24} e^{ - i h_{2} x } 
+ \widetilde{c}^2_{24} \left( \widetilde{c}^2_{34} e^{ - i h_{4} x } + \widetilde{s}^2_{34} e^{ - i h_{3} x } \right) \right] 
\nonumber \\
&+&
\sin 2\theta_{34} 
\widetilde{c}_{24} \widetilde{c}_{34} \widetilde{s}_{34} \left( e^{ - i h_{4} x } - e^{ - i h_{3} x } \right),  
\nonumber \\
S_{S S} ^{(0)} 
&=&
s^2_{34} \left( \widetilde{s}^2_{34} e^{ - i h_{4} x } + \widetilde{c}^2_{34} e^{ - i h_{3} x } \right)  
+ c^2_{34} \left[ \widetilde{s}^2_{24} e^{ - i h_{2} x } 
+ \widetilde{c}^2_{24} \left( \widetilde{c}^2_{34} e^{ - i h_{4} x } + \widetilde{s}^2_{34} e^{ - i h_{3} x } \right) \right] 
\nonumber \\
&-&
\sin 2\theta_{34} 
\widetilde{c}_{24} \widetilde{c}_{34} \widetilde{s}_{34} \left( e^{ - i h_{4} x } - e^{ - i h_{3} x } \right). 
\label{S-elements-0th-nubar-1}
\end{eqnarray}
\begin{eqnarray} 
S_{\mu \tau} ^{(0)} 
&=& 
e^{ i \widetilde{\phi}_{24} } e^{ - i ( \delta + \phi_{34} ) } 
\biggl\{ 
c_{34} \widetilde{s}_{24} \widetilde{c}_{34} \widetilde{s}_{34} \left( e^{ - i h_{4} x } - e^{ - i h_{3} x } \right) 
+ 
s_{34} \widetilde{c}_{24} \widetilde{s}_{24} 
\left[ - e^{ - i h_{2} x } 
+ \left( \widetilde{c}^2_{34} e^{ - i h_{4} x } + \widetilde{s}^2_{34} e^{ - i h_{3} x } \right)  \right] 
\biggr\} 
\nonumber \\
&=&
\left[ S_{\tau \mu} \right]^{T*}, 
\nonumber \\
S_{\mu S} ^{(0)} 
&=&
e^{ i \widetilde{\phi}_{24} } 
\biggl\{ 
- 
s_{34} \widetilde{s}_{24} \widetilde{c}_{34} \widetilde{s}_{34} \left( e^{ - i h_{4} x } - e^{ - i h_{3} x } \right) 
+ c_{34} 
\widetilde{c}_{24} \widetilde{s}_{24} 
\left[ - e^{ - i h_{2} x } 
+ \left( \widetilde{c}^2_{34} e^{ - i h_{4} x } + \widetilde{s}^2_{34} e^{ - i h_{3} x } \right)  \right] 
\biggr\} 
\nonumber \\
&=&
\left[ S_{S \mu} \right]^{T*}, 
\nonumber \\
S_{\tau S} ^{(0)} 
&=&
e^{ i ( \delta + \phi_{34} ) } 
\biggl\{ c_{34} s_{34} 
\left[ \widetilde{s}^2_{24} e^{ - i h_{2} x } 
+ ( \widetilde{c}^2_{24} \widetilde{c}^2_{34} - \widetilde{s}^2_{34} ) e^{ - i h_{4} x } 
+ ( \widetilde{c}^2_{24} \widetilde{s}^2_{34} - \widetilde{c}^2_{34} ) e^{ - i h_{3} x } \right] 
\nonumber \\
&+& 
\cos 2\theta_{34} 
\widetilde{c}_{24} \widetilde{c}_{34} \widetilde{s}_{34} \left( e^{ - i h_{4} x } - e^{ - i h_{3} x } \right) 
\biggr\} 
= 
\left[ S_{S \tau} \right]^{T*}. 
\label{S-elements-0th-nubar-2}
\end{eqnarray}

The non-vanishing $S_{\alpha \beta}^{(1)}$ matrix elements are given, using the $\bar{S}_{ij}^{(1)}$ matrix elements given in eq.~\eqref{barS-elements-1st-nubar}, as 
\begin{eqnarray} 
S_{\mu \mu} ^{(1)} 
&=&
e^{ - i \widetilde{\phi}_{24} } \widetilde{c}_{24} \widetilde{s}_{24}  
\left\{ - e^{ i ( \delta + \widetilde{\phi}_{34} ) } \widetilde{s}_{34} \bar{S}_{23} + \widetilde{c}_{34} \bar{S}_{24} \right\} 
+ e^{ i \widetilde{\phi}_{24} } \widetilde{c}_{24} \widetilde{s}_{24}  
\left\{ - e^{ - i ( \delta + \widetilde{\phi}_{34} ) } \widetilde{s}_{34} \bar{S}_{32} + \widetilde{c}_{34} \bar{S}_{42} \right\},  
\nonumber \\
S_{\tau \tau} ^{(1)} 
&=& 
\left( s^2_{34} \widetilde{c}_{24} \widetilde{s}_{24} \widetilde{s}_{34} 
- c_{34} s_{34} \widetilde{s}_{24} \widetilde{c}_{34} \right) 
\left\{ 
e^{ i ( \delta + \phi_{34} ) } e^{ - i \widetilde{\phi}_{24} } 
\bar{S}_{23} 
+ e^{ - i ( \delta + \phi_{34} ) } e^{ i \widetilde{\phi}_{24} } 
\bar{S}_{32} 
\right\} 
\nonumber \\
&-&
\left( 
c_{34} s_{34} \widetilde{s}_{24} \widetilde{s}_{34} 
+ s^2_{34} \widetilde{c}_{24} \widetilde{s}_{24} \widetilde{c}_{34} 
\right) 
\left\{ 
e^{ - i \widetilde{\phi}_{24} } \bar{S}_{24} 
+ e^{ i \widetilde{\phi}_{24} } \bar{S}_{42} 
\right\}, 
\nonumber \\
S_{S S} ^{(1)} 
&=&
\left( c^2_{34} \widetilde{c}_{24} \widetilde{s}_{24} \widetilde{s}_{34} 
+ c_{34} s_{34} \widetilde{s}_{24} \widetilde{c}_{34} \right) 
\left\{ 
e^{ i ( \delta + \phi_{34} ) } e^{ - i \widetilde{\phi}_{24} } 
\bar{S}_{23} 
+ 
e^{ - i ( \delta + \phi_{34} ) } e^{ i \widetilde{\phi}_{24} } 
\bar{S}_{32} 
\right\} 
\nonumber \\
&+& 
\left( c_{34} s_{34} \widetilde{s}_{24} \widetilde{s}_{34} 
- c^2_{34} \widetilde{c}_{24} \widetilde{s}_{24} \widetilde{c}_{34} \right) 
\left\{ e^{ - i \widetilde{\phi}_{24} } \bar{S}_{24} 
+ e^{ i \widetilde{\phi}_{24} } \bar{S}_{42} \right\}.  
\label{S-elements-1st-nubar-1}
\end{eqnarray}
\begin{eqnarray} 
S_{e \mu} ^{(1)} 
&=&
\widetilde{c}_{24} \bar{S}_{12} 
+ \widetilde{s}_{24} e^{ - i \widetilde{\phi}_{24} } 
\left\{ - e^{ i ( \delta + \widetilde{\phi}_{34} ) } \widetilde{s}_{34} \bar{S}_{13} + \widetilde{c}_{34} \bar{S}_{14} \right\} 
= 
\left[ S_{\mu e} ^{(1)} \right] ^{T*}, 
\nonumber \\
S_{e \tau} ^{(1)} 
&=&
- e^{ - i ( \delta + \phi_{34} ) } e^{ i \widetilde{\phi}_{24} } s_{34} \widetilde{s}_{24} \bar{S}_{12} 
+ \left( c_{34} \widetilde{c}_{34} 
- 
s_{34} \widetilde{c}_{24} \widetilde{s}_{34} \right) \bar{S}_{13} 
+ e^{ - i ( \delta + \phi_{34} ) } 
\left( 
c_{34} \widetilde{s}_{34}  
+ s_{34} \widetilde{c}_{24} \widetilde{c}_{34} 
\right) 
\bar{S}_{14} 
\nonumber \\
&=& 
\left[ S_{\tau e} ^{(1)} \right] ^{T*}, 
\nonumber \\
S_{e S} ^{(1)} 
&=&
- e^{ i \widetilde{\phi}_{24} } c_{34} \widetilde{s}_{24} \bar{S}_{12}  
- e^{ i ( \delta + \phi_{34} ) } 
\left( s_{34} \widetilde{c}_{34} 
+ 
c_{34} \widetilde{c}_{24} \widetilde{s}_{34} \right) \bar{S}_{13} 
- 
\left( s_{34} \widetilde{s}_{34} 
- c_{34} \widetilde{c}_{24} \widetilde{c}_{34} \right) 
\bar{S}_{14} 
= 
\left[ S_{S e} ^{(1)} \right] ^{T*},  
\nonumber \\
S_{\tau \mu} ^{(1)} 
&=&
c_{34} \widetilde{c}_{24} 
\left\{ \widetilde{c}_{34} \bar{S}_{32} + e^{ i ( \delta + \widetilde{\phi}_{34} ) } \widetilde{s}_{34} \bar{S}_{42} \right\} 
\nonumber \\
&& 
\hspace{-8mm}
+ e^{ - i \widetilde{\phi}_{24} } e^{ i ( \delta + \phi_{34} ) } 
\left[ 
e^{ i \widetilde{\phi}_{24} } s_{34} \widetilde{c}^2_{24} 
\left\{ - e^{ - i ( \delta + \widetilde{\phi}_{34} ) } \widetilde{s}_{34} \bar{S}_{32} + \widetilde{c}_{34} \bar{S}_{42} \right\} 
- e^{ - i \widetilde{\phi}_{24} } s_{34} \widetilde{s}^2_{24} 
\left\{ - e^{ i ( \delta + \widetilde{\phi}_{34} ) } \widetilde{s}_{34} \bar{S}_{23} + \widetilde{c}_{34} \bar{S}_{24} \right\} 
\right] 
\nonumber \\
&=& 
\left[ S_{\mu \tau} ^{(1)} \right] ^{T*},  
\nonumber \\
S_{S \mu} ^{(1)} 
&=&
- e^{ - i ( \delta + \phi_{34} ) } s_{34} \widetilde{c}_{24} 
\left\{ \widetilde{c}_{34} \bar{S}_{32} + e^{ i ( \delta + \widetilde{\phi}_{34} ) } \widetilde{s}_{34} \bar{S}_{42} \right\} 
\nonumber \\
&& 
\hspace{-8mm}
+
c_{34} e^{ - i \widetilde{\phi}_{24} } 
\left[ e^{ i \widetilde{\phi}_{24} } \widetilde{c}^2_{24} 
\left\{ - e^{ - i ( \delta + \widetilde{\phi}_{34} ) } \widetilde{s}_{34} \bar{S}_{32} + \widetilde{c}_{34} \bar{S}_{42} \right\} 
- e^{ - i \widetilde{\phi}_{24} } \widetilde{s}^2_{24} 
\left\{ - e^{ i ( \delta + \widetilde{\phi}_{34} ) } \widetilde{s}_{34} \bar{S}_{23} + \widetilde{c}_{34} \bar{S}_{24} \right\} 
\right] 
=  
\left[ S_{\mu S} ^{(1)} \right] ^{T*}, 
\nonumber \\
S_{S \tau} ^{(1)} 
&=&
e^{ - i ( \delta + \phi_{34} ) } 
\biggl\{ 
e^{ i ( \delta + \phi_{34} ) } e^{ - i \widetilde{\phi}_{24} } 
\left( c_{34} s_{34} \widetilde{c}_{24} \widetilde{s}_{24} \widetilde{s}_{34} 
- c^2_{34} \widetilde{s}_{24} \widetilde{c}_{34} \right) 
\bar{S}_{23} 
\nonumber \\
&+&
e^{ - i ( \delta + \phi_{34} ) } e^{ i \widetilde{\phi}_{24} } 
\left( c_{34} s_{34} \widetilde{c}_{24} \widetilde{s}_{24} \widetilde{s}_{34} 
+ s^2_{34} \widetilde{s}_{24} \widetilde{c}_{34} \right) 
\bar{S}_{32} 
\nonumber \\
&& 
\hspace{-6mm} 
- 
e^{ - i \widetilde{\phi}_{24} } 
\left( c_{34} s_{34} \widetilde{c}_{24} \widetilde{s}_{24} \widetilde{c}_{34} 
+ c^2_{34} \widetilde{s}_{24} \widetilde{s}_{34} \right) 
\bar{S}_{24} 
- e^{ i \widetilde{\phi}_{24} } 
\left( c_{34} s_{34} \widetilde{c}_{24} \widetilde{s}_{24} \widetilde{c}_{34} 
- s^2_{34} \widetilde{s}_{24} \widetilde{s}_{34} \right) 
\bar{S}_{42} 
\biggr\} 
= 
\left[ S_{\tau S} ^{(1)} \right] ^{T*}.  
\label{S-elements-1st-nubar-2}
\end{eqnarray}

\newpage

\section{The oscillation probability in the anti-neutrino channel} 
\label{sec:probability-summary-nubar}

The matrix structure of the $S$ matrix is given in eq.~\eqref{flavorS-1st-nubar}, from which the higher order correction terms have the characteristic patterns. In the $\bar{\nu}_{\mu}$ row channels, the three channels, $P (\bar{\nu}_{\mu} \rightarrow \bar{\nu}_{\mu})$, $P (\bar{\nu}_{\mu} \rightarrow \bar{\nu}_{\tau})$ and $P (\bar{\nu}_{\mu} \rightarrow \bar{\nu}_{S})$, have first-order corrections. Whereas $P (\bar{\nu}_{\mu} \rightarrow \bar{\nu}_{e})$ starts at the leading second order expression. In the $\bar{\nu}_{e}$ row channels, $P(\bar{\nu}_{e} \rightarrow \bar{\nu}_{e} )$ has no first order, but the second-order correction. $P(\bar{\nu}_{e} \rightarrow \bar{\nu}_{\mu} )$, $P(\bar{\nu}_{e} \rightarrow \bar{\nu}_{\tau} )$, and $P(\bar{\nu}_{e} \rightarrow \bar{\nu}_{S} )$, start at the leading second order terms. 

\subsection{The oscillation probability in the $\bar{\nu}_{\mu}$ row} 
\label{sec:numu-bar-row}

The expression of $P(\bar{\nu}_{\mu} \rightarrow \bar{\nu}_{\mu} )$ is given in eq.~\eqref{Pmumu-1st-nubar} to first order. We only show the probabilities of the other two channels, with $P (\bar{\nu}_{\mu} \rightarrow \bar{\nu}_{S})$ not shown. 
\begin{eqnarray} 
&&
P (\bar{\nu}_{\mu} \rightarrow \bar{\nu}_{\tau}) 
= 
\vert S_{32}^{(0)} \vert^2 
+ 2 \mbox{Re} \left[ ( S_{32}^{(0)} )^* S_{32}^{(1)} \right] 
\nonumber \\
&=&
c^2_{34} \widetilde{s}^2_{24} \sin^2 2\widetilde{\theta}_{34} 
\sin^2 \frac{ ( h_{4} - h_{3} ) x }{2} 
\nonumber \\
&+& 
s^2_{34} \sin^2 2\widetilde{\theta}_{24} 
\biggl\{ 
\widetilde{c}^2_{34} \sin^2 \frac{ ( h_{4} - h_{2} ) x }{2} 
+ \widetilde{s}^2_{34} \sin^2 \frac{ ( h_{3} - h_{2} ) x }{2} 
- \widetilde{c}^2_{34} \widetilde{s}^2_{34} \sin^2 \frac{ ( h_{4} - h_{3} ) x }{2} 
\biggr\} 
\nonumber \\
&+& 
c_{34} s_{34} \widetilde{s}_{24} 
\sin 2\widetilde{\theta}_{24} \sin 2\widetilde{\theta}_{34} 
\biggl\{ 
\cos 2\widetilde{\theta}_{34} \sin^2 \frac{ ( h_{4} - h_{3} ) x }{2} 
+ \sin^2 \frac{ ( h_{4} - h_{2} ) x }{2} - \sin^2 \frac{ ( h_{3} - h_{2} ) x }{2} 
\biggr\} 
\nonumber \\
&+& 
c^2_{34} s_{34} \widetilde{s}_{24} \sin 2\widetilde{\theta}_{34} 
\biggl[ 
\left( c_{34} \sin 2\widetilde{\theta}_{24} \widetilde{c}^2_{34} 
- s_{34} \widetilde{s}_{24} \cos 2\widetilde{\theta}_{24} 
\sin 2\widetilde{\theta}_{34} \right) \frac{ |b| }{2E} 
\nonumber \\
&& 
\hspace{-4mm} 
\times 
\frac{1}{ ( h_{3} - h_{2} ) } 
\biggl\{ 
\sin^2 \frac{ ( h_{4} - h_{3} ) x }{2} 
- \sin^2 \frac{ ( h_{4} - h_{2} ) x }{2} 
+ \sin^2 \frac{ ( h_{3} - h_{2} ) x }{2} 
\biggr\} 
\nonumber \\
&-& 
\left( c_{34} \sin 2\widetilde{\theta}_{24} \widetilde{s}^2_{34} 
+ s_{34} \widetilde{s}_{24} \cos 2\widetilde{\theta}_{24} \sin 2\widetilde{\theta}_{34} 
\right) 
\frac{ |b| }{2E} 
\nonumber \\
&& 
\hspace{-4mm} 
\times 
\frac{1}{ ( h_{4} - h_{2} ) } 
\biggl\{ 
\sin^2 \frac{ ( h_{4} - h_{3} ) x }{2} 
+ \sin^2 \frac{ ( h_{4} - h_{2} ) x }{2} 
- \sin^2 \frac{ ( h_{3} - h_{2} ) x }{2} 
\biggr\} 
\biggr] 
\nonumber \\
&+& 
c_{34} s^2_{34} \sin 2\widetilde{\theta}_{24} 
\biggl[ 
\left( - c_{34} \widetilde{c}^2_{34} \sin 2\widetilde{\theta}_{24} 
+ s_{34} \widetilde{s}_{24} \sin 2\widetilde{\theta}_{34} 
\cos 2\widetilde{\theta}_{24} \right) 
\frac{ |b| }{2E} 
\nonumber \\
&\times&
\frac{ 1 }{ ( h_{3} - h_{2} ) } 
\biggl\{ 
- \widetilde{c}^2_{34} \sin^2 \frac{ ( h_{4} - h_{3} ) x }{2}
+ \widetilde{c}^2_{34} \sin^2 \frac{ ( h_{4} - h_{2} ) x }{2}
+ ( 1 + \widetilde{s}^2_{34} ) \sin^2 \frac{ ( h_{3} - h_{2} ) x }{2} 
\biggr\} 
\nonumber \\
&-& 
\left( c_{34} \widetilde{s}^2_{34} \sin 2\widetilde{\theta}_{24} 
+ s_{34} \widetilde{s}_{24} \sin 2\widetilde{\theta}_{34} 
\cos 2\widetilde{\theta}_{24} \right) 
\frac{ |b| }{2E} 
\nonumber \\
&\times&
\frac{1}{ ( h_{4} - h_{2} ) } 
\biggl\{ 
( 1 + \widetilde{c}^2_{34} ) \sin^2 \frac{ ( h_{4} - h_{2} ) x }{2}
- \widetilde{s}^2_{34} \sin^2 \frac{ ( h_{4} - h_{3} ) x }{2}
+ \widetilde{s}^2_{34} \sin^2 \frac{ ( h_{3} - h_{2} ) x }{2} 
\biggr\} 
\biggr]. 
\nonumber \\
\label{Pmutau-1st-nubar}
\end{eqnarray} 
\begin{eqnarray} 
&& 
P (\bar{\nu}_{\mu} \rightarrow \bar{\nu}_{e}) 
= 
\vert S_{e \mu} ^{(1)} \vert ^2 
\nonumber \\
&=& 
\sin^2 2\theta_{14}
\left( \frac{ \Delta m^2_{41} }{2E} \right)^2 
\biggl[ 
\widetilde{c}^2_{24} \sin^2 ( \widetilde{\theta}_{24} - \theta_{24} ) 
\frac{ 1 }{ ( h_{2} - h_{1} )^2 } 
\sin^2 \frac{ ( h_{2} - h_{1} ) x }{2}
\nonumber \\
&+&
\widetilde{s}^2_{24} \widetilde{s}^4_{34} 
\cos^2 ( \widetilde{\theta}_{24} - \theta_{24} ) 
\frac{ 1 }{ ( h_{3} - h_{1} )^2 } 
\sin^2 \frac{ ( h_{3} - h_{1} ) x }{2}
\nonumber \\
&+&
\widetilde{s}^2_{24} \widetilde{c}^4_{34} 
\cos^2 ( \widetilde{\theta}_{24} - \theta_{24} ) 
\frac{ 1 }{ ( h_{4} - h_{1} )^2 } 
\sin^2 \frac{ ( h_{4} - h_{1} ) x }{2} 
\nonumber \\
&-& 
\widetilde{c}_{24} \widetilde{s}_{24} \widetilde{s}^2_{34} 
\sin ( \widetilde{\theta}_{24} - \theta_{24} ) 
\cos ( \widetilde{\theta}_{24} - \theta_{24} ) 
\frac{ 1 }{ ( h_{2} - h_{1} ) ( h_{3} - h_{1} ) } 
\nonumber \\
&\times&
\biggl\{ - \sin^2 \frac{ ( h_{3} - h_{2} ) x }{2} 
+ \sin^2 \frac{ ( h_{3} - h_{1} ) x }{2} 
+ \sin^2 \frac{ ( h_{2} - h_{1} ) x }{2} \biggr\} 
\nonumber \\
&-&
\widetilde{c}_{24} \widetilde{s}_{24} \widetilde{c}^2_{34} 
\sin ( \widetilde{\theta}_{24} - \theta_{24} ) 
\cos ( \widetilde{\theta}_{24} - \theta_{24} ) 
\frac{ 1 }{ ( h_{2} - h_{1} ) ( h_{4} - h_{1} ) } 
\nonumber \\
&\times&
\biggl\{ - \sin^2 \frac{ ( h_{4} - h_{2} ) x }{2} 
+ \sin^2 \frac{ ( h_{4} - h_{1} ) x }{2} 
+ \sin^2 \frac{ ( h_{2} - h_{1} ) x }{2} \biggr\} 
\nonumber \\
&+& 
\widetilde{s}^2_{24} \widetilde{c}^2_{34} \widetilde{s}^2_{34} 
\cos^2 ( \widetilde{\theta}_{24} - \theta_{24} ) 
\frac{ 1 }{ ( h_{4} - h_{1} ) ( h_{3} - h_{1} ) } 
\nonumber \\
&\times&
\biggl\{ 
- \sin^2 \frac{ ( h_{4} - h_{3} ) x }{2} 
+ \sin^2 \frac{ ( h_{4} - h_{1} ) x }{2} 
+ \sin^2 \frac{ ( h_{3} - h_{1} ) x }{2} 
\biggr\} 
\biggr].  
\label{P-mue-2nd-nubar}
\end{eqnarray}

With the expression of $P (\bar{\nu}_{\mu} \rightarrow \bar{\nu}_{S})$ (computed but not shown), unitarity in the first order, $P(\bar{\nu}_{\mu} \rightarrow \bar{\nu}_{\mu}) ^{(1)} + P (\bar{\nu}_{\mu} \rightarrow \bar{\nu}_{\tau}) ^{(1)} + P (\bar{\nu}_{\mu} \rightarrow \bar{\nu}_{S}) ^{(1)} = 0$, is explicitly verified. 

\subsection{The oscillation probability in the $\bar{\nu}_{e}$ row} 
\label{sec:nue-bar-row}

Due to $T$ invariance, $P(\bar{\nu}_{e} \rightarrow \bar{\nu}_{\mu} ) = P(\bar{\nu}_{\mu} \rightarrow \bar{\nu}_{e} )$, and hence it is already computed in the above, eq.~\eqref{P-mue-2nd-nubar}. Here, we present only $P (\bar{\nu}_{e} \rightarrow \bar{\nu}_{e})$ and $P( \bar{\nu}_{e} \rightarrow \bar{\nu}_{\tau} )$, leaving $P( \bar{\nu}_{e} \rightarrow \bar{\nu}_{S} )$ not shown. 
$P (\bar{\nu}_{e} \rightarrow \bar{\nu}_{e})$ has the second-order correction, and the rest start at the second order. 
Unitarity to the second-order, 
$P( \bar{\nu}_{e} \rightarrow \bar{\nu}_{e} )^{(2)} 
+ P( \bar{\nu}_{e} \rightarrow \bar{\nu}_{\mu} )^{(2)}
+ P( \bar{\nu}_{e} \rightarrow \bar{\nu}_{\tau} )^{(2)} 
+ P( \bar{\nu}_{e} \rightarrow \bar{\nu}_{S} )^{(2)} = 0$, 
is explicitly verified. 
\begin{eqnarray} 
&&
P(\bar{\nu}_{e} \rightarrow \bar{\nu}_{e}) 
= 
| S_{ee}^{(0)} |^2 
+ 2 \mbox{Re} \left[ ( S_{ee}^{(0)} )^* S_{ee}^{(2)} \right] 
\nonumber \\
&=& 
1 - \left( \frac{ \Delta m^2_{41} }{2E} \right)^2 \sin^2 2\theta_{14} 
\biggl[ 
\sin^2 ( \widetilde{\theta}_{24} - \theta_{24} ) 
\frac{1}{ ( h_{2} - h_{1} )^2 }
\sin^2 \frac{ ( h_{2} - h_{1} ) x  }{2} 
\nonumber \\
&+&
\cos^2 ( \widetilde{\theta}_{24} - \theta_{24} ) 
\left\{ 
\widetilde{s}^2_{34} 
\frac{1}{ ( h_{3} - h_{1} )^2 } \sin^2 \frac{ ( h_{3} - h_{1} ) x  }{2} 
+ \widetilde{c}^2_{34} 
\frac{1}{ ( h_{4} - h_{1} )^2 } \sin^2 \frac{ ( h_{4} - h_{1} ) x  }{2} 
\right\} 
\biggr]. 
\nonumber \\
\label{P-ee-2nd-nubar}
\end{eqnarray}
\begin{eqnarray} 
&& 
P( \bar{\nu}_{e} \rightarrow \bar{\nu}_{\tau} ) 
= 
\vert S_{\tau e} ^{(1)} \vert ^2 
\nonumber \\
&=& 
\sin^2 2\theta_{14} 
\left( \frac{ \Delta m^2_{41} }{2E} \right)^2 
\biggl[ 
s^2_{34} \widetilde{s}^2_{24} \sin^2 ( \widetilde{\theta}_{24} - \theta_{24} ) 
\frac{ 1 }{ ( h_{2} - h_{1} )^2 } 
\sin^2 \frac{ ( h_{2} - h_{1} ) x }{2}
\nonumber \\
&+&
\left( c_{34} \widetilde{c}_{34} - s_{34} \widetilde{c}_{24} \widetilde{s}_{34} \right)^2 
\widetilde{s}^2_{34} 
\cos^2 ( \widetilde{\theta}_{24} - \theta_{24} ) 
\frac{ 1 }{ ( h_{3} - h_{1} )^2 } 
\sin^2 \frac{ ( h_{3} - h_{1} ) x }{2} 
\nonumber \\
&+&
\left( c_{34} \widetilde{s}_{34} + s_{34} \widetilde{c}_{24} \widetilde{c}_{34} \right)^2 
\widetilde{c}^2_{34} \cos^2 ( \widetilde{\theta}_{24} - \theta_{24} ) 
\frac{ 1 }{ ( h_{4} - h_{1} )^2 } 
\sin^2 \frac{ ( h_{4} - h_{1} ) x }{2}
\nonumber \\
&-& 
s_{34} \widetilde{s}_{24} \widetilde{s}_{34} 
\left( c_{34} \widetilde{c}_{34} - s_{34} \widetilde{c}_{24} \widetilde{s}_{34} \right) 
\cos ( \widetilde{\theta}_{24} - \theta_{24} ) 
\sin ( \widetilde{\theta}_{24} - \theta_{24} ) 
\frac{1}{ ( h_{2} - h_{1} ) ( h_{3} - h_{1} ) } 
\nonumber \\
&\times& 
\biggl\{ - \sin^2 \frac{ ( h_{3} - h_{2} ) x }{2} 
+ \sin^2 \frac{ ( h_{3} - h_{1} ) x }{2} 
+ \sin^2 \frac{ ( h_{2} - h_{1} ) x }{2} \biggr\} 
\nonumber \\
&+& 
s_{34} \widetilde{s}_{24} \widetilde{c}_{34} 
\left( c_{34} \widetilde{s}_{34} + s_{34} \widetilde{c}_{24} \widetilde{c}_{34} \right) 
\cos ( \widetilde{\theta}_{24} - \theta_{24} ) 
\sin ( \widetilde{\theta}_{24} - \theta_{24} ) 
\frac{1}{ ( h_{2} - h_{1} ) ( h_{4} - h_{1} ) } 
\nonumber \\
&\times& 
\biggl\{ - \sin^2 \frac{ ( h_{4} - h_{2} ) x }{2} 
+ \sin^2 \frac{ ( h_{4} - h_{1} ) x }{2} 
+ \sin^2 \frac{ ( h_{2} - h_{1} ) x }{2} \biggr\} 
\nonumber \\
&-& 
\widetilde{c}_{34} \widetilde{s}_{34} 
\left[ \left( c^2_{34} - s^2_{34} \widetilde{c}^2_{24} \right) \widetilde{c}_{34} \widetilde{s}_{34} + c_{34} s_{34} \widetilde{c}_{24} \cos 2\widetilde{\theta}_{34} 
\right] 
\cos^2 ( \widetilde{\theta}_{24} - \theta_{24} ) 
\frac{1}{ ( h_{4} - h_{1} ) ( h_{3} - h_{1} ) } 
\nonumber \\
&\times&
\biggl\{ 
- \sin^2 \frac{ ( h_{4} - h_{3} ) x }{2} 
+ \sin^2 \frac{ ( h_{4} - h_{1} ) x }{2} 
+ \sin^2 \frac{ ( h_{3} - h_{1} ) x }{2} 
\biggr\} 
\biggr]. 
\label{P-etau-2nd-nubar}
\end{eqnarray}

\section{Understanding Tables~\ref{tab:suppression-nu} and~\ref{tab:suppression-nubar}}
\label{sec:understanding-tables}

One can identify the two types of physically relevant region for the SA resonance. They are characterized as ``on-peak'' and ``off-peak'' around the SA resonance. To explain what we mean by on-peak and off-peak, we take the 3-4 resonance with the expression of the matter-affected angle given in eq.~\eqref{matter-theta34}. By on-peak we mean $( \hat{\lambda}_{4} - \hat{\lambda}_{3} )^2 \ll ( \widetilde{c}_{24} \sin 2\theta_{34} |b| )^2$ so that $\sin 2\widetilde{\theta}_{34} \approx 1$. In contrast, by off-peak we mean $( \hat{\lambda}_{4} - \hat{\lambda}_{3} )^2 \gsim ( \widetilde{c}_{24} \sin 2\theta_{34} |b| )^2$ so that $\sin 2\theta_{34}$ in the numerator of $\sin 2\widetilde{\theta}_{34}$ can act as a suppression factor, as explained in eq.~\eqref{prefactor-mumu} using the expression of 
$P (\bar{\nu}_{\mu} \rightarrow \bar{\nu}_{\mu})$.\footnote{
If we evaluate the matter-affected angles at the edge of off-peak $( \hat{\lambda}_{4} - \hat{\lambda}_{3} )^2 = ( \widetilde{c}_{24} \sin 2\theta_{34} |b| )^2$, $\tan 2\widetilde{\theta}_{34} = 1$. It means $\widetilde{\theta}_{14} = \pi/8$, or $\widetilde{s}^2_{34} = 0.15$, which means that $\widetilde{s}^2_{34}$ can provide further suppression factor at off-peak. However, in our estimate given in   Tables~\ref{tab:suppression-nu} and~\ref{tab:suppression-nubar}, we disregard this suppression to give a conservative estimate. }

At off-peak of the SA resonance, generally speaking, the probability has the two terms, the leading and the next to leading order terms. In the simplest case they are the zeroth-order term, and the interference between zeroth and first-order terms. In some cases, $P(\nu_{\mu} \rightarrow \nu_{\mu})$ and $P(\nu_{e} \rightarrow \nu_{e})$ in the neutrino channel, the interference terms are the one between the zeroth and second-order terms, due to the absence of the first-order $S$ matrix elements, as shown in eq.~\eqref{flavorS-nu-0th-2nd}. We note here that the terminology of the zeroth, first, and second orders refers the perturbation series as formulated in section~\ref{sec:calculation-barS}. 
In the exceptional cases, the leading term of the probability takes the form of the first-order amplitude squared. These are the cases of 
$P(\nu_{\mu} \rightarrow \nu_{e})$ and $P(\nu_{\mu} \rightarrow \nu_{\tau})$ in the neutrino channel, and $P (\bar{\nu}_{\mu} \rightarrow \bar{\nu}_{e})$ and $P (\bar{\nu}_{e} \rightarrow \bar{\nu}_{\tau})$ in the anti-neutrino channel. It occurs because $S_{e \mu}^{(0)} = S_{\tau \mu}^{(0)} =0$ in the neutrino channel, and $S_{e \mu}^{(0)} = S_{\tau e}^{(0)} = 0$ in the anti-neutrino channel, as can be seen in eqs.~\eqref{flavorS-nu-0th-2nd} and~\eqref{flavorS-1st-nubar}, respectively. 

Though a line by line explanation is possible in each channel, it would become too lengthy, and the simplest way to confirm the results of Tables~\ref{tab:suppression-nu} and~\ref{tab:suppression-nubar} is to follow the similar analysis as we have done in $P (\bar{\nu}_{\mu} \rightarrow \bar{\nu}_{\mu})$. 
If one executes this exercise one will found that the pre-factors in $P(\bar{\nu}_{\mu} \rightarrow \bar{\nu}_{\tau})$ have richer varieties. There exist four pre-factors in the interference terms in eq.~\eqref{Pmutau-1st-nubar}, but each of them has the two clusters of terms with the suppression factors $s^2_{24} s^2_{34}$ and $s^2_{24} s^4_{34}$. In such cases only the least suppression factor is tabulated in  Tables~\ref{tab:suppression-nu} and~\ref{tab:suppression-nubar}. 

\subsection{A strange-looking case} 
\label{sec:strange-looking} 

The seemingly strange-looking feature in $P(\nu_{e} \rightarrow \nu_{\tau})$, the zeroth- and first-order terms have the same suppression factor $s^2_{34}$ as given in eq.~\eqref{P-etau-nu}, occurs by accident. 
The zeroth-order flavor basis $S$ matrix element $S_{\tau e}^{(0)}$ picks up the factor $e^{ - i ( \delta + \phi_{34} ) } s_{34}$ which comes from the 3-4 rotation that needed to transform the bar-basis $S$ matrix to the flavor-basis one as in eq.~\eqref{flavor-bar-S}. On the other hand, $S_{\tau e}^{(1)} = c_{34} \left( \widetilde{c}_{14} \bar{S}_{31} ^{(1)} + \widetilde{s}_{14} \bar{S}_{34} ^{(1)} \right) $ does not pick up that $s_{34}$, but instead the both $\bar{S}_{31} ^{(1)}$ and $\bar{S}_{34} ^{(1)}$ have $s_{34}$ that comes from $A_{31}$ and $A_{34}$ in $\bar{H}_{1}$, see \eqref{Aij-summary-nu}. That is, $s_{34}$ in $S_{\tau e}^{(0)}$ and $s_{34}$ in $S_{\tau e}^{(1)}$ have the different origins, but they are the same $s_{34}$, and hence the both zeroth- and first-order (interference) parts of the probability have the identical factor $s^2_{34}$.

\end{document}